\documentstyle[12pt,epsf]{article}

\epsfverbosetrue

\newcommand{\ewdueup}[2]{\setlength{\epsfxsize}{#2}\epsfbox[10 30 640
  300]{#1}}
\newcommand{\ewxy}[2]{\setlength{\epsfxsize}{#2}\epsfbox[10 30 640
  590]{#1}}

\def\g{\gamma} \def\k{\kappa}  
   
\def\l{\left} \def\r{\right} \def\ord#1{{\cal O}\l(#1\r)}

 \def\kQ{\k_Q} \def\kQp{\k_{Q'}}

\def\Mb{M_{\Lambda_b}} \def\Mc{M_{\Lambda_c}}
\newcommand{\dl}{\stackrel{\leftarrow}{D}}
\newcommand{\dr}{\stackrel{\rightarrow}{D}} \def\tGm{\tilde\Gamma^\mu}
\newcommand{\plus}{\makebox[15pt][c]{$+$}}
\newcommand{\minus}{\makebox[15pt][c]{$-$}}
\newcommand{\ds}{\displaystyle} \newcommand{\be}{\begin{equation}}
  \newcommand{\nn}{\nonumber} \newcommand{\ee}{\end{equation}}
\newcommand{\bea}{\begin{eqnarray}} \newcommand{\ba}{\begin{array}}
    \newcommand{\ea}{\end{array}} \newcommand{\eea}{\end{eqnarray}}
\newcommand{\errr}[2]{\raisebox{0.08em}{\scriptsize
    {$\;\begin{array}{@{}l@{}}
        \plus\makebox[0.9em][r]{#1} \\[-0.12em]
        \minus\makebox[0.9em][r]{#2}
                       \end{array}$}}}

                 \newcommand{\err}[2]{\raisebox{0.08em}{\scriptsize
                     {$\;\begin{array}{@{}l@{}}
                         \plus\makebox[0.55em][r]{#1} \\[-0.12em]
                         \minus\makebox[0.55em][r]{#2}
                       \end{array}$}}}
                 \newcommand{\er}[2]{\raisebox{0.08em}{\scriptsize
                     {$\;\begin{array}{@{}l@{}}
                         \plus\makebox[0.15em][r]{#1} \\[-0.12em]
                         \minus\makebox[0.15em][r]{#2}
                       \end{array}$}}}

\newcommand{\one}{1\:\!\!\!{\rm{I}}}              
\newcommand{\ff}{form factors}
\newcommand{\pslash}{\rlap{p}{\kern0.1em\hbox{/}}}
\newcommand{\qslash}{\rlap{q}{\kern0.1em\hbox{/}}}

\begin{document}
\begin{titlepage}

\renewcommand{\thefootnote}{\fnsymbol{footnote}}

\begin{flushright}
Edinburgh 96/31\\
CPT-97/P.3459\\
UG-DFM-1/97\\
SHEP 97/14\\
hep-lat/9709028
\end{flushright}  
\vspace*{5mm}

\begin{center}
  {\Huge First Lattice Study\\
  of Semileptonic Decays of $\Lambda_b$ and $\Xi_b$ Baryons.}\\[10mm]
  {\large\it UKQCD Collaboration}\\[3mm]
  
{\bf  K.C.~Bowler$^1$, R.D.~Kenway$^1$, L.~Lellouch$^2$, 
J.~Nieves$^3$, O.~Oliveira$^1$~\footnote{Present address: 
Departamente de Fisica, Universidade de Coimbra, 3000 Coimbra, Portugal}, 
D.G.~Richards$^1$, C.T.~Sachrajda$^4$, N.~Stella$^4$
and P.~Ueberholz$^1$~\footnote{Present address: Department of 
Physics, University of Wuppertal, Wuppertal D-42097, Germany} }\\[2em]
$^1$Department of Physics \& Astronomy, The University of Edinburgh,
Edinburgh EH9~3JZ, UK\\[2mm]
$^2$Centre de Physique Th\'eorique\footnote{
Unit\'e Propre de Recherche 7061}, CNRS Luminy, Case 907
F-13288 Marseille Cedex 9, France\\[2mm]
$^3$Departamento de Fisica Moderna, Avenida Fuentenueva, 18071 Granada,
Spain\\[2mm]
$^4$Department of Physics and Astronomy, University of Southampton,
Highfield, Southampton SO17 1BJ, UK

\end{center}
\hfill
\pagebreak
\begin{abstract}
  We present the results of the first lattice study of 
  semileptonic decays of baryons containing a $b$-quark. Predictions
  for the decay distributions are given and the Isgur-Wise functions
  for heavy baryons are computed, for values of the velocity transfer
  up to about $\omega = 1.2$.  The computations are performed on a
  $24^3\times 48$ lattice at $\beta=6.2$ 
  using the Sheikholeslami-Wohlert action in
  the quenched approximation.
\end{abstract}

\vspace{2cm}

\setcounter{footnote}{0}
\renewcommand{\thefootnote}{\arabic{footnote}}

\end{titlepage}

\section{Introduction}

The discovery of the $\Lambda_b$ baryon at LEP~\cite{LEP} and the
observation of its semileptonic decay~\cite{LEPSEMIL} makes a study of
the weak interactions of heavy baryons on the lattice timely.  A
knowledge of the strong interaction effects in semileptonic decays is
necessary for the determination of the $V_{cb}$ element of the CKM
matrix from the experimentally measured rates and distributions. Up to
now $|V_{cb}|$ has been measured from the inclusive and exclusive
decays of heavy mesons.

In this paper we present the results of the first non-perturbative
computation of the semileptonic decays of heavy baryons performed
using lattice QCD, encouraged by our previous results on the
spectroscopy~\cite{noi}. The main purpose of this first study is to
establish whether such a calculation is feasible, and to identify the
principal sources of systematic uncertainties and statistical
fluctuations. Nevertheless, in spite of the exploratory character of
this investigation, we are able to determine the main features
exhibited by the six \ff\ which enter in the decay amplitude for the
processes $\Lambda_b \to \Lambda_c +l\bar{\nu} $ and $\Xi_b \to \Xi_c
+l\bar{\nu}$, and hence to extract a considerable amount of
phenomenologically interesting information.

Using Heavy Quark Symmetry it is possible to relate the six \ff\  to a
unique baryonic Isgur-Wise function, $\xi(\omega)$~\cite{tutto}, where
$\omega= v\cdot v'$, and $v$ and $v'$ are the four-velocities of the
initial and final state baryons. We compute $\xi(\omega)$, and by
studying the dependence of the form factors on the mass (or masses) of
the heavy quark, $m_Q$, we are able to use Heavy Quark Effective Theory
(HQET) to give an estimate of the ${\cal{O}}(1/m_Q)$ corrections to the
infinite-mass results (for a review and references to the
original literature see Ref.~\cite{NEUBERT}). The dependence of the
Isgur-Wise function on the masses of the light quarks is more uncertain
because of our limited data. In particular, we performed the
simulations at two values of the light-quark mass (a little larger and
a little smaller than that of the strange quark). Although the results
for these masses are very encouraging, a significant uncertainty is
introduced when the results are extrapolated, as a function of the
masses of the light quarks, to the chiral limit.  We present the
results for the Isgur-Wise function for massless light quarks (relevant
for the decay $\Lambda_b \to \Lambda_c +l\bar{\nu}$), as well as for the
case in which one light quark is massless and other is the strange
quark, relevant for the process $\Xi_b \to \Xi_c +l\bar{\nu}$. One of the
main goals of a future simulation will be to determine, in detail, the
dependence of $\xi(\omega)$ on the masses of the light quarks, and
hence to reduce the uncertainty due to the extrapolation of the results
to physical masses.

We obtain for the decay rate integrated over the
range $\omega\in[1,1.2]$
\bea 
\int_1^{1.2} d\omega\,\frac{d\Gamma}{d\omega}(\Lambda_b\to
\Lambda_c +l\bar{\nu}) &=& 1.4\er{5}{4}\ |V_{cb}|^2\, 10^{13}\,s^{-1}  \nn\\
\int_1^{1.2} d\omega\,\frac{d\Gamma}{d\omega}
(\Xi_b\to \Xi_c +l\bar{\nu}) &=& 1.6\er{4}{5}\ |V_{cb}|^2
\,10^{13}\,s^{-1} .
\eea
This range of $\omega$ corresponds to that for which we have the most
reliable results. We also obtain the slopes of the
differential-decay-rate form factors which are to be compared to the
slopes obtained by performing fits to experimental results for
$d\Gamma/d\omega$ versus $\omega$, for $\omega$ near 1, when such
results become available.  We find:
\be
\rho^2_{\cal B} = 1.1\pm 1.0
\ee
for $\Lambda_b\to\Lambda_c\ell\bar\nu$ decays
and
\be
\rho^2_{\cal B} = 1.4\pm 0.8
\ee
for $\Xi_b\to\Xi_c\ell\bar\nu$ decays. 
A more detailed discussion is presented in 
Section~\ref{subsec:decay}.

The plan of the paper is the following: in Section 2 we recapitulate the
theoretical framework for the definition of the baryonic Isgur-Wise
function. Section 3 gives details of the
simulation and Section 4 presents the results of the numerical
analysis of both two- and three-point functions.  We study the
dependence of the Isgur-Wise function on the velocity transfer and on
the masses of the heavy and light quarks in Section 5.  In Section 6,
we consider the phenomenological implications of our results for the
baryonic Isgur-Wise function and give estimates of differential and 
partially-integrated decay
rates.  The main body of the paper is accompanied by three appendices
where we lay out some lengthy parts of the calculation and of the
analysis.

For the reader who is not interested in the details of the computation
we have attempted to write Sections 2 (theoretical framework) and 6
(results and implications for phenomenology) in a self-contained way.

\section{Theoretical background}
\label{teo}

In this section we outline the theoretical framework needed to study
the semileptonic decays of $\Lambda_b$ and $\Xi_b$ baryons. The two
baryons differ in their light flavour content, but are identical in all
other quantum numbers and, in particular, they can be described in the
same manner within the framework of HQET.  For simplicity we will
only consider the $\Lambda_b$ decay.

The non-perturbative strong interaction effects in the exclusive
semileptonic decay of the $\Lambda_b$ are contained in the matrix
elements of the $V-A$ weak current, $J_{\mu}=\bar
c\gamma_{\mu}(1-\gamma_5) b$, which can be written in terms of six
invariant \ff , $f_i, g_i$ with $i=1,2,3$, as follows:
\bea
\langle \Lambda_c^{(s)}(p')| J_{\mu} | \Lambda_b^{(r)}(p)\rangle=
\bar{u}^{(s)}_c(p') & \left[  \gamma_{\mu}(f_1-\gamma_5 g_1)+
  i\sigma_{\mu\nu}q^{\nu}(f_2-\gamma_5 g_2) \right.\nonumber\\
& \left. + iq_{\mu}(f_3-\gamma_5
  g_3)\right]u^{(r)}_b(p). \label{FFQCD} 
\eea
In Eq.~(\ref{FFQCD}), the momenta and polarisations of the initial and
final baryons have been explicitly indicated.  The \ff\  are functions
of $q^2$, where $q$ is the 4-momentum transfer ($q=p'-p$).  The
decomposition above is convenient since, if the lepton masses are
neglected, only the dominant \ff\  $f_1$ and $g_1$ contribute to the
rate.  Since both the quarks destroyed and created by $J_\mu$ are
heavy, HQET provides a useful guide to the study of the
form factors.  In addition, the $\Lambda_b$ baryon has a particularly
simple structure in that it is composed of a heavy quark and light
degrees of freedom with zero total angular momentum, so that Heavy
Quark Symmetry has considerable predictive power. Expression
(\ref{FFQCD}) can be rewritten in terms of the velocity variables
$v_{\mu}=p_{\mu}/M_{\Lambda_b}$ and $v'_{\mu}=p'_{\mu}/M_{\Lambda_c}$
using a different set of form factors, which are functions of the
velocity transfer $\omega=v\cdot v'$:
\bea 
&\langle \Lambda_c^{(s)}(v')| \bar{c} \gamma_{\mu}(1-\gamma_5) b|
\Lambda_b^{(r)}(v)\rangle =\nn \\
& \bar{u}^{(s)}_c(v') \left[ \gamma_{\mu}(F_1-\gamma_5 G_1)+
  v_{\mu}(F_2-\gamma_5 G_2)+ v'_{\mu}(F_3-\gamma_5
  G_3)\right]u^{(r)}_b(v).
\label{FFHQET}
\eea

In principle, we do not need to use HQET at all but can evaluate
the matrix elements in Eq.~(\ref{FFQCD}) directly. However, as the inverse
lattice spacing in our calculation is about $2.9$~GeV, we cannot use
the physical $b$-quark mass in our computation.  Instead, we evaluate
the matrix element in Eq.~(\ref{FFQCD}) for a variety of ``$b$" and
``$c$" quark masses in the region of the physical charm quark mass,
and extrapolate the results to the physical values.  HQET provides
us with the theoretical framework to perform this extrapolation.  The
hadronic \ff\  $F_i, G_i$ can be expanded in inverse powers of the
heavy-quark masses; the non-perturbative QCD effects in the
coefficients of this expansion are
universal, mass-independent functions of the velocity transfer. This
is achieved by constructing an operator product expansion of the weak
currents. The analysis at leading order \cite{tutto} establishes the
important result that all the baryonic form factors are described by a
single universal function called the (baryonic) Isgur-Wise
function, $\xi(\omega,\mu)$, \footnote{We draw the reader's attention
to the fact that the Isgur-Wise function which describes baryonic
weak matrix elements is different from that entering in mesonic
transitions.  For simplicity we shall use the name Isgur-Wise
function to refer to the baryonic one, whenever this does not lead
to ambiguities. } 
\be \langle \Lambda_c^{(s)}(v')| \bar{c}
\gamma_{\nu}(1-\gamma_5) b | \Lambda_b^{(r)}(v)\rangle =
\xi(\omega,\mu)\bar{u}^{(s)}_c(v')\gamma_{\nu}(1-\gamma_5) u^{(r)}_b(v), 
\ee 
which is
normalized to the identity at $\omega=1$ and where $\mu$ is the scale
at which $\xi$ is renormalized.  The relation between the
six \ff\  and the renormalized Isgur-Wise function 
is given at leading order by the
expressions \cite{NEUBERT,Nrad} 
\be F_i(\omega) =\hat{C}_i(\omega) \xi^{\rm ren}(\omega),
\qquad G_i(\omega) =\hat{C}^5_i(\omega) \xi^{\rm ren}(\omega), 
\label{iwdef}
\ee
where the scale dependence of the Isgur-Wise function
$\xi(\omega,\mu)$ is reabsorbed into the definition of the short
distance coefficients $\hat{C}^{(5)}_i(\omega)$.  The coefficient
functions $\hat C_i^{(5)}$ are known up to order $\alpha_s^2(z\log
z)^n$, where $z=m_c/m_b$ is the ratio of the heavy-quark masses and
$n=0,1,2$.
The perturbative expansion of $\hat
C_1$ and $\hat C_1^5$ is of the form 
\be\hat C_1^{(5)} = 1 +
\delta_1^{(5)}(\omega) \alpha_s + ...  
\label{c1pertexp}
\ee
whereas the expansion of
$\hat C_{2,3}^{(5)}$ starts at ${\cal{O}}(\alpha_s)$.  The perturbative
corrections are rather small since the coefficients of $\alpha_s$ are
typically of order 1, for our range of masses and values of $\omega$,
so that $\hat C_{2,3}^{(5)}$ are much smaller than 
$\hat C_1^{(5)}$ ($\hat C_{2,3}^{(5)}\sim 0.2\,\hat C_1^{(5)}$).

We also attempt below to obtain some information on the $1/m_Q$
corrections to the \ff . In order to include these corrections it is
convenient to define a new function (see Ref.~\cite{NEUBERT}) 
\be
\hat\xi_{QQ'}(\omega)=\xi^{\rm ren} + \Big( \frac{\bar\Lambda}{2m_Q} +
\frac{\bar\Lambda}{2m_Q'}\Big)\left[ 2\chi^{\rm ren}(\omega) +
  \frac{\omega -1}{\omega+1}\xi^{\rm ren}(\omega)\right]
\label{newIW}
\ee 
which is also scale-independent and normalized at zero recoil,
since the function $\chi(\omega)$, arising from the higher-dimension
operators in the HQET Lagrangian, vanishes at $\omega=1$.

$\bar\Lambda$ in Eq.~(\ref{newIW}) is the binding energy of the heavy
quark in the corresponding $\Lambda$-baryon,
\be
\bar\Lambda=M_{\Lambda_b}-m_b = M_{\Lambda_c} -m_c
\ ,\label{Lambdabar}
\ee
up to $1/m_Q$ corrections.  The quark masses in
Eq.~(\ref{Lambdabar}) are generally taken to be pole masses, which
contain renormalon ambiguities of ${\cal{O}}(\Lambda_{QCD})$. For the
form factors, the ambiguities due to $\bar\Lambda$ are cancelled by
those arising from the higher-order terms in the perturbative series
for the coefficients $\hat C_i^{(5)}(\omega)$. In practice, we only
know the coefficients $\hat C_i^{(5)}(\omega)$ up to one-loop order,
and we implicitly assume that $\bar\Lambda$ is obtained from some
physical quantity with a similar precision.  As we note in 
Section~\ref{phenom}, some of the freedom of what to assign to the
coefficients and what to the power corrections cancels in the
prediction for the physical form factors obtained from those calculated
on the lattice.

$\hat\xi_{QQ'}$ is not a universal function and its dependence on the
flavour of the heavy quarks must be studied in detail (see 
Section~\ref{sec:hqdep}).
The relation between the new function $\hat\xi_{QQ'}(\omega)$ and the
form factors is given by
\bea
F_i(\omega) &=& N_i(\omega) \hat\xi_{QQ'}(\omega)+\ord{1/
m_{Q^{(')}}^2}, \nn\\
G_i(\omega) &=&N^5_i(\omega) \hat\xi_{QQ'}(\omega)+\ord{1/
m_{Q^{(')}}^2},
\label{doppione}
\eea
where the coefficients $N_i^{(5)}$ contain both radiative and $1/m_Q$
corrections. The exact expressions are~\cite{NEUBERT}  
\bea
N_1(\omega)&=&\hat C_1(\bar\omega)\left[ 1 + \frac{2}{\omega +1} 
\Big( \frac{\bar\Lambda}{2m_Q}+ \frac{\bar\Lambda}{2m_{Q'}}\Big)\right], \nn\\
N_2(\omega)&=&\hat C_2(\bar\omega)
\left(1+ \frac{2\omega}{\omega+1}\frac{\bar\Lambda}{2 m_Q}\right)-
\Big[\hat C_1(\bar\omega)+\hat C_3(\bar\omega)\Big]\frac{2}{\omega+1}
\frac{\bar\Lambda}{2m_{Q'}}\, , \nn\\
N_3(\omega)&=&\hat C_3(\bar\omega)
\left(1+ \frac{2\omega}{\omega+1}\frac{\bar\Lambda}{2 m_{Q'}}\right)-
\Big[\hat C_1(\bar\omega)+\hat C_2(\bar\omega)\Big]\frac{2}{\omega+1}
\frac{\bar\Lambda}{2m_Q}\, ,\nn\\ &&
\label{1-overm-exp}
\\
N_1^5(\omega)&=&\hat C_1^5(\bar\omega)\, ,\nn\\
N_2^5(\omega)&=&\hat C_2^5(\bar\omega)
\left(1+ \frac{2}{\omega+1}\frac{\bar\Lambda}{2 m_{Q'}}+
\frac{\bar{\Lambda}}{m_Q}\right)\nn\\
&-&\Big[\hat C_1^5(\bar\omega)+\hat C_3^5(\bar\omega)\Big]\frac{2}{\omega+1}
\frac{\bar\Lambda}{2{m_{Q'}}}\, ,\nn\\
N_3^5(\omega)&=&\hat C_3^5(\bar\omega)
\left(1+ \frac{2}{\omega+1}\frac{\bar\Lambda}{2 m_{Q}}+
\frac{\bar{\Lambda}}{m_{Q'}}\right)\nn\\
&+&
\Big[\hat C_1^5(\bar\omega)-\hat C_2^5(\bar\omega)\Big]\frac{2}{\omega+1}
\frac{\bar\Lambda}{2{m_Q}}\, .\nn
\eea
where the ``velocity transfer of the free quark" $\bar\omega$ is given by
\be
\bar\omega=\omega + (\omega-1)\left( \frac{\bar{\Lambda}}{m_Q} +
\frac{\bar{\Lambda}}{m_{Q'}} \right).
\ee
At zero recoil, Luke's theorem \cite{luke} protects the quantities
$\sum_i F_i(1)$ and $G_1(1)$ from ${\cal{O}}(1/m_Q)$ corrections:
\bea
\sum_i F_i(1)=\sum_i \hat C_i(1)+{\cal{O}}(1/m_Q^2) &\ \ \to\ \ &
 \frac{\sum_i F_i(1)}
{N_{\rm sum}(1)}=1 + {\cal{O}}(1/m_Q^2)
\nn \\
G_1(1)=\hat C_1^5(1) + {\cal{O}}(1/m_Q^2) &\ \ \to \ \ & 
\frac{G_1(1)}{N_1^5(1)}=1
+{\cal{O}}(1/m_Q^2).
\label{normcond}\eea
For degenerate quark masses, $m_Q={m_{Q'}}$, the coefficient 
\bea
N_{\rm sum}(\omega)=\sum_i N_i(\omega)=
\hat C_1(\bar\omega) 
&+&\hat C_2(\bar\omega)
\left(1+ \frac{2(\omega -1)}{\omega+1}\frac{\bar\Lambda}{2 m_Q}\right)
\nn\\
&+&\hat C_3(\bar\omega)
\left(1+ \frac{2(\omega -1)}{\omega+1}\frac{\bar\Lambda}{2 m_{Q'}}\right),
\label{Nsum}
\eea
equals one at zero recoil, as required by vector current conservation.

The coefficients (\ref{1-overm-exp}) and (\ref{Nsum}) depend on the
heavy-quark masses and on the value of the baryonic binding energy
$\bar\Lambda$.  Different choices for the definition and values of the
quark masses to be used in these expressions lead to differences which
are of ${\cal{O}}(\alpha_s^2)$ and ${\cal{O}}(1/m_Q^2)$, and hence are
formally of the size of the terms we are neglecting.  Moreover, we
find that different choices of the quark mass lead to negligible
differences in the form factors for $F_1$, $\sum_i F_i$ and $G_1$.
This is not the case for the form factors $F_2, F_3$ and $G_2, G_3$,
however, for which the coefficient functions are zero at tree level
and for which the ${\cal{O}}(1/m_Q)$ terms represent a major
contribution.  In these cases we take $m_Q=M_{\Lambda}-\bar\Lambda$,
and $\bar\Lambda$ is obtained from a fit of the theoretical prediction
for the \ff\ to the lattice results (see Section~\ref{sec:1overm}). In
all other cases, where the results are insensitive to the choice of
the quark mass, we take
\be
m_Q=\frac{a^{-1}}{4}(3M_{V}+M_{P})-200\ {\rm MeV}
\label{formulamass}
\ee
where $M_{V}$ and $M_{P}$ are the masses of the vector and pseudoscalar
heavy-light mesons, as measured very precisely from a previous simulation
performed  on the same set of configurations. $200$ MeV is an estimate of
the spin-averaged mesonic binding energy \cite{CG}.
For the lattice spacing, as discussed in greater detail in Section
\ref{lattice}, we take 
\be
a^{-1}=2.9\pm 0.2\ {\rm GeV}.
\label{eq:ainv}
\ee
The values of the quark masses obtained in this way are presented in 
 Table~\ref{tabquarkmass}, and the coefficients $N_i^{(5)}$ and $\hat C_i$
which are used in the rest of this paper are reported in 
Tables~\ref{tab:rad} and 
\ref{tab:rad1} for $\Lambda_{QCD}=250$ MeV and $n_f=0$ 
(as may be more appropriate for a quenched calculation). 

\begin{table}
\begin{center}
\begin{tabular}{c|ccc}
$\kappa_Q $&$\ \ \ \ \ M_P\,a\ \ \ \ \ \  $&$\ \ \ \ \ M_V\,a 
\ \ \ \ \ \ $&$\ \ \ \ \ m_Q$(GeV)\\
\hline
0.121 & $ 0.874\er{4}{3} $&$ 0.896\er{5}{4} $& 2.38 \\
0.125 & $ 0.773\er{3}{3} $&$ 0.799\er{4}{3} $& 2.10 \\
0.129 & $ 0.665\er{3}{3} $&$ 0.696\er{4}{4} $& 1.80 \\
0.133 &  $ 0.547\er{3}{3} $&$ 0.588\er{4}{5} $& 1.48 \\
\end{tabular}
\end{center}
\caption{\em Values of the quark masses in physical units, for 
\protect$a^{-1}=2.9\protect$ GeV and \protect$\bar\Lambda=200\protect$ MeV, 
as obtained from the
pseudoscalar and vector meson masses in lattice units. ($\kappa_Q$
is the heavy quark's hopping parameter.) \label{tabquarkmass}}
\end{table}

\begin{table}
\begin{center}
\begin{tabular}{c|c|c|cccc}
$\kappa_Q\to\kappa_{Q'}$ &$\omega$&$\bar\omega$&$N_1^5(\omega)=
\hat{C}_1^{5}(\bar\omega)$
&$\hat{C}_2(\bar\omega)$&$\hat{C}_1(\bar\omega)$&$N_{\rm sum}(\omega)$\\
\hline
$0.121\to 0.121$& 1.0& 1.0& 0.961 &0.019 &0.961 &1.0\\
$0.121\to 0.125$& 1.0& 1.0& 0.965 &0.024 &0.965 &1.01 \\
$0.121\to 0.129$& 1.0& 1.0& 0.971 &0.03 &0.971 &1.02 \\
$0.121\to 0.133$& 1.0& 1.0& 0.98 &0.04 &0.98 &1.04 \\
\hline
$0.125\to 0.121$& 1.0& 1.0& 0.956 &0.016 &0.956 &0.992 \\
$0.125\to 0.125$& 1.0& 1.0& 0.959 &0.02 &0.959 &1.0\\
$0.125\to 0.129$& 1.0& 1.0& 0.965 &0.027 &0.965 &1.01 \\
$0.125\to 0.133$& 1.0& 1.0& 0.973 &0.037 &0.973 &1.03 \\
\hline
$0.129\to 0.121$& 1.0& 1.0& 0.949 &0.011 &0.949 &0.982 \\
$0.129\to 0.125$& 1.0& 1.0& 0.952 &0.016 &0.952 &0.989 \\
$0.129\to 0.129$& 1.0& 1.0& 0.957 &0.022 &0.957 &1.0\\
$0.129\to 0.133$& 1.0& 1.0& 0.964 &0.031 &0.964 &1.02 \\
\hline
$0.133\to 0.121$& 1.0& 1.0& 0.938 &0.0044 &0.938 &0.967 \\
$0.133\to 0.125$& 1.0& 1.0& 0.941 &0.0086 &0.941 &0.973 \\
$0.133\to 0.129$& 1.0& 1.0& 0.945 &0.015 &0.945 &0.984 \\
$0.133\to 0.133$& 1.0& 1.0& 0.952 &0.024 &0.952 &1.0\\
\hline
$0.121\to 0.121$& 1.1& 1.15 & 0.932 &0.018 &0.932 &0.968 \\
$0.121\to 0.125$& 1.1& 1.15 & 0.937 &0.022 &0.937 &0.977 \\
$0.121\to 0.129$& 1.1& 1.16 & 0.944 &0.028 &0.944 &0.99 \\
$0.121\to 0.133$& 1.1& 1.17 & 0.955 &0.037 &0.955 &1.01 \\
\hline
$0.125\to 0.121$& 1.1& 1.15 & 0.927 &0.014 &0.927 &0.96 \\
$0.125\to 0.125$& 1.1& 1.16 & 0.931 &0.018 &0.931 &0.969 \\
$0.125\to 0.129$& 1.1& 1.16 & 0.938 &0.024 &0.938 &0.981 \\
$0.125\to 0.133$& 1.1& 1.17 & 0.948 &0.033 &0.948 &1.0\\
\hline
$0.129\to 0.121$& 1.1& 1.16 & 0.92 &0.01 &0.92 &0.95 \\
$0.129\to 0.125$& 1.1& 1.16 & 0.924 &0.014 &0.924 &0.958 \\
$0.129\to 0.129$& 1.1& 1.17 & 0.93 &0.02 &0.93 &0.97 \\
$0.129\to 0.133$& 1.1& 1.17 & 0.94 &0.028 &0.94 &0.99 \\
\hline
$0.133\to 0.121$& 1.1& 1.17 & 0.91 &0.0039 &0.91 &0.935 \\
$0.133\to 0.125$& 1.1& 1.17 & 0.913 &0.0077 &0.913 &0.943 \\
$0.133\to 0.129$& 1.1& 1.17 & 0.919 &0.013 &0.919 &0.954 \\
$0.133\to 0.133$& 1.1& 1.18 & 0.929 &0.021 &0.929 &0.973
\end{tabular}
\end{center}
\caption{\em Radiative and \protect${\cal{O}}(1/m_Q)\protect$ 
correction factors for \protect$\omega=1.0\protect$ and 
\protect$ 1.1\protect$. ($\kappa_Q$ and $\kappa_{Q'}$ are the
initial and final heavy quark hopping parameters.) \label{tab:rad}}
\end{table}

\begin{table}
\begin{center}
\begin{tabular}{c|c|c|cccc}
$\kappa_Q\to\kappa_{Q'}$ &$\omega$&$\bar\omega$&$N_1^5(\omega)=
\hat{C}_1^{5}(\bar\omega)$
&$\hat{C}_2(\bar\omega)$&$\hat{C}_1(\bar\omega)$&$N_{\rm sum}(\omega)$\\
\hline
$0.121\to 0.121$& 1.2& 1.3 & 0.904 &0.016 &0.904 &0.937 \\
$0.121\to 0.125$& 1.2& 1.31 & 0.909 &0.02 &0.909 &0.946 \\
$0.121\to 0.129$& 1.2& 1.32 & 0.917 &0.025 &0.917 &0.96 \\
$0.121\to 0.133$& 1.2& 1.33 & 0.929 &0.034 &0.929 &0.981 \\
\hline
$0.125\to 0.121$& 1.2& 1.31 & 0.898 &0.013 &0.898 &0.929 \\
$0.125\to 0.125$& 1.2& 1.31 & 0.903 &0.017 &0.903 &0.938 \\
$0.125\to 0.129$& 1.2& 1.32 & 0.911 &0.022 &0.911 &0.951 \\
$0.125\to 0.133$& 1.2& 1.34 & 0.923 &0.03 &0.923 &0.972 \\
\hline
$0.129\to 0.121$& 1.2& 1.32 & 0.891 &0.0091 &0.891 &0.919 \\
$0.129\to 0.125$& 1.2& 1.32 & 0.896 &0.013 &0.896 &0.928 \\
$0.129\to 0.129$& 1.2& 1.33 & 0.903 &0.018 &0.903 &0.941 \\
$0.129\to 0.133$& 1.2& 1.35 & 0.915 &0.026 &0.915 &0.961 \\
\hline
$0.133\to 0.121$& 1.2& 1.33 & 0.881 &0.0034 &0.881 &0.905 \\
$0.133\to 0.125$& 1.2& 1.34 & 0.886 &0.0069 &0.886 &0.913 \\
$0.133\to 0.129$& 1.2& 1.35 & 0.893 &0.012 &0.893 &0.925 \\
$0.133\to 0.133$& 1.2& 1.36 & 0.904 &0.019 &0.904 &0.945 \\
\hline
$0.121\to 0.121$& 1.3& 1.45 & 0.876 &0.015 &0.876 &0.907 \\
$0.121\to 0.125$& 1.3& 1.46 & 0.882 &0.018 &0.882 &0.916 \\
$0.121\to 0.129$& 1.3& 1.48 & 0.89 &0.023 &0.89 &0.93 \\
$0.121\to 0.133$& 1.3& 1.5 & 0.903 &0.031 &0.903 &0.951 \\
\hline
$0.125\to 0.121$& 1.3& 1.46 & 0.871 &0.012 &0.871 &0.899 \\
$0.125\to 0.125$& 1.3& 1.47 & 0.876 &0.015 &0.876 &0.909 \\
$0.125\to 0.129$& 1.3& 1.49 & 0.884 &0.02 &0.884 &0.922 \\
$0.125\to 0.133$& 1.3& 1.51 & 0.897 &0.028 &0.897 &0.943 \\
\hline
$0.129\to 0.121$& 1.3& 1.48 & 0.864 &0.0083 &0.864 &0.89 \\
$0.129\to 0.125$& 1.3& 1.49 & 0.869 &0.012 &0.869 &0.898 \\
$0.129\to 0.129$& 1.3& 1.5 & 0.877 &0.016 &0.877 &0.911 \\
$0.129\to 0.133$& 1.3& 1.52 & 0.889 &0.024 &0.889 &0.932 \\
\hline
$0.133\to 0.121$& 1.3& 1.5 & 0.854 &0.003 &0.854 &0.875 \\
$0.133\to 0.125$& 1.3& 1.51 & 0.859 &0.0062 &0.859 &0.884 \\
$0.133\to 0.129$& 1.3& 1.52 & 0.866 &0.011 &0.866 &0.896 \\
$0.133\to 0.133$& 1.3& 1.54 & 0.878 &0.018 &0.878 &0.916
\end{tabular}
\end{center}
\caption{\em Radiative and \protect${\cal{O}}(1/m_Q)\protect$ 
correction factors for 
\protect$\omega= 1.2\protect$ and $1.3$. ($\kappa_Q$ and $\kappa_{Q'}$ are the
initial and final heavy quark hopping parameters.) 
\label{tab:rad1}}
\end{table}

Below, we will proceed as follows. Since $N_1^5$ and $N_{\rm sum}$ are
close to unity and the \ff\ $G_1(\omega)$ and $\sum_i F_i(\omega)$ are
insensitive to the value of $\bar\Lambda$, we can use the measured
values of these \ff\ to determine the Isgur-Wise function reliably.
For the individual vector \ff\ $F_1, F_2$ and $F_3$ (and similarly for
$G_2$ and $G_3$) the situation is different since the ${\cal{O}}(1/m_Q)$
corrections are significant. From these \ff , using the Isgur-Wise
function already obtained, we determine the $\bar\Lambda$ parameter.
Finally we use the combined results to determine the \ff\ for the
physical $b\to c$ decays.


\section{Details of the simulation}
\label{lattice}

Our calculation is performed using 60 $SU(3)$ gauge field configurations
generated on a $24^3\times48$ lattice at $\beta=6.2$, using the hybrid
over-relaxed algorithm described in Ref.~\cite{a_units}. 
Since we are studying the decays of quarks whose masses are large in
lattice units, we must control discretisation errors.
In order to reduce these errors, we use an $\ord{a}$-improved
fermion action
originally proposed by Sheikholeslami and Wohlert\cite{SW}, given
by
\be S_F^{SW} = S_F^W - i\frac{\kappa}{2}\sum_{x,\mu,\nu}\bar{q}(x)
         F_{\mu\nu}(x)\sigma_{\mu\nu}q(x),
\label{sclover}
\ee
where $S_F^W$ is the Wilson action:
\bea
S_F^W = \sum _x \Biggl(\bar{q}(x)q(x)\Biggr.
             &-&\kappa\sum _\mu\Bigl[
            \bar{q}(x)(1- \gamma _\mu )U_\mu (x) q(x+\hat\mu )
\Bigr.\nn\\
&+&\Bigl.\Biggl.\ \bar{q}(x+\hat\mu )(1 + \gamma _\mu )
   U^\dagger _\mu(x)
            q(x)\Bigr]\Biggr)
\label{sfw}\ .
\eea
The use of the SW action reduces discretisation errors from
${\cal{O}}(ma)$ to ${\cal{O}}(\alpha_s\,ma)$  
\cite{SW,HEATLIE} provided one also uses
``improved'' operators, for example, those  obtained 
by ``rotating'' the field of the
heavy quark, Q:
\be
Q(x)\longrightarrow (1-\frac{1}{2}\gamma\cdot\dr)\,Q(x)\ .
\ee
Thus, to obtain an $\ord{a}$-improved evaluation of the matrix element
in  Eq.~(\ref{FFHQET}), we use a ``rotated'' improved current
\be
J^\mu\equiv\bar Q'(x)\tGm\,Q(x)
\label{improp}
\ ,
\ee
where
\be
\tGm=(1+\frac{1}{2}\g\cdot\dl)\,(1-\gamma^{\mu})\gamma_5\,
(1-\frac{1}{2}\gamma\cdot\dr).
\label{gamimprop}
\ee

The gauge field configurations and the light quark propagators were
generated on the 64-node i860 Meiko Computing Surface at the University
of Edinburgh.  The heavy quark propagators were computed using the
Cray T3D, also at Edinburgh.

In order to enhance the signal for the baryon correlation functions,
the light and heavy quark propagators have been computed using the
Jacobi smearing method \cite{SMEARING}, at both the sink and the source
(SS).  Since smearing is not a Lorentz-invariant operation, it alters
some of the transformation properties of computed quantities. In a
previous publication \cite{noi}, we have shown that such an effect is
evident in  two-point baryonic correlators at non-zero momentum.  In
Appendix \ref{app1}, we include a study of the smearing effects  for SS
three-point functions.

Statistical errors are obtained from a bootstrap
procedure~\cite{EFRON}.  This involves the creation of 1000 bootstrap
samples from the original set of 60 configurations by randomly
selecting 60 configurations per sample (with replacement). Statistical
errors are then obtained from the central 68\% of the corresponding
bootstrap distributions as detailed in Ref.~\cite{qlhms}.


\subsection{Correlators and details of the analysis}

In order to study semileptonic decays of the type $\Lambda_b
\rightarrow \Lambda_c \,l \,\bar{\nu}_l$ on the lattice, we consider
the following three-point correlators:
\be
(C(t_x,t_y))^{Q\rightarrow Q'}_{\mu}=
\sum_{\vec{x}}\sum_{\vec{y}} e^{-i\vec{p} '\cdot\vec{x}}
e^{-i\vec{q}\cdot\vec{y}} \langle {{\cal O}}^{Q'}(x)
(J_{\mu}(y))^{Q\rightarrow Q'} \bar{{\cal
 O}}^{{Q}}(0) \rangle  \label{main3p} 
\ee
where the spinorial indices are implicit.
The operator 
\be
{\cal O}^Q(x)=\epsilon_{abc}(l_1^{aT}{\cal C}\gamma_5 l_2^{b})Q^c 
\label{oqdef}\ee
is the interpolating operator for the $\Lambda$ baryon, and the
current which mediates the decay of a heavy quark $Q$ into a second
heavy quark $Q'$ is given in Eq.~(\ref{improp}). $l_1$ and $l_2$ represent
light quark fields. 
The three-point function $(C(t_x,t_y))_\mu^{Q\to Q'}$ can be
written in terms of quark propagators, as
\be 
(C(t_x,t_y))_\mu^{Q\to Q'} = -\langle \sum_{\vec
  y}(\Sigma^{c'd}(0,y;t_x,\vec p)_{Q'}
\Gamma_{\mu}S^{dc'}_{Q}(y,0))e^{-i\vec q\cdot\vec y}\rangle, 
\ee
where $S^{dc}_{Q}(y,0)$ is the propagator of the $Q$ quark from $y$
to the origin in the presence of a background field configuration,
\be
 \Sigma^{cd}(0,y;t_x,\vec p)_{Q'}=\epsilon_{abc}\epsilon_{a'b'c'}
 \sum_{\vec{x}} e^{-i\vec{p}'\cdot\vec{x}}
 {\rm Tr} \Big[ S_{l_1}^{aa'\ T}(x,0){\cal C}\gamma_5
 S_{l_2}^{bb'}(x,0)\gamma_5{\cal C}\Big] S_{Q'}^{cd}(x,y) 
 \label{ext}
\ee
where $T$ represents the transpose in spinor-space,
$\langle ..\rangle$ denotes the average over gluon configurations,
and $a,b,c,d,a',b',c'$ are colour indices.  $S_{l_1}$ and $S_{l_2}$
are the propagators of the two light quarks.  The extended propagator
(\ref{ext}) can be evaluated using the standard source method reviewed
in Ref.~\cite{sourcemethod}.
 
In the limit of large $t_x$ and $t_x-t_y$ (in the forward part of the
lattice), where the ground state contributions to the
correlation function should dominate, we can rewrite the correlator as
follows~\footnote{This and the following expressions are only correct
  in the case in which local operators are used. The actual case of
  smeared-smeared correlators, which is discussed in Appendix
  \ref{app1}, is more complicated but conceptually similar.}
\be
 (C(t_x,t_y))_\mu^{Q\to Q'} = \frac{Z Z'}{4E E'} e^{-E'(t_x-t_y)} e^{-Et_y}
 \left( (\pslash' + M') {\cal F}_{\mu}^{Q',Q}
 (\pslash + M)  \right)
 \label{local},
\ee
where
\be
 \ba{rclrcl}
 E&=& \sqrt{M^{2} + |{\vec p|}^{2}}
       & E' &=& \sqrt{M^{'\, 2} + |\vec p - \vec q|^2}  \\
 M &=& M_{\Lambda(Q)}    & M' &=& M_{\Lambda(Q')} \\
 p_{\mu} &=& (E,\vec{p}) &\qquad  p'_{\mu} &=&(E',\vec{p}-\vec{q})  \\
 Z &=& Z(\Lambda(Q),|\vec{p}|) \qquad & Z' 
&=& Z(\Lambda(Q'),|\vec{p}-\vec{q}|) .
\ea
\label{allthedefinition}
\ee
The weak matrix element can be written in terms of the six lattice \ff
,
\be
 \langle \Lambda_{Q'}^{(r)}(\vec{p}^{\,\,\prime}) | 
 (J_{\mu}(0,\vec{0}))^{Q\rightarrow Q'}| 
 \Lambda_{Q}^{(s)}(\vec{p}) \rangle =
 \bar{u}_{Q'}^{(r)}(\vec{p}^{\,\,\prime}){\cal F}_{\mu}^{Q',Q}(p',p) 
 u_{Q}^{(s)}(\vec{p})
\ee
\bea
{\rm with} \qquad
 {\cal F}_{\mu}^{Q',Q}(p',p) &=& 
\left(F^L_1(\omega)\gamma_{\mu} + 
 F^L_2 (\omega) v_{\mu} + F^L_3 (\omega) v_{\mu}^{\prime}\right) \nonumber \\
 &- &  \left(G^L_1(\omega)\gamma_{\mu} + 
 G^L_2 (\omega) v_{\mu} + G^L_3 (\omega) v_{\mu}^{\prime}\right)\gamma_5 
 \label{3p}
\eea
where $u_{Q(Q')}^{(s),(r)}$ are the spinors of the heavy baryons and
$s$ and $r$ are helicity indices. The lattice \ff\  $F^L_i$ and $G^L_i$
are related to the physical \ff\  $F_i$ and $G_i$ through the
renormalisation constants of the lattice vector and axial current,
respectively. This will be discussed in some detail below.

\section{Details of the numerical analysis}
\label{ana}

In this section we describe our procedure for the extraction of the
form factors from the lattice correlation functions.

\subsection{Analysis of two-point correlation functions at non-zero momentum}
\label{sec:2pt}

The wave-function factors $Z,Z'$ and the energies which appear in
Eq.~(\ref{local}) can be obtained from the analysis of the
appropriate two-point correlation functions:
\begin{equation}
C^Q_2(\vec{p},t)=
\sum_{\vec{x}} e^{-i\vec{p}\cdot\vec{x}}\langle{\cal O}^Q(\vec{x},t) 
\overline{{\cal O}}^Q(\vec{0},0)\rangle.
\label{defC2}
\end{equation}
This correlator was evaluated for four values of the heavy-quark
hopping parameter corresponding to masses around that of the charm
quark, and for three combinations of the light-quark masses, as shown in
Table~(\ref{tab:Q_masses}). We have computed the correlator for
momenta up to $|\vec{p}|=2p_{\rm min}$, where $p_{\rm
  min}=2\pi/La=\pi/12a$ is the minimum non-zero momentum allowed on
our lattice. For the analysis, equivalent momenta have been averaged
to reduce the statistical fluctuations.

\begin{table}
\begin{center}
\begin{tabular}{c|cccc}
 $\kappa_{l1}/\kappa_{l2}$&\multicolumn{4}{c}{$\kappa_Q$}\\ \hline
 0.14144/0.14144 & 0.121 & 0.125 & 0.129 &0.133 \\
 0.14144/0.14226 & - &  - & 0.129 &  -  \\ 
 0.14226/0.14226 &  - &-  & 0.129 &  -  
 \end{tabular}
 \end{center}
\caption{\em Quark hopping parameter 
combinations used in the calculation of baryon
two-point functions. ($\kappa_{l1}$ and $\kappa_{l2}$ are the
two light-quark hopping parameters and $\kappa_Q$, that of the
heavy quark.)\label{tab:Q_masses}}
\end{table}

For the actual case of smeared-smeared operators (see Appendix
\ref{app1}), for large time separations and imposing antiperiodic
boundary conditions in time, the correlator $C_2^Q$ becomes
\bea
&C_2^Q(t,\vec{p})=
\nn\\
&Z_s^2(|\vec{p}|) \l\{e^{-Et} \left[
\frac{E+M-\alpha^2(E-M)}{4E}\one +\frac{E+M+\alpha^2(E-M)}{4E}\gamma_0 -
\frac{2\alpha}{4E}\vec{p}\cdot\vec{\gamma} \right] \nn \right.\\
&- e^{-E(T-t)} \left.\left[
\frac{E+M-\alpha^2(E-M)}{4E}\one -\frac{E+M+\alpha^2(E-M)}{4E}\gamma_0 -
\frac{2\alpha}{4E}\vec{p}\cdot\vec{\gamma} \right]\right\},
\label{SS}
\eea
where $p^{\mu}$ is defined in (\ref{allthedefinition}),
$\tilde{p}^{\mu} = (E,-\vec{p})$ is the 4-momentum of the antibaryon
propagating in the backward part of the lattice, and $Z_s, \alpha$ are the 
amplitudes of the smeared operator (defined in Eqs.~(\ref{J2s}) and 
(\ref{redefinition})\,).

At zero momentum, the analysis is particularly simple since the only
non-zero components are independent of the unphysical amplitude
$\alpha$. Details of this case are given in Ref.~\cite{noi}.  At
finite momentum, we have used both the diagonal and the off-diagonal
components of the spinorial matrix to extract the energy from the
exponential fall-off. The values of the amplitudes $Z_s$ and $\alpha$
were obtained by fitting separately the contributions proportional to
the identity, to $\gamma_0$ and to $\gamma_i$, and by taking suitable
linear combinations of the overall factors, as explained in the
appendix of Ref.~\cite{noi}. The results of these fits, for those mass
combinations which are relevant to the present study, and for momenta
up to $|\vec{p}|=\sqrt{2}\pi/12a$ are reported in
Tables~\ref{tabenergy} and \ref{tabzeta}.

\begin{table}\begin{center}\begin{tabular}{ll|ccc}
$\kappa_{l1}/\kappa_{l2}$ &$ \kappa_Q$ &$\vec{p}=(0,0,0)$&$\vec{p}=
(\frac{2\pi}{La},0,0)$
 &$\vec{p}=(\frac{2\pi}{La},\frac{2\pi}{La},0)$\\ \hline
$0.14144/0.14144 $&$ 0.121 $&$ 1.138\er{7}{7} $&$  1.167\er{9}{9}$&$
  1.192\err{13}{12}$\\
                  &$ 0.125 $&$ 1.040\er{6}{6}$&$  1.072\er{9}{9}$&$
  1.099\err{13}{12}$\\
                  &$ 0.129 $&$ 0.938\er{6}{6}$&$  0.973\er{8}{7}$&$
  1.002\err{13}{13}$\\
                  &$ 0.133 $&$ 0.829\er{6}{6}$&$  0.868\er{8}{7}$&$
  0.901\err{14}{13}$\\ \hline
$0.14144/0.14226 $&$ 0.129 $&$ 0.910\er{8}{7}$&$  0.943\er{9}{9}$&$
  0.972\err{12}{14}$\\ \hline
$0.14226/0.14226 $&$ 0.129 $&$ 0.876\er{9}{8}$&$  0.914\err{10}{10}$&$
0.948\err{17}{17}$\\ \hline
${\rm chiral}/{\rm chiral}$&$ 0.129 $&$0.807\err{15}{10} $&$ $&$ $\\ \hline
${\rm chiral}/{\rm strange}$&$ 0.129 $&$0.853\err{14}{6} $&$ $&$ $
\end{tabular}
\end{center}
\caption{\em Energies of the $\Lambda$-baryon in lattice units 
for all the momenta and quark hopping parameters
relevant to the present study.\label{tabenergy}}
\end{table}

\begin{table}
\begin{center}
\begin{tabular}{c|c|cc|cc}
& \multicolumn{1}{c|}{$\vec{p}=(0,0,0)$}
& \multicolumn{2}{c|}{$\vec{p}=(p_{\rm min},0,0)$}
& \multicolumn{2}{c}{$\vec{p}=(p_{\rm min},p_{\rm min},0)$}\\
\hline
$\kappa_Q$ &$Z_s^2\times 10^4$
&$Z_s^2\times 10^4$&$ \alpha $&$Z_s^2\times 10^4$&$ \alpha $\\
\hline
\multicolumn{6}{c}{$(\kappa_{l1},\kappa_{l2})=(0.14144,0.14144)$}\\
\hline
0.121&$4.44\err{48}{38}$ 
&$2.85\err{40}{32}$&$0.66\er{7}{7}$&$1.96\err{42}{32}$&$0.56\er{9}{8}$\\
0.125&$4.41\err{48}{37}$ 
&$2.86\err{40}{32}$&$0.70\er{7}{7}$&$1.95\err{40}{32}$&$0.60\err{10}{8}$\\
0.129&$4.35\err{45}{36}$ 
&$2.84\err{35}{30}$&$0.77\er{7}{7}$&$1.94\err{40}{32}$&$0.66\err{11}{9}$\\
0.133&$4.17\err{41}{35}$ 
&$2.75\err{34}{28}$&$0.83\er{7}{8}$&$1.87\err{40}{31}$&$0.73\err{13}{11}$\\
\hline
\multicolumn{6}{c}{$(\kappa_{l1},\kappa_{l2})=(0.14144,0.14226)$}\\
\hline
0.129& $4.02\err{42}{36}$ 
&$2.56\err{36}{29}$&$ 0.76\er{8}{8}$&$1.68\err{39}{30}$&$ 0.68\err{13}{11}$\\
\hline
\multicolumn{6}{c}{$(\kappa_{l1},\kappa_{l2})=(0.14226,0.14226)$}\\
\hline
0.129&$3.77\err{87}{75}$ 
&$2.42\err{38}{32}$&$0.71\er{9}{8}$&$1.60\err{45}{33}$&$0.65\err{15}{12}$
\end{tabular}
\end{center}
\caption{\em Amplitudes $Z_s$  and $\alpha$ obtained from the analysis of the
finite momentum two-point functions.\label{tabzeta}}
\end{table}

The case corresponding to $\kappa_{l1}=\kappa_{l2}=0.14144$ is further
studied in detail to check the precision with which 
the dispersion relations are satisfied as the momentum of the baryon is
increased.
It is nowadays customary \cite{slope,semilape} to replace the
continuum dispersion relation (cdr)
\be
a^2E^2=a^2m^2 +p^2a^2
\label{cdr}
\ee
with the so-called  lattice dispersion relation (ldr)
\be
a^2E^2=a^2m^2 +\sin^2(pa),
\label{ldr}
\ee
which is suggested by the form of the free fermionic propagator on a
discrete lattice. For the heavy-quark masses in Table~\ref{tabenergy}
the two dispersion relations yield essentially indistinguishable
results for the energy at $|\vec{p}|=p_{\rm min}$. In addition, the
theoretical predictions coincide with the measured values, confirming
that the systematic effects at this low value of the momentum are
negligible. For momentum $|\vec{p}|=\sqrt{2}p_{\rm min}$, we note that
the predicted value is about $1\%$ larger than the measured one,
although always compatible within one sigma.  
Given this result, it seems that the correction obtained with the lattice
dispersion relation goes in the right direction.  However, much more
precise data are needed to draw a firm conclusion on this issue.

Finally, for the conversion of our values for masses and energies into
physical units we need an estimate of the inverse lattice spacing in GeV.
Following Ref.~\cite{noi}, we use the value given in Eq.~(\ref{eq:ainv}).
The error in Eq.~(\ref{eq:ainv}) is large enough to encompass all
our estimates for $a^{-1}$ from quantities such as $m_\rho$, $f_\pi$,
$m_N$, the  string tension $\sqrt{K}$ and the hadronic scale $R_0$ 
discussed in \cite{Sommer}. 


\subsection{Three-point functions and lattice form factors}

In this subsection we explain our procedure for extracting the
form factors from the computed three-point (and two-point) correlation
functions. We have computed the three-point functions for the mass
combinations tabulated in Table~\ref{tab:Q_masses3pt}. In order to study
the dependence of the form factors on the masses of the heavy quarks,
we have computed the correlation functions for all combinations of
$\kQ$ and $\kQp$ taken from 0.121, 0.125, 0.129, 0.133, but with the
light-quark masses fixed by $\kappa_{l1}=\kappa_{l2}=0.14144$ (which
is close to that of the strange quark).  On the other hand, the
dependence on the light-quark masses was studied by keeping fixed the
mass of the initial and final heavy quarks
$\kappa_{Q}=\kappa_{Q'}=0.129$ which is very close to that of the
charm quark, and considering the three light hopping-parameter
combinations $\kappa_{l1}=\kappa_{l2}=0.14144$, $\kappa_{l1}=0.14226,
\kappa_{l2}=0.14144$, and $\kappa_{l1}=\kappa_{l2}=0.14226$. In 
light of the encouraging results obtained below, we envisage the
possibility of repeating the calculation on a larger sample of heavy-
and light-quark masses.
  
\begin{table}
\begin{center}
\begin{tabular}{c|cccc}
$\kappa_{l1}/\kappa_{l2}$&
\multicolumn{4}{c}{$\kappa_Q\ \ \to \ \ \kappa_{Q'}$}\\ \hline
\multicolumn{5}{c}{degenerate transitions}\\
\hline
0.14144/ &&&\\
0.14144 &
 $ 0.121\to 0.121 $&$ 0.125\to 0.125$ &$ 0.129\to 0.129 $&$0.133 \to 0.133 $\\
\hline
0.14144/ &&&\\
0.14226 &  - & $0.129\to 0.129$ & - & -\\
\hline
0.14226/ &&&\\
0.14226 &  - & $0.129 \to 0.129$ & - & - \\
\hline
\multicolumn{5}{c}{non-degenerate transitions}\\
\hline
$0.14144/$ & &$ 0.121\to 0.125$ &$ 0.121\to 0.129 $&$0.121 \to 0.133 $\\
$0.14144$  &$ 0.125\to 0.121 $&-&$ 0.125\to 0.129 $&$0.135 \to 0.133 $\\
&$ 0.129\to 0.121 $&$ 0.129\to 0.125$ &-&$0.133 \to 0.133 $\\ 
&$ 0.133\to 0.121 $&$ 0.133\to 0.125$ &$ 0.133\to 0.129 $&-
\end{tabular}
\end{center}
\caption{\em Quark hopping parameter 
combinations used in the calculation of baryon
three-point functions.
($\kappa_{l1}$ and $\kappa_{l2}$ are the
two light-quark hopping parameters while $\kappa_Q$ and $\kappa_{Q'}$
are those of the initial and final
heavy quarks.)\label{tab:Q_masses3pt}}
\end{table}
 
We measure the six \ff\  for the different quark masses from the
three-point functions (\ref{main3p}), whose expression, for large
values of $t_x$ and $t_x-t_y$ in the forward part of the lattice
and for local interpolating operators was given in Eq.~(\ref{local}). The
more complicated case of SS correlators, which is explained in detail
in Appendix \ref{app1}, can be written schematically as follows:
\bea
(C(t))^{Q\rightarrow Q'}_{\mu}&=
&\frac{ZZ'}{16E E'}e^{-E'(T/2-t)}e^{-Et} \times 
\left[
{\cal V}_{\mu}(\alpha,\beta,M,M',\vec{p},\vec{p}^{\;\prime})\ F^L_1(\omega) 
\right.\nn\\
&+&{\cal W}(\alpha,\beta,M,M',\vec{p},\vec{p}^{\;\prime})
\Big(v_{\mu}F^L_2(\omega)+v'_{\mu}F^L_3(\omega)\Big) \nn\\
&-&{\cal A}_{\mu}(\alpha,\beta,M,M',\vec{p},\vec{p}^{\;\prime})
\ G^L_1(\omega) \nn\\
&+&\left.{\cal B}(\alpha,\beta,M,M',\vec{p},\vec{p}^{\;\prime})
\Big(v_{\mu}G^L_2(\omega)+v'_{\mu}G^L_3(\omega)\Big)  \right]
\label{main3pL}
\eea
where we have set $t_x=T/2$ (which is the value we chose in our
computations), and have renamed the remaining time variable $t_y\to
t$.  The matrices ${\cal V,W,A,B}$ are different linear combinations
of the Dirac matrices, with coefficients which depend on the arguments
shown.  The overall normalisation factor, as well as the four matrices
${\cal V,W,A,B}$ can be fully reconstructed from the two-point
functions at the corresponding values of the masses and momenta, as
illustrated in Section \ref{sec:2pt}.

We will now consider separately the two cases in which we keep either
the axial or vector currents from the $V-A$ current which mediates the
semileptonic weak decays of heavy baryons. The following discussion
is based on equations~(\ref{SSF}) and (\ref{SSB}) in
Appendix~\ref{app1}.


\subsubsection{Analysis of the axial \ff }
\label{sec:anaA}

The freedom to choose the Lorentz index $\mu$ and the spinor components
appropriately allows us to extract the required form factors
efficiently~\footnote{Alternatively we could use suitable projection
operators in spinor space.}.
It is convenient to think of the four-by-four spinorial matrix as
subdivided into four two-by-two matrices, as explained in Appendix
\ref{app2}. We find that

\begin{enumerate}
\item
{\em current indices $\mu=i$ with $i=1,2,3$}:
the large components\footnote{Here and in the following, we refer to
  the numerical coefficients of the form factors in
  Eq.~(\ref{main3pL}) as {\em large} if they are proportional to the
  energy or mass of the baryon, and as small if they are proportional
  to the spatial momentum $p$.} of ${\cal A}_{i}$ are located in the
top (bottom) diagonal submatrix in the forward (backward) part of the
lattice. The large components of ${\cal B}$ are
located in the top-right (bottom-left) corner submatrix. Thus the 
contributions of the form factor $G_1^L$ can be separated from that
of $G_{2,3}^L$.
\item
{\em current index $\mu=0$}:
the large components of both ${\cal A}_0$ and ${\cal B}$ are located
in the top-right (bottom-left) corner matrices in the
forward (backward) part of the lattice. So the three \ff\  contribute to
the same spinorial components, making the extraction of the \ff\  very
uncertain.  Therefore, the equations obtained from the current with
index $\mu=0$ will not be considered in the analysis below.
\end{enumerate}

We conclude that from an analysis of the correlators with index
$\mu=1,2,3$, it is possible to obtain a clean straightforward
determination of the \ff .  We consider the asymptotic form of
Eq.~(\ref{main3pL}), fitting for either $G^L_1$ or $G^L_{2,3}$
separately, depending on the particular spinorial component under
study.

\begin{figure}
\begin{picture}(140,80)
\put(-7,0){\ewdueup{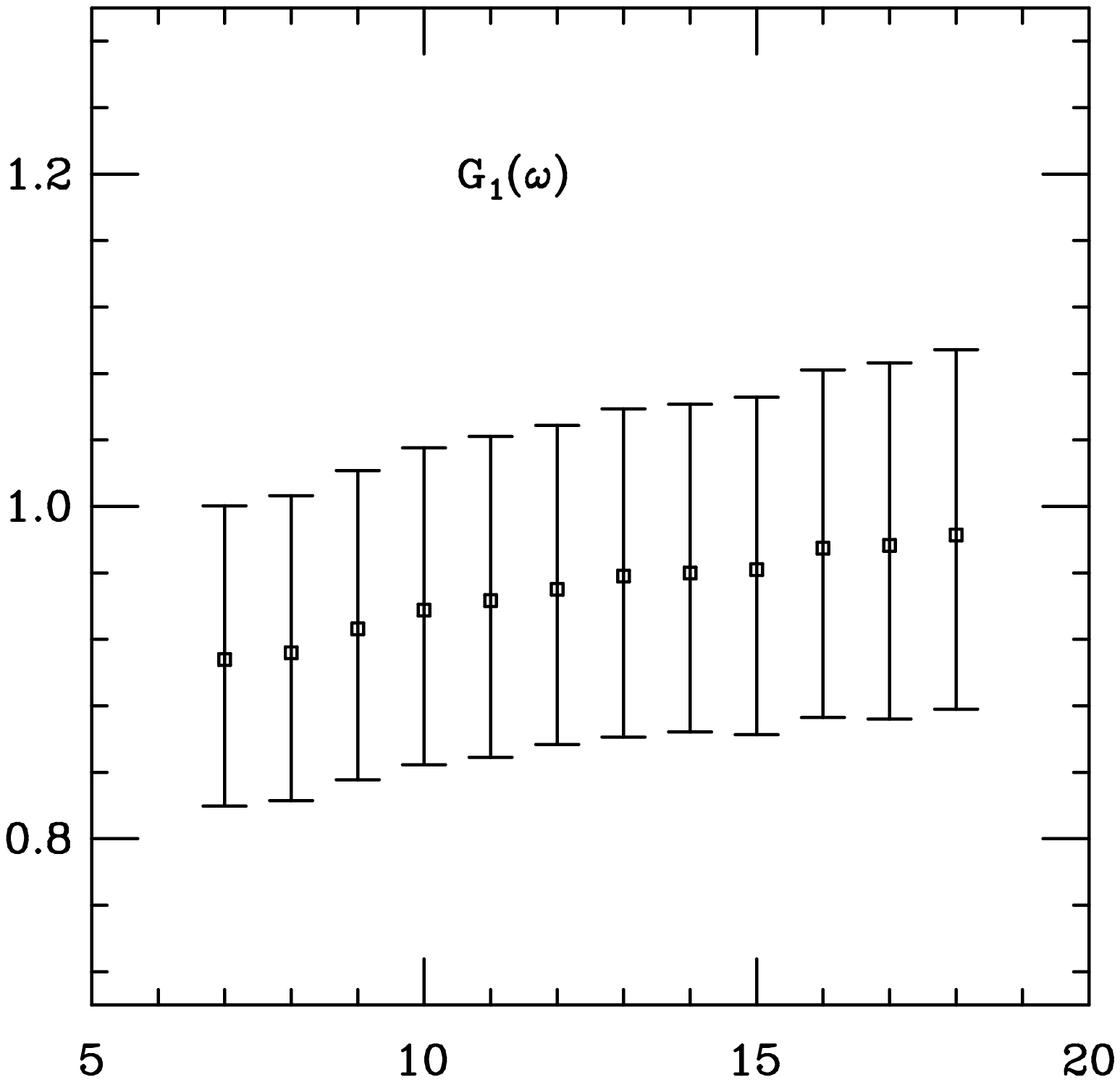}{90mm}}
\put(33,0){{\bf A}}
\put(63,0){\ewdueup{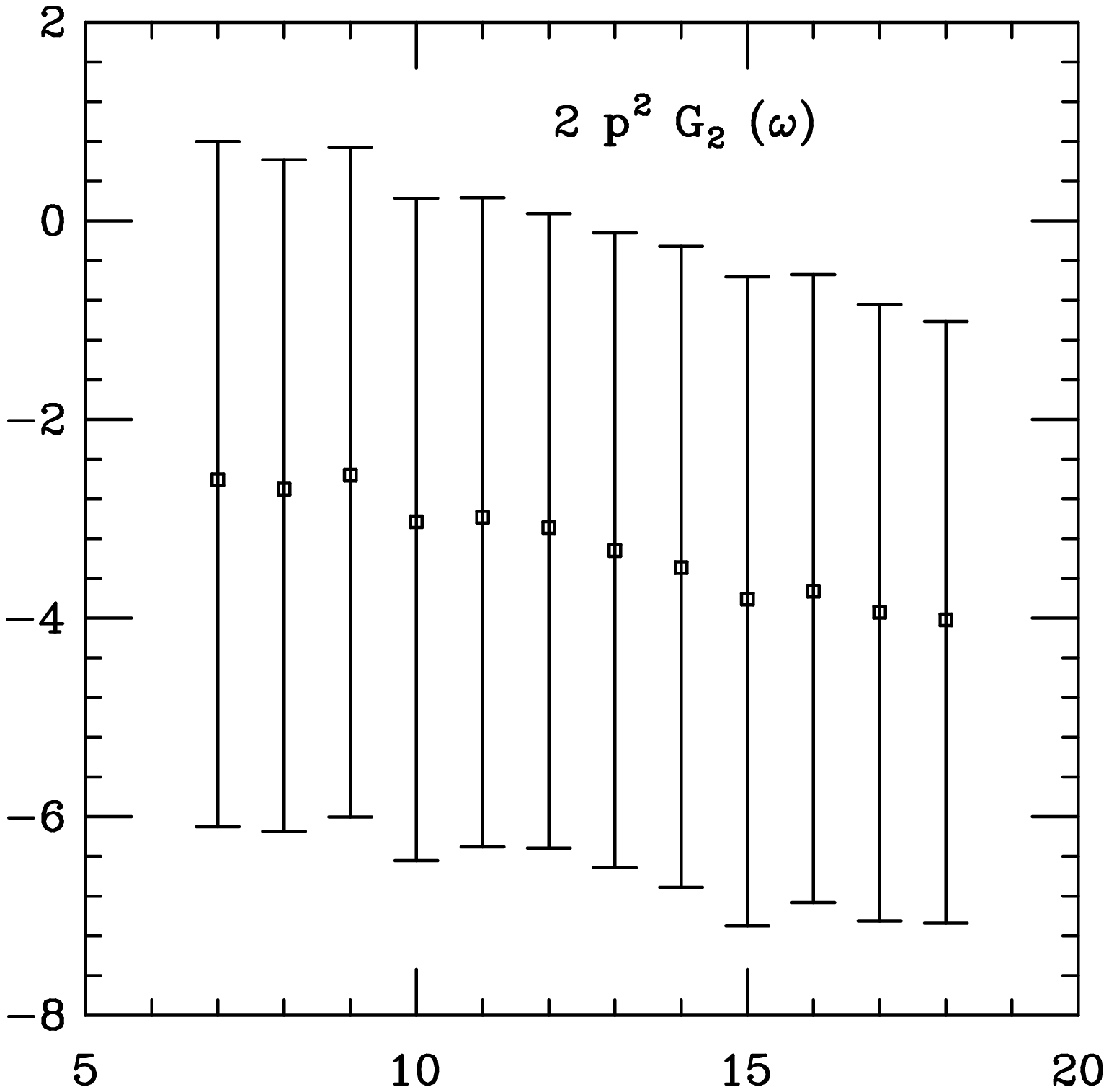}{90mm}}
\put(103,0){{\bf B}}
\end{picture}
\begin{picture}(140,80)
\put(-7,0){\ewdueup{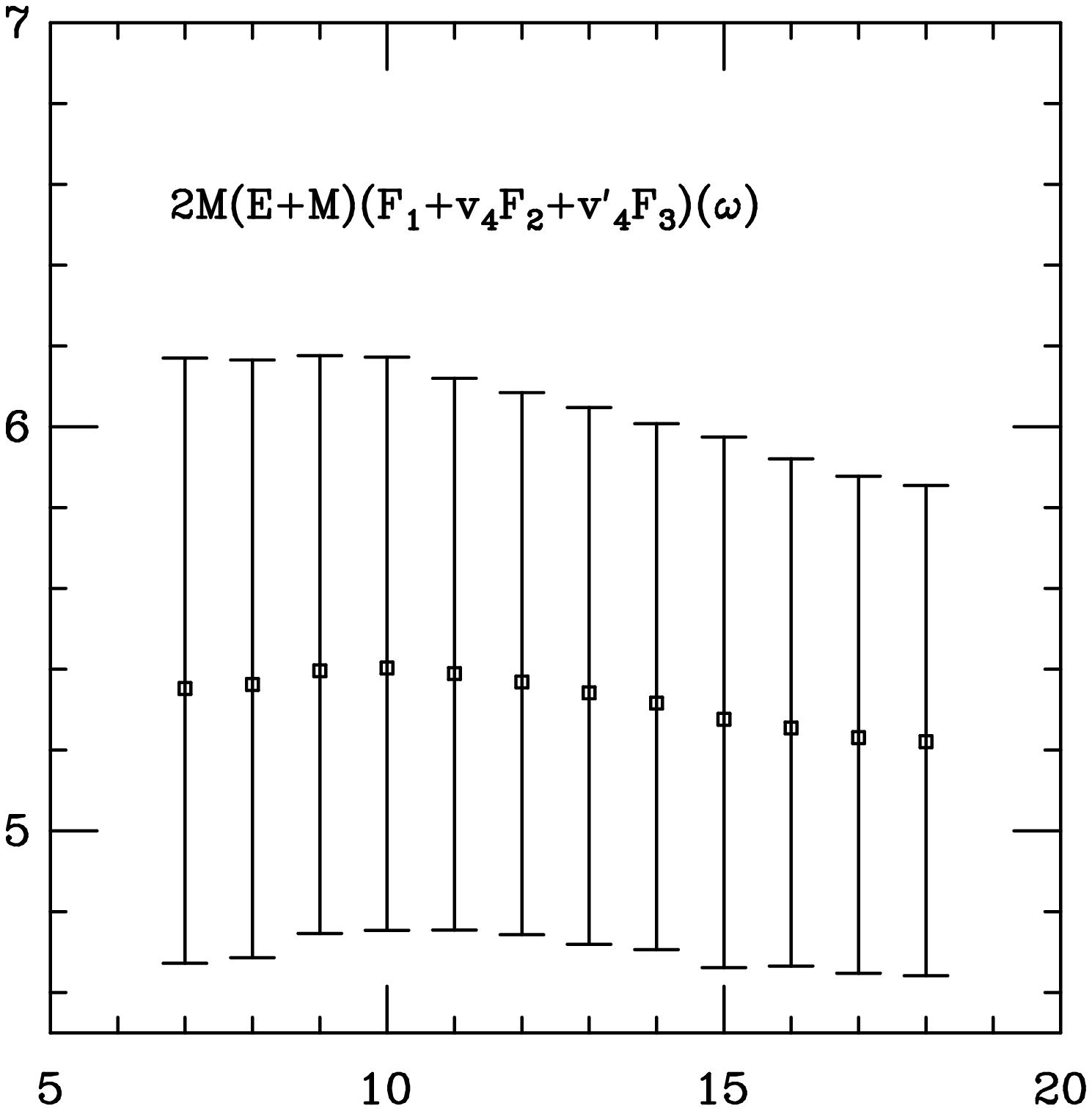}{90mm}}
\put(33,0){{\bf C}}
\put(63,0){\ewdueup{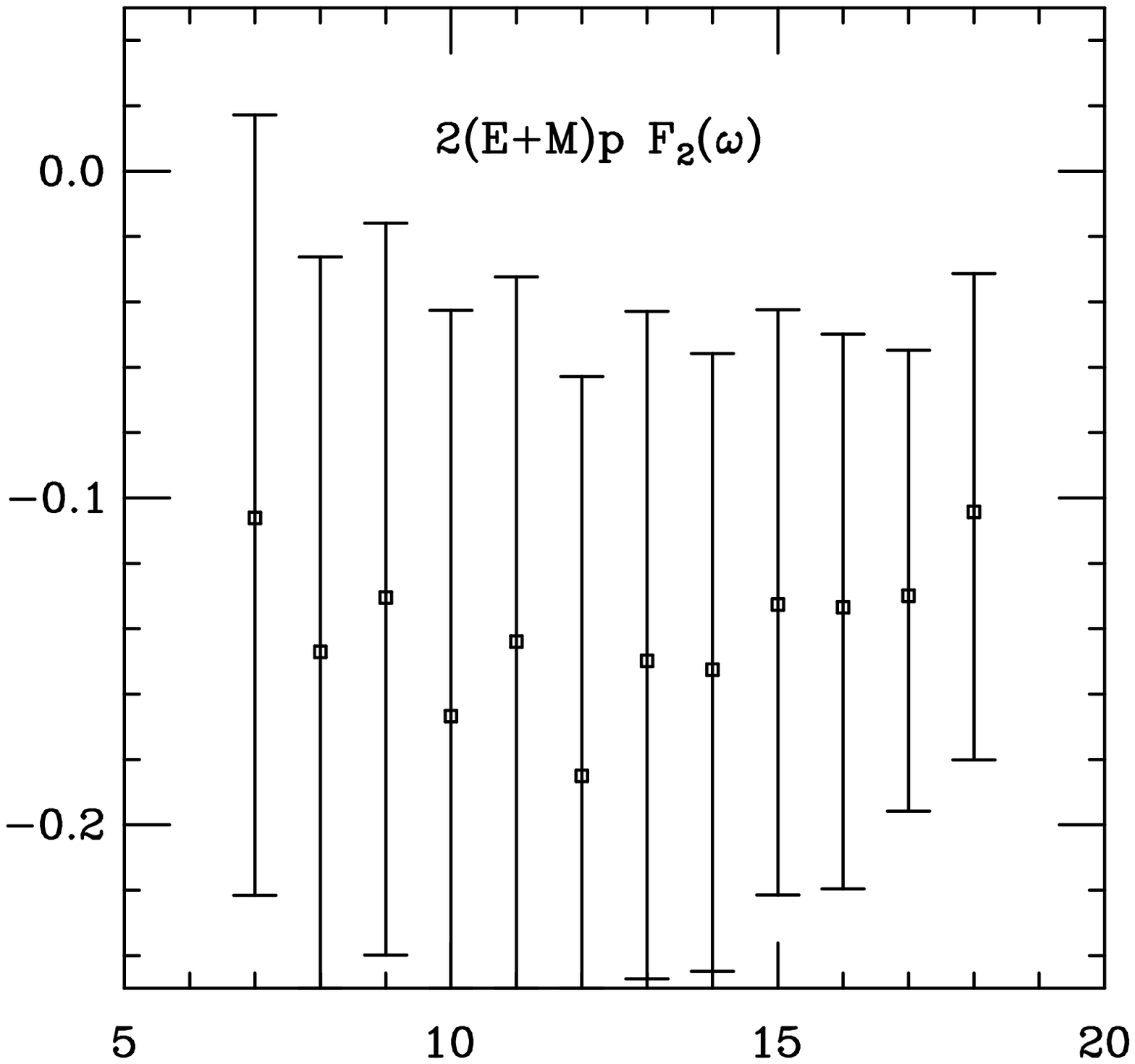}{90mm}}
\put(103,0){{\bf D}}
\end{picture}
\caption{\em Examples of the plateaux used in the fit to (from left to
right, top to bottom) \protect$G_1(\omega)\protect$, \protect$
G_2(\omega)\protect$, \protect$\sum_i F_i(\omega)\protect$ and
\protect$ F_2(\omega) \protect$.
\protect$\kappa_{l1}=\kappa_{l2}=0.14144\protect$,
\protect$\kappa_Q=\kappa_{Q'}=0.121\protect$, the initial particle is
at rest and the final particle has momentum $(p', 0, 0)$, where
$p'=2\pi/La$.}
\label{figplat}
\end{figure}

Once the time-dependent factor in (\ref{main3pL}) has been divided
out, we observe long and stable plateaux, centered around $t=12$.  As
an example we exhibit in Figures~\ref{figplat}A and \ref{figplat}B,
the plateaux for the form factors $G^L_1$ and $G^L_2$, for one set of
masses and choice of momenta.  We have fitted $G^L_1$ for 3, 5 and 7
timeslices.  The central values of the fits are insensitive to the
choice of fitting interval, and in order to avoid
possible contaminations due to exited states, we have decided to
restrict the fitting range to three time slices centered around
$t=12$. The $\chi^2_{\rm dof}$ for these fits were always very
reasonable, ranging from $\chi^2_{\rm dof}\sim 0.5$ to $\chi^2_{\rm
  dof}\sim 2$.

As was discussed in Section \ref{teo}, the \ff\  are related to the
physical renormalized Isgur-Wise function through a multiplicative
renormalisation which takes into account the short-distance QCD
corrections. Furthermore an additional renormalisation constant
($Z_A$) must be introduced in order to relate the lattice improved
axial current (\ref{gamimprop}) to the continuum one.  Perturbative
and non-perturbative calculations of $Z_A$ are available in the
literature \cite{GMPHS,jonivar}, but it can also be estimated
non-perturbatively in this computation.  As can be seen from
Eqs.~(\ref{normcond}), $G_1$ reduces to the Isgur-Wise
function, at zero recoil, and it is free from ${\cal{O}}(1/m_Q)$ corrections
because of Luke's theorem.  If we make the reasonable assumption that
${\cal{O}}(1/m_Q^2)$ corrections are small, a measure of $G^L_1(1)/N^5_1(1)$
will give us a non-perturbative estimate of $Z_A^{-1}$. For degenerate 
transitions,
\be
\ba{cccc}
\ds{\frac{N^5_1(1)}{G^L_1(1)}}=0.98\er{8}{8} &{\rm at}\ \ &\kappa_{l1}=
\kappa_{l2}=0.14144, {\rm and}\ \kappa_h=0.133 \\
&&&\\
\ds{\frac{N^5_1(1)}{G^L_1(1)}}=0.97\er{8}{8} &{\rm at}\ \ &\kappa_{l1}=
\kappa_{l2}=0.14144, {\rm and}\ \kappa_h=0.129 \\
&&&\\
\ds{\frac{N^5_1(1)}{G^L_1(1)}}=0.98\er{9}{8} &{\rm at}\ \ &\kappa_{l1}=
\kappa_{l2}=0.14144, {\rm and}\ \kappa_h=0.125 \\
&&&\\
\ds{\frac{N^5_1(1)}{G^L_1(1)}}=0.98\err{10}{9} &{\rm at}\ \ &\kappa_{l1}=
\kappa_{l2}=0.14144, {\rm and}\ \kappa_h=0.121 \, ,
\ea
\label{zainv}
\ee
to be compared with the non-perturbative estimate of 
$Z_A$, obtained for light quarks \cite{jonivar}
\be
Z_A^{\rm non-pert} = 1.04\er{1}{1} .
\label{zabest}
\ee
The coefficients $N^5_1$ used in Eqs.~(\ref{zainv}) were computed
from expressions (\ref{1-overm-exp}). Within the statistical precision
of our calculation, we have no evidence of discretisation errors
nor of 
${\cal{O}}(1/m_Q^2)$ corrections (assuming that these two effects do not 
partially cancel, 
which is very unlikely give the range of masses we consider). Indeed,
we observe that the four values in (\ref{zainv}) are
exceptionally stable with the quark mass.

In Tables~\ref{tabG}, \ref{tabGG} and \ref{tabG_1}, we present our 
results, for the quantity
\be
\hat\xi_{QQ'}(\omega)=\frac{G^L_1(\omega)}{G^L_1(1)}\times 
\frac{N_1^5(1)}{N^5_1(\omega)}
\label{G1norm}
\ee
for all the quark masses
and for initial and final momenta up to $|\vec{p}|,|\vec{p}^{\,\,\prime}|=
\sqrt{2}$.
It follows from the above discussion that this quantity is independent of
the lattice renormalisation constant $Z_A$.

\begin{table}
\begin{center}
\begin{tabular}{l|cc|cc|cc|cc}
\multicolumn{9}{c}{$\kappa_{l1}=0.14144,\ \ \ \ \
\kappa_{l2}=0.14144 $}\\
\hline
$\vec{p},\vec{p}^{\,\,\prime}$&\multicolumn{2}{c|}{$0.121\to
0.121$}&\multicolumn{2}{c|}{$0.121\to
0.125$}&\multicolumn{2}{c|}{$0.121\to
0.129$}&\multicolumn{2}{c}{$0.121\to 0.133$}\\ $[p_{\rm min}]$
&$\omega$&$\hat\xi_{QQ'}$&$\omega$&$\hat\xi_{QQ'}$&$\omega$
&$\hat\xi_{QQ'}$&$\omega$&$\hat\xi_{QQ'}$
\\
\hline
&&&&&&&&\\[-9.5pt]$\!{\parbox{1.5cm}{(0,0,0),
(1,0,0)}}\!$&$1.026$&$0.97\!\!\er{3}{3}$&$1.030$
&$0.96\!\!\er{3}{3}$&$1.037$&$0.94\!\!\er{4}{3}$&$1.048$&$0.91\!\!\er{4}{4}$\\
&&&&&&&&\\[-9.5pt]$\!{\parbox{1.5cm}{(0,0,0),
(1,1,0)}}\!$&$1.050$&$0.95\!\!\er{6}{8}$&$1.060$
&$0.93\!\!\er{7}{8}$&$1.073$&$0.91\!\!\er{8}{7}$&$1.09$&$0.90\!\!\er{7}{8}$\\
&&&&&&&&\\[-9.5pt]$\!{\parbox{1.5cm}{(1,0,0),
(0,1,0)}}\!$&$1.052$&$0.94\!\!\err{11}{12}$&$1.057$
&$0.92\!\!\err{10}{10}$&$1.064$&$0.91\!\!\err{12}{9}$
&$1.07$&$0.89\!\!\err{10}{8}$\\
&&&&&&&&\\[-9.5pt]$\!{\parbox{1.5cm}{(1,0,0),
(1,0,0)}}\!$&$1.000$&$1.02\!\!\err{10}{11}$&$1.000$
&$0.95\!\!\err{11}{12}$&$1.001$&$0.94\!\!\err{12}{12}$
&$1.003$&$0.93\!\!\err{12}{12}$\\
&&&&&&&&\\[-9.5pt]$\!{\parbox{1.5cm}{(1,0,0),
(-1,0,0)}}\!$&$1.10$&$0.81\!\!\er{9}{8}$&$1.11$&$0.76\!\!\er{9}{8}$
&$1.13$&$0.74\!\!\er{9}{7}$&$1.15$&$0.70\!\!\er{8}{7}$\\
&&&&&&&&\\[-9.5pt]$\!{\parbox{1.5cm}{(1,0,0), (0,0,0)}}\!$ &-&-
&$1.026$&$0.95\!\!\er{4}{5}$&$1.026$&$0.95\!\!\er{5}{4}$&$1.026$
&$0.95\!\!\er{5}{4}$\\
&&&&&&&&\\[-9.5pt]$\!{\parbox{1.5cm}{(1,0,0),
(0,1,1)}}\!$&$1.077$&$0.95\!\!\er{2}{2}$&$1.09$
&$0.92\!\!\err{17}{16}$&$1.10$&$0.88\!\!\err{18}{15}$
&$1.12$&$0.80\!\!\err{17}{15}$
\end{tabular}
\end{center}
\caption{\em Estimates of the function $\hat\xi_{QQ'}$ as
obtained from the axial form factor $G_1$. 
All the transitions, corresponding to initial heavy $\kappa=0.121$ and $0.125$
and initial and final momenta up to 
\protect$ |\protect\vec{p}\,|,|\protect\vec{p}^{\;\protect\prime}|=
\protect\sqrt{2} p_{\rm min}\protect$ 
are shown.
Statistical errors in $\omega$ are in the last digit or beyond.}\label{tabG}
\end{table}

\begin{center}
\begin{tabular}{l|cc|cc|cc|cc}
\multicolumn{9}{c}{$\kappa_{l1}=0.14144,\ \ \ \ \
\kappa_{l2}=0.14144 $}\\
\hline
$\vec{p},\vec{p}^{\,\,\prime}$&\multicolumn{2}{c|}{$0.125\to 0.125$}
&\multicolumn{2}{c|}{$0.125\to 0.121$}
&\multicolumn{2}{c|}{$0.125\to 0.129$}&\multicolumn{2}{c}{$0.125\to 0.133$}\\
$[p_{\rm min}]$   &$\omega$&$\hat\xi_{QQ'}$&$\omega$&$\hat\xi_{QQ'}$
&$\omega$&$\hat\xi_{QQ'}$&$\omega$&$\hat\xi_{QQ'}$ \\
\hline
&&&&&&&&\\[-9.5pt]$\!{\parbox{1.5cm}{(0,0,0), (1,0,0)}}\!$&$1.030$
&$0.95\!\!\er{3}{3}$&$1.026$&$0.99\!\!\er{2}{3}$&$1.037$
&$0.94\!\!\er{4}{3}$&$1.048$&$0.91\!\!\err{5}{4}$\\
&&&&&&&&\\[-9.5pt]$\!{\parbox{1.5cm}{(0,0,0), (1,1,0)}}\!$
&$1.060$&$0.95\!\!\er{6}{8}$&$1.050$&$0.96\!\!\er{6}{8}$
&$1.073$&$0.93\!\!\er{8}{7}$&$1.09$&$0.89\!\!\er{8}{7}$\\
&&&&&&&&\\[-9.5pt]$\!{\parbox{1.5cm}{(1,0,0), (0,1,0)}}\!$
&$1.062$&$0.91\!\!\err{10}{10}$&$1.057$&$0.92\!\!\err{10}{10}$
&$1.069$&$0.89\!\!\err{11}{8}$&$1.08$&$0.87\!\!\err{10}{8}$\\
&&&&&&&&\\[-9.5pt]$\!{\parbox{1.5cm}{(1,0,0), (1,0,0)}}\!$
&$1.000$&$1.02\!\!\err{10}{11}$&$1.000$&$0.97\!\!\err{11}{12}$
&$1.000$&$0.95\!\!\err{13}{11}$&$1.002$&$0.94\!\!\err{13}{12}$\\
&&&&&&&&\\[-9.5pt]$\!{\parbox{1.5cm}{(1,0,0), (-1,0,0)}}\!$
&$1.12$&$0.76\!\!\er{8}{8}$&$1.11$&$0.75\!\!\err{10}{8}$
&$1.14$&$0.71\!\!\er{9}{6}$&$1.16$&$0.67\!\!\er{8}{6}$\\
&&&&&&&&\\[-9.5pt]$\!{\parbox{1.5cm}{(1,0,0), (0,0,0)}}\!$
& -&-                       &$1.030$&$0.94\!\!\er{5}{5}$
&$1.030$&$0.94\!\!\er{6}{4}$&$1.030$&$0.94\!\!\er{5}{4}$\\
&&&&&&&&\\[-9.5pt]$\!{\parbox{1.5cm}{(1,0,0), (0,1,1)}}\!$
&$1.09$&$0.90\!\!\err{17}{16}$&$1.08$&$0.91\!\!\err{17}{16}$
&$1.11$&$0.85\!\!\err{18}{15}$&$1.13$&$0.78\!\!\err{17}{14}$\\
\multicolumn{9}{c}{ }\\
\multicolumn{9}{c}{ }\\
\multicolumn{9}{c}{Table \ref{tabG}: {\em (continued)}}
\end{tabular}
\end{center}

\begin{table}
\begin{center}
\begin{tabular}{l|cc|cc|cc|cc}
\multicolumn{9}{c}{$\kappa_{l1}=0.14144,\ \ \ \ \ \kappa_{l2}=0.14144 $}\\
\hline
$\vec{p},\vec{p}^{\,\,\prime} $&\multicolumn{2}{c|}{$0.133\to 0.133$}
&\multicolumn{2}{c|}{$0.133\to 0.121$}
&\multicolumn{2}{c|}{$0.133\to 0.125$}&\multicolumn{2}{c}{$0.133\to 0.129$}\\
$[p_{\rm min}]$    &$\omega$&$\hat\xi_{QQ'}$
&$\omega$&$\hat\xi_{QQ'}$&$\omega$&$\hat\xi_{QQ'}$&$\omega$&$\hat\xi_{QQ'}$ \\
\hline
&&&&&&&&\\[-9.5pt]$\!{\parbox{1.5cm}{(0,0,0), (1,0,0)}}\!$
&$1.048$&$0.91\!\!\er{3}{3}$&$1.026$&$1.02\!\!\er{2}{3}$
&$1.030$&$1.00\!\!\er{3}{3}$&$1.037$&$0.96\!\!\er{3}{3}$\\
&&&&&&&&\\[-9.5pt]$\!{\parbox{1.5cm}{(0,0,0), (1,1,0)}}\!$
&$1.09$&$0.82\!\!\er{6}{8}$&$1.050$&$0.96\!\!\er{7}{7}$
&$1.08$&$0.97\!\!\er{7}{7}$&$1.073$&$0.93\!\!\er{6}{7}$\\
&&&&&&&&\\[-9.5pt]$\!{\parbox{1.5cm}{(1,0,0), (0,1,0)}}\!$
&$1.10$&$0.82\!\!\er{9}{8}$&$1.074$&$0.88\!\!\err{10}{9}$
&$1.07$&$0.87\!\!\err{10}{8}$&$1.09$&$0.85\!\!\er{9}{8}$\\
&&&&&&&&\\[-9.5pt]$\!{\parbox{1.5cm}{(1,0,0), (1,0,0)}}\!$
&$1.000$&$1.02\!\!\err{12}{13}$&$1.003$&$0.98\!\!\err{12}{11}$
&$1.002$&$0.98\!\!\err{13}{12}$&$1.001$&$0.97\!\!\err{12}{13}$\\
&&&&&&&&\\[-9.5pt]$\!{\parbox{1.5cm}{(1,0,0), (-1,0,0)}}\!$
&$1.20$&$0.57\!\!\er{6}{6}$&$1.15$&$0.69\!\!\err{10}{7}$
&$1.16$&$0.67\!\!\er{9}{7}$&$1.173$&$0.63\!\!\er{7}{6}$\\
&&&&&&&&\\[-9.5pt]$\!{\parbox{1.5cm}{(1,0,0), (0,0,0)}}\!$
&- &-                       &$1.048$&$0.90\!\!\er{6}{6}$
&$1.048$&$0.91\!\!\er{6}{5}$&$1.048$
&$0.92\!\!\er{5}{5}$\\
&&&&&&&&\\[-9.5pt]$\!{\parbox{1.5cm}{(1,0,0), (0,1,1)}}\!$
&$1.15$&$0.71\!\!\err{14}{13}$&$1.100$&$0.83\!\!\err{16}{14}$
&$1.111$&$0.80\!\!\err{16}{13}$&$1.13$&$0.76\!\!\err{15}{14}$
\end{tabular}
\end{center}
\caption{\em Estimates of the function $\hat\xi_{QQ'}$ as
obtained from the axial form factor $G_1$. 
All the transitions, corresponding to initial heavy $\kappa=0.133$ and $0.129$
and initial and final momenta up to 
\protect$ |\protect\vec{p}\, |,|\protect\vec{p}^{\;\protect\prime}|=
\protect\sqrt{2} p_{\rm min}\protect$ 
are shown.
Statistical errors in $\omega$ are in the last digit or beyond.}\label{tabGG}
\end{table}

\begin{center}
\begin{tabular}{l|cc|cc|cc|cc}
\multicolumn{9}{c}{$\kappa_{l1}=0.14144,\ \ \ \ \ \kappa_{l2}=0.14144 $}\\
\hline
$\vec{p},\vec{p}^{\,\,\prime} $  &\multicolumn{2}{c|}{$0.129\to 0.129$}
&\multicolumn{2}{c|}{$0.129\to 0.121$}
&\multicolumn{2}{c|}{$0.129\to 0.125$}
&\multicolumn{2}{c}{$0.129\to 0.133$}\\
$[p_{\rm min}]$     &$\omega$&$\hat\xi_{QQ'}$
&$\omega$&$\hat\xi_{QQ'}$&$\omega$&$\hat\xi_{QQ'}$&$\omega$&$\hat\xi_{QQ'}$ \\
\hline
&&&&&&&&\\[-9.5pt]$\!{\parbox{1.5cm}{(0,0,0), (1,0,0)}}\!$
&$1.037$&$0.94\!\!\er{3}{3}$&$1.026$&$1.01\!\!\er{3}{3}$
&$1.030$&$0.98\!\!\er{3}{3}$&$1.048$&$0.91\!\!\er{4}{4}$\\
&&&&&&&&\\[-9.5pt]$\!{\parbox{1.5cm}{(0,0,0), (1,1,0)}}\!$
&$1.073$&$0.93\!\!\er{6}{8}$&$1.050$&$0.97\!\!\er{8}{7}$
&$1.060$&$0.96\!\!\er{7}{7}$&$1.09$&$0.88\!\!\er{6}{8}$\\
&&&&&&&&\\[-9.5pt]$\!{\parbox{1.5cm}{(1,0,0), (0,1,0)}}\!$
&$1.076$&$0.88\!\!\er{9}{9}$&$1.064$&$0.91\!\!\err{11}{9}$
&$1.069$&$0.89\!\!\err{10}{8}$&$1.09$&$0.84\!\!\er{8}{8}$\\
&&&&&&&&\\[-9.5pt]$\!{\parbox{1.5cm}{(1,0,0), (1,0,0)}}\!$
&$1.000$&$1.02\!\!\err{11}{11}$&$1.001$&$0.98\!\!\err{12}{11}$
&$1.000$&$0.97\!\!\err{13}{12}$&$1.001$&$0.96\!\!\err{13}{13}$\\
&&&&&&&&\\[-9.5pt]$\!{\parbox{1.5cm}{(1,0,0), (-1,0,0)}}\!$
&$1.15$&$0.68\!\!\er{7}{6}$&$1.13$&$0.72\!\!\er{10}{7}$
&$1.14$&$0.71\!\!\er{9}{7}$&$1.17$&$0.63\!\!\er{6}{6}$\\
&&&&&&&&\\[-9.5pt]$\!{\parbox{1.5cm}{(1,0,0), (0,0,0)}}\!$
                            &-
&-&$1.037$&$0.92\!\!\er{6}{5}$&$1.037$&$0.92\!\!\er{6}{4}$
&$1.037$&$0.92\!\!\er{4}{4}$\\
&&&&&&&&\\[-9.5pt]$\!{\parbox{1.5cm}{(1,0,0), (0,1,1)}}\!$
&$1.11$&$0.81\!\!\err{16}{14}$&$1.09$&$0.89\!\!\err{17}{14}$
&$1.10$&$0.86\!\!\err{17}{14}$&$1.13$&$0.74\!\!\err{14}{14}$\\
\multicolumn{9}{c}{ }\\
\multicolumn{9}{c}{ }\\
\multicolumn{9}{c}{Table \ref{tabGG}: {\em (continued)}}
\end{tabular}
\end{center}

\begin{table}
\begin{center}
\begin{tabular}{cc}
\begin{tabular}{l|cc|cc}
\multicolumn{5}{c}{$0.129\to 0.129$}\\
\hline
$\vec{p},\vec{p}^{\,\,\prime} $ &\multicolumn{2}{c|}{$\kappa_{l1}=
0.14144,\ \kappa_{l2}=0.14226 $} 
&\multicolumn{2}{c}{$\kappa_{l1}=
0.14226,\ \kappa_{l2}=0.14226 $}\\
$[p_{\rm min}]$     &$\omega$&$\hat\xi_{QQ'}$ &$\omega$&$\hat\xi_{QQ'}$\\
\hline
&&\\[-9.5pt]$\!{\parbox{1.5cm}{(0,0,0), (1,0,0)}}\!$
&$1.040$&$0.92\!\!\er{3}{4}$ &$1.043$&$0.92\!\!\er{5}{6}$\\
&&\\[-9.5pt]$\!{\parbox{1.5cm}{(0,0,0), (1,1,0)}}\!$
&$1.078$&$0.95\!\!\er{6}{9}$  &$1.084$&$0.98\!\!\err{7}{10}$\\
&&\\[-9.5pt]$\!{\parbox{1.5cm}{(1,0,0), (0,1,0)}}\!$
&$1.081$&$0.85\!\!\err{12}{11}$ &$1.088$&$0.80\!\!\err{18}{15}$\\
&&\\[-9.5pt]$\!{\parbox{1.5cm}{(1,0,0), (1,0,0)}}\!$
&$1.000$&$1.02\!\!\err{11}{11}$ &$1.000$&$0.85\!\!\err{24}{25}$\\
&&\\[-9.5pt]$\!{\parbox{1.5cm}{(1,0,0), (-1,0,0)}}\!$
&$1.16$&$0.68\!\!\er{9}{9}$ &$1.18$&$0.72\!\!\err{13}{12}$\\
&&\\[-9.5pt]$\!{\parbox{1.5cm}{(1,0,0), (0,1,1)}}\!$
&$1.121$&$0.77\!\!\err{20}{19}$ &$1.131$&$0.59\!\!\err{27}{25}$\\
\end{tabular}
\end{tabular}
\end{center}
\caption{\em Estimates of the function $\hat\xi_{QQ'}$ as
obtained from the axial form factor $G_1$. 
All the degenerate transitions $\kappa_Q=\kappa_{Q'}=0.129$ and light masses
corresponding to $\kappa_{l1}, \kappa_{l2}=0.14144, 0.14226$ and
$0.14226,0.14226$ with initial and final momenta up to 
\protect$ |\protect\vec{p}\, |,|\protect\vec{p}^{\;\protect\prime}|=
\protect\sqrt{2} p_{\rm min}\protect$ 
are shown.
Statistical errors in $\omega$ are in the last digit or beyond.\label{tabG_1}}
\end{table}

We observe that: 
\begin{itemize}

\item Our determinations of $\hat\xi_{QQ'}(\omega)$ have statistical errors 
ranging from $3$ to $15\%$. As a general rule, we note that for the same
heavy-quark masses, errors increase with the momentum of the final 
baryon. Furthermore, errors are amplified as the light quarks approach the
chiral limit.

\item The estimates of $\hat\xi_{QQ'}(\omega)$ obtained from transitions
with final momentum $|\vec{p}^{\,\,\prime}|=\sqrt{2}p_{\rm min}$ 
are in agreement
with those of similar $\omega$ obtained from channels with lower momentum.
However, since these points are more affected by discretisation errors,
we do not include them in our determination of the Isgur-Wise function.

\end{itemize}

\begin{figure}
\begin{picture}(140,70)
\put(-5,0){\ewxy{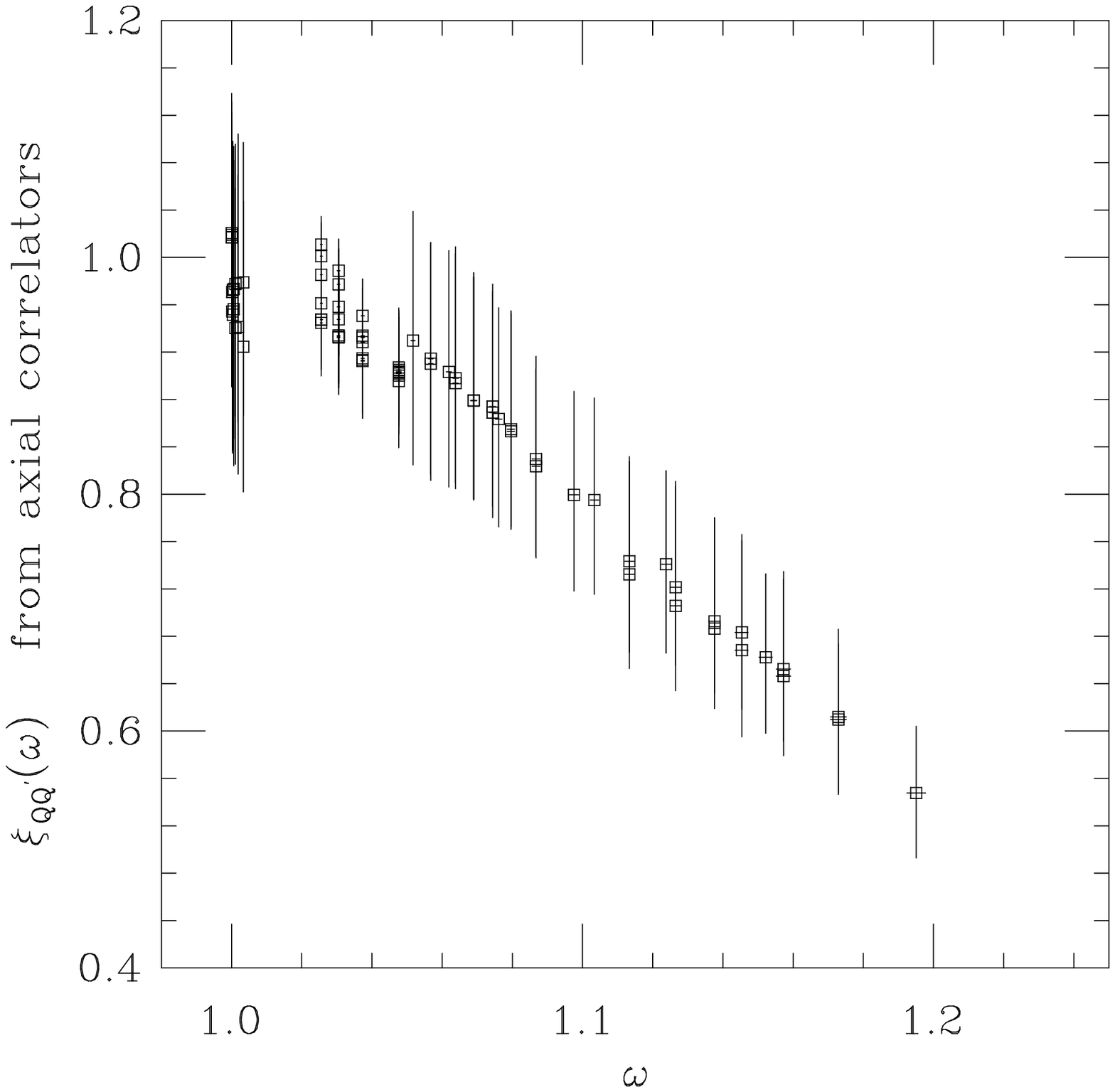}{90mm}}
\put(65,0){\ewxy{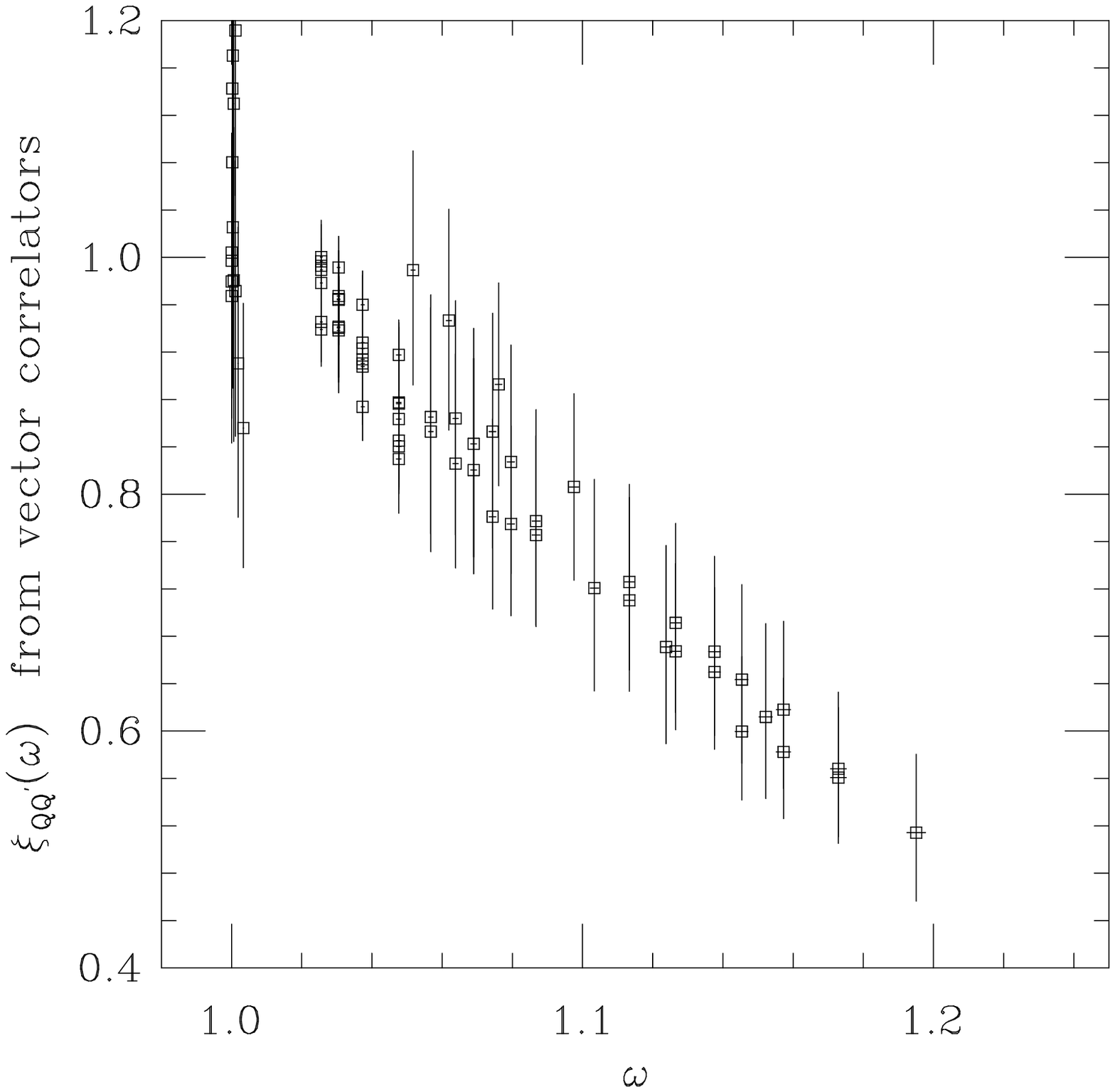}{90mm}}
\end{picture}
\caption{\em $\hat\xi_{QQ'}(\omega)$, as obtained from axial 
(left) and vector (right) current matrix elements.}
\label{fig:IWG_1}
\end{figure}

The estimates reported in Tables~\ref{tabG} and \ref{tabGG} are also plotted in
Figure~\ref{fig:IWG_1}, for all the degenerate and non-degenerate
transitions, at $\kappa_{l1}=\kappa_{l2}=0.14144$. We have included
all the points obtained from baryons either at rest or with momenta
$|\vec{p}|,|\vec{p}^{\ \prime}| = 
p_{\rm min}$.  The interpretation of the dependence of
$\hat\xi_{QQ'}(\omega)$ on the velocity transfer is postponed to Section
\ref{sec:hqdep}.

We conclude this section with a comment on the determination of the
suppressed form factors $G^L_2$ and $G^L_3$. As can be observed from
Figure~\ref{figplat}B, the signal is very noisy and compatible with
zero. Because of this feature, which is common to most of the momentum
channels, we are unable to determine these form factors, within the
available statistics.


\subsubsection{Analysis of the vector \ff\  }
\label{sec:anaV}

\label{anavec}
The analysis of the vector \ff\   proceeds along the same lines as those
followed for the axial \ff , with some significant differences.
It is interesting to measure not only the three \ff\ $F^L_i$, with 
$i=1,2, 3$, separately, but also their sum
because of the normalisation condition (\ref{normcond}). Indeed, as we
will see below, it proves to be easier to determine the sum precisely than
the individual form factors.

We now study the different spinor components of the two matrices
${\cal V}_\mu,{\cal W}$ for different values of the index $\mu$.  
As was the case
for the axial current, the discussion is based on
equations~(\ref{SSF}) and (\ref{SSB}) in Appendix~\ref{app1}.
We find that
\begin{enumerate}

\item {\em for indices $\mu=i$ with $i=1,2,3$}: the large components
  of ${\cal V}_{i}$ are located in the top-right (bottom-left) 
  submatrix in the forward (backward) part of the lattice, whereas
  the non-vanishing components of ${\cal W}$ are located along the
  diagonal.  Thus, the \ff\  $F^L_1$ and $F^L_{2,3}$ give separate
  contributions to different spinorial components. However, as shown
  in Appendix \ref{app2}, spinorial components located in the
  off-diagonal submatrices are always proportional to the amplitudes
  $\alpha$ and $\beta$ and thus are noisier than those located in the
  diagonal submatrices. From this feature we can already anticipate
  that the determination of $F^L_1$ is going to be less precise than
  that of $G^L_1$.

\item {\em for current indices $\mu=0$}: the large components of both
  ${\cal V}_\mu$ and ${\cal W}$ are located in the top (bottom)-diagonal
  submatrix in the forward (backward) part of the lattice. In this
  case one obtains the linear combination of the three form factors
  $F_1^L+ v_0\,F_2^L + v_0'\,F_3^L$ which is approximately equal to
  $F_1 + F_2 + F_3$ at the low momenta which we are using. These
  channels have a very clean and precise signal, which is the reason
  for the accurate determination of the sum of the form factors.

\end{enumerate}

In view of the above discussion, it has proved to be advantageous to consider
equations both with $\mu=i$ and with $\mu=0$, minimizing the
full covariance matrix with respect to the three \ff , and reconstructing
their sum. 
Examples of plateaux obtained from channels with $\mu=0$ and with
$\mu=i$ are shown in Figures~\ref{figplat}C and \ref{figplat}D,
respectively.  As expected, the relative statistical error is smaller
in the first case, and we find that the stability and symmetry of the
plateau is also better.  In extracting the values of the form factors
we have again restricted the fitting range to three time slices centered
around $t=12$, and we find very reasonable values of $\chi^2_{\rm
  dof}$.  Confirming the indications of the preliminary discussion
above, we measure $F^L_1, F^L_2$ and $F^L_3$ individually with
relatively large statistical errors, ranging from $6$ to $20\%$,
whereas the errors on their sum are never more than $10\%$.

In order to extract the form factors from the correlation functions
measured on the lattice it is necessary to determine the
renormalisation constant $Z_V$ which relates the lattice and physical
vector currents. This constant has previously been determined
non-perturbatively, by studying the matrix elements of the charge
operator between meson states, and the results are reviewed in
Appendix \ref{sec:zv}.  
Noting that for degenerate transitions between baryon states at rest,
$Z_V\, V_0^L$ is the charge operator, we find here that
\be
1=Z_V( \sum_i F^L_i(1))\ .
\label{zvdef}\ee
The numerical values of $Z_V$ obtained in this way are:
\be
\ba{cccc}
Z_V=0.84\er{7}{6} &{\rm at}\ \ &\kappa_{l1}=\kappa_{l2}=0.14144\ 
{\rm and}\ \kappa_h=0.133 \\
Z_V=0.85\er{8}{7} &{\rm at}\ \ &\kappa_{l1}=\kappa_{l2}=0.14144\ 
{\rm and}\ \kappa_h=0.129 \\
Z_V=0.87\er{9}{7} &{\rm at}\ \ &\kappa_{l1}=\kappa_{l2}=0.14144\ 
{\rm and}\ \kappa_h=0.125 \\
Z_V=0.88\er{9}{8} &{\rm at}\ \ &\kappa_{l1}=\kappa_{l2}=0.14144\ 
{\rm and}\ \kappa_h=0.121 
\ea
\label{zvinv}
\ee
which are in agreement with the other non-perturbative estimates presented
in Appendix \ref{sec:zv}, obtained with a different method. 
Below we will use $1/\sum_i F_i(1)$ to normalise the lattice vector current.

In Tables~\ref{tabV}, \ref{tabVV} and \ref{tabVF}, we present our
results for the quantity \be
\hat\xi_{QQ'}(\omega)=\frac{F^L_1(\omega)+F^L_2(\omega)+F^L_3(\omega)}
{\sum_i F^L_i(1)}\frac{N_{\rm sum}(1)}{N_{\rm sum}(\omega)}
\label{F1norm}\ee
for all the combinations of quark masses and for initial and final
momenta up to $|\vec{p}|,|\vec{p}^{\,\,\prime}|=p_{\rm min}$.  The results
reported in Tables~\ref{tabV} and \ref{tabVV} are also plotted in
Figure~\ref{fig:IWG_1}, for all the degenerate and non-degenerate
transitions, at $\kappa_{l1}=\kappa_{l2}=0.14144$.  The study of the
dependence of $\hat\xi_{QQ'}(\omega)$ on the velocity transfer is again
postponed to Section \ref{sec:hqdep}.

\begin{table}
\begin{center}
\begin{tabular}{l|cc|cc|cc|cc}
\multicolumn{9}{c}{$\kappa_{l1}=
0.14144,\ \ \ \ \ \kappa_{l2}=0.14144 $}\\
\hline
$\vec{p},\vec{p}^{\,\,\prime} $ &\multicolumn{2}{c|}{$0.121\to 0.121$}
&\multicolumn{2}{c|}{$0.121\to 0.125$}&\multicolumn{2}{c|}{$0.121\to 0.129$}
&\multicolumn{2}{c}{$0.121\to 0.133$}\\
$[p_{\rm min}]$   &$\omega$&$\hat\xi_{QQ'}$&$\omega$&$\hat\xi_{QQ'}$
&$\omega$&$\hat\xi_{QQ'}$&$\omega$&$\hat\xi_{QQ'}$ \\
\hline
&&&&&&&&\\[-9.5pt]$\!{\parbox{1.5cm}{(0,0,0), (1,0,0)}}\!$&$1.026 $
&$1.00\er{3}{3}$&$1.030 $&$0.94\er{4}{5}$&$1.037 $&$0.91\er{3}{4}$
&$1.048 $&$0.84\er{4}{4}$\\
&&&&&&&&\\[-9.5pt]$\!{\parbox{1.5cm}{(1,0,0), (0,1,0)}}\!$&$1.052 $
&$0.99\err{10}{9}$&$1.057 $&$0.87\err{10}{10}$&$1.064 $&$0.83\er{9}{9}$
&$1.074 $&$0.79\er{9}{8}$\\
&&&&&&&&\\[-9.5pt]$\!{\parbox{1.5cm}{(1,0,0), (1,0,0)}}\!$&$1.000 $
&$1.00\err{10}{11}$&$1.000 $&$1.08\err{12}{13}$&$1.001 $
&$0.97\err{11}{13}$&$1.003 $&$0.86\err{11}{12}$\\
&&&&&&&&\\[-9.5pt]$\!{\parbox{1.5cm}{(1,0,0), (-1,0,0)}}\!$
&$1.103 $&$0.73\er{9}{9}$&$1.113 $&$0.74\er{8}{7}$&$1.127 $
&$0.67\er{7}{7}$&$1.145 $&$0.61\er{6}{6}$\\
&&&&&&&&\\[-9.5pt]$\!{\parbox{1.5cm}{(1,0,0), (0,0,0)}}\!$
                      &-&-&$1.026 $&$0.99\er{4}{5}$&$1.026 $
&$0.98\er{4}{4}$&$1.026 $
&$0.95\er{3}{4}$\\
\hline
$\vec{p},\vec{p}^{\,\,\prime} $  &\multicolumn{2}{c|}{$0.125\to 0.125$}
&\multicolumn{2}{c|}{$0.125\to 0.121$}
&\multicolumn{2}{c|}{$0.125\to 0.129$}&\multicolumn{2}{c}{$0.125\to 0.133$}\\
$[p_{\rm min}]$  &$\omega$&$\hat\xi_{QQ'}$
&$\omega$&$\hat\xi_{QQ'}$&$\omega$&$\hat\xi_{QQ'}$&$\omega$&$\hat\xi_{QQ'}$ \\
\hline
&&&&&&&&\\[-9.5pt]$\!{\parbox{1.5cm}{(0,0,0), (1,0,0)}}\!$
&$1.030 $&$0.99\er{2}{3}$&$1.026 $&$0.94\er{2}{3}$&$1.037 $
&$0.92\er{3}{4}$&$1.048 $&$0.85\er{4}{4}$\\
&&&&&&&&\\[-9.5pt]$\!{\parbox{1.5cm}{(1,0,0), (0,1,0)}}\!$
&$1.062 $&$0.95\er{9}{9}$&$1.057 $&$0.86\err{10}{10}$&$1.069 $
&$0.83\er{9}{9}$&$1.080 $&$0.78\er{8}{8}$\\
&&&&&&&&\\[-9.5pt]$\!{\parbox{1.5cm}{(1,0,0), (1,0,0)}}\!$&$1.000 $
&$1.00\err{11}{12}$&$1.000 $&$1.14\err{13}{14}$&$1.000 $
&$1.03\err{11}{14}$&$1.002 $&$0.91\err{11}{13}$\\
&&&&&&&&\\[-9.5pt]$\!{\parbox{1.5cm}{(1,0,0), (-1,0,0)}}\!$
&$1.124 $&$0.68\er{8}{8}$&$1.113 $&$0.72\er{8}{7}$&$1.138 $
&$0.66\er{7}{6}$&$1.157 $&$0.59\er{5}{6}$\\
&&&&&&&&\\[-9.5pt]$\!{\parbox{1.5cm}{(1,0,0), (0,0,0)}}\!$&-&-
&                      $1.030 $&$0.94\er{5}{6}$&$1.030 $
&$0.97\er{4}{4}$&$1.030 $&$0.94\er{4}{4}$
\end{tabular}
\end{center}
\caption{\em Estimates of the function $\hat\xi_{QQ'}$ as
obtained from the vector form factor combination $\sum_i F_i$. 
All the transitions, corresponding to initial heavy $\kappa=0.121$ and $0.125$
and initial and final momenta up to 
\protect$ |\protect\vec{p}\, |,|\protect\vec{p}^{\;\protect\prime}|=
\protect\sqrt{2} p_{\rm min}\protect$ 
are shown.
Statistical errors in $\omega$ are in the last digit or beyond.}\label{tabV}
\end{table}

\begin{table}
\begin{center}
\begin{tabular}{l|cc|cc|cc|cc}
\multicolumn{9}{c}{$\kappa_{l1}=
0.14144,\ \ \ \ \ \kappa_{l2}=0.14144 $}\\
\hline
$\vec{p},\vec{p}^{\,\,\prime} $  &\multicolumn{2}{c|}{$0.129\to 0.129$}
&\multicolumn{2}{c|}{$0.129\to 0.121$}&\multicolumn{2}{c|}{$0.129\to 0.125$}
&\multicolumn{2}{c}{$0.129\to 0.133$}\\
$[p_{\rm min}]$  &$\omega$&$\hat\xi_{QQ'}$&$\omega$&$\hat\xi_{QQ'}$
&$\omega$&$\hat\xi_{QQ'}$&$\omega$&$\hat\xi_{QQ'}$ \\
\hline
&&&&&&&&\\[-9.5pt]$\!{\parbox{1.5cm}{(0,0,0), (1,0,0)}}\!$&$1.037 $
&$0.96\er{3}{3}$&$1.026 $&$1.00\er{2}{2}$&$1.030 $&$0.97\er{2}{3}$
&$1.048 $&$0.83\er{5}{5}$\\
&&&&&&&&\\[-9.5pt]$\!{\parbox{1.5cm}{(1,0,0), (0,1,0)}}\!$&$1.076 $
&$0.90\er{8}{8}$&$1.064 $&$0.87\err{10}{10}$&$1.069 $&$0.85\err{10}{9}$
&$1.087 $&$0.77\er{8}{8}$\\
&&&&&&&&\\[-9.5pt]$\!{\parbox{1.5cm}{(1,0,0), (1,0,0)}}\!$&
$1.000 $&$0.98\err{11}{12}$&$1.001 $&
$1.19\err{13}{14}$&$1.000 $&$1.17\err{13}{14}$&
$1.001 $&$0.98\err{13}{14}$\\
&&&&&&&&\\[-9.5pt]$\!{\parbox{1.5cm}{(1,0,0), (-1,0,0)}}\!$&$1.152 $
&$0.62\er{8}{7}$&$1.127 $&$0.70\er{8}{7}$&$1.138 $&$0.68\er{8}{7}$
&$1.173 $&$0.57\er{6}{5}$\\
&&&&&&&&\\[-9.5pt]$\!{\parbox{1.5cm}{(1,0,0), (0,0,0)}}\!$&-&-
&                      $1.037 $&$0.93\er{6}{6}$&$1.037 $&$0.93\er{5}{6}$
&$1.037 $&$0.91\er{4}{4}$\\
\hline
$\vec{p},\vec{p}^{\,\,\prime} $   &\multicolumn{2}{c|}{$0.133\to 0.133$}
&\multicolumn{2}{c|}{$0.133\to 0.121$}&\multicolumn{2}{c|}{$0.133\to 0.125$}
&\multicolumn{2}{c}{$0.133\to 0.129$}\\
$[p_{\rm min}]$ &$\omega$&$\hat\xi_{QQ'}$&$\omega$&$\hat\xi_{QQ'}$&$\omega$
&$\hat\xi_{QQ'}$&$\omega$&$\hat\xi_{QQ'}$ \\
\hline
&&&&&&&&\\[-9.5pt]$\!{\parbox{1.5cm}{(0,0,0), (1,0,0)}}\!$&$1.048 $
&$0.92\er{3}{3}$&$1.026 $&$1.00\er{2}{2}$&$1.030 $&$0.96\er{2}{2}$
&$1.037 $&$0.88\er{2}{3}$\\
&&&&&&&&\\[-9.5pt]$\!{\parbox{1.5cm}{(1,0,0), (0,1,0)}}\!$&$1.098 $
&$0.82\er{8}{8}$&$1.074 $&$0.86\err{10}{9}$&$1.080 $&$0.84\err{10}{9}$
&$1.087 $&$0.79\er{9}{9}$\\
&&&&&&&&\\[-9.5pt]$\!{\parbox{1.5cm}{(1,0,0), (1,0,0)}}\!$&$1.000 $
&$0.97\err{12}{12}$&$1.003 $&$1.21\err{13}{15}$&$1.002 $&$1.20\err{13}{15}$
&$1.001 $&$1.13\err{13}{14}$\\
&&&&&&&&\\[-9.5pt]$\!{\parbox{1.5cm}{(1,0,0), (-1,0,0)}}\!$&$1.195 $
&$0.53\er{6}{6}$&$1.145 $&$0.66\er{8}{7}$&$1.157 $&$0.63\er{7}{6}$
&$1.173 $&$0.58\er{6}{6}$\\
&&&&&&&&\\[-9.5pt]$\!{\parbox{1.5cm}{(1,0,0), (0,0,0)}}\!$&-&-
&                      $1.048 $&$0.88\er{7}{7}$&$1.048 $&$0.88\er{6}{6}$
&$1.048 $&$0.87\er{5}{5}$\\
\end{tabular}
\end{center}
\caption{\em Estimates of the function $\hat\xi_{QQ'}$ as
obtained from the vector form factor combination $\sum_i F_i$. 
All the transitions, corresponding to initial heavy $\kappa=0.133$ and $0.129$
and initial and final momenta up to 
\protect$ |\protect\vec{p}\, |,|\protect\vec{p}^{\;\protect\prime}|=
\protect\sqrt{2} p_{\rm min}\protect$ 
are shown.
Statistical errors in $\omega$ are in the last digit or beyond.}\label{tabVV}
\end{table}

\begin{table}
\begin{center}
\begin{tabular}{l|cc|cc}
\multicolumn{5}{c}{$0.129\to 0.129$}\\
\hline
$\vec{p},\vec{p}^{\,\,\prime} $ &\multicolumn{2}{c|}{$\kappa_{l1}=
0.14144,\ 
\kappa_{l2}=0.14226 $}
&\multicolumn{2}{c}{$\kappa_{l1}=0.14226,\ 
\kappa_{l2}=0.14226 $}\\
$[p_{\rm min}]$    &$\omega$&$\hat\xi_{QQ'}$ &$\omega$&$\hat\xi_{QQ'}$\\
\hline
&&\\[-9.5pt]$\!{\parbox{1.5cm}{(0,0,0), (1,0,0)}}\!$
&$1.040$&$0.95\er{3}{4}$ &$1.043$&$0.93\er{5}{5}$\\
&&\\[-9.5pt]$\!{\parbox{1.5cm}{(1,0,0), (0,1,0)}}\!$
&$1.081$&$0.87\err{10}{11}$ &$1.088$&$0.77\err{14}{13}$\\
&&\\[-9.5pt]$\!{\parbox{1.5cm}{(1,0,0), (1,0,0)}}\!$
&$1.000$&$0.86\err{16}{15}$ &$1.000$&$0.63\err{27}{26}$\\
&&\\[-9.5pt]$\!{\parbox{1.5cm}{(1,0,0), (-1,0,0)}}\!$
&$1.16$&$0.62\er{9}{8}$ &$1.18$&$0.64\err{13}{10}$
\end{tabular}
\end{center}
\caption{\em Estimates of the function $\hat\xi_{QQ'}$ as
obtained from the form factor $\sum_i F_i$. 
All the degenerate transitions $\kappa_Q=\kappa_{Q'}=0.129$ and light masses
corresponding to $\kappa_{l1}, \kappa_{l2}=0.14144, 0.14226$ and
$0.14226,0.14226$ with initial and final momenta up to 
\protect$ |\protect\vec{p}\, |,|\protect\vec{p}^{\protect\,\,\prime}|=
\protect\sqrt{2} p_{\rm min}\protect$ 
are shown.
Statistical errors in $\omega$ are in the last digit or beyond.\label{tabVF}}
\end{table}

\begin{figure}
\begin{picture}(140,70)
\put(-7,0){\ewxy{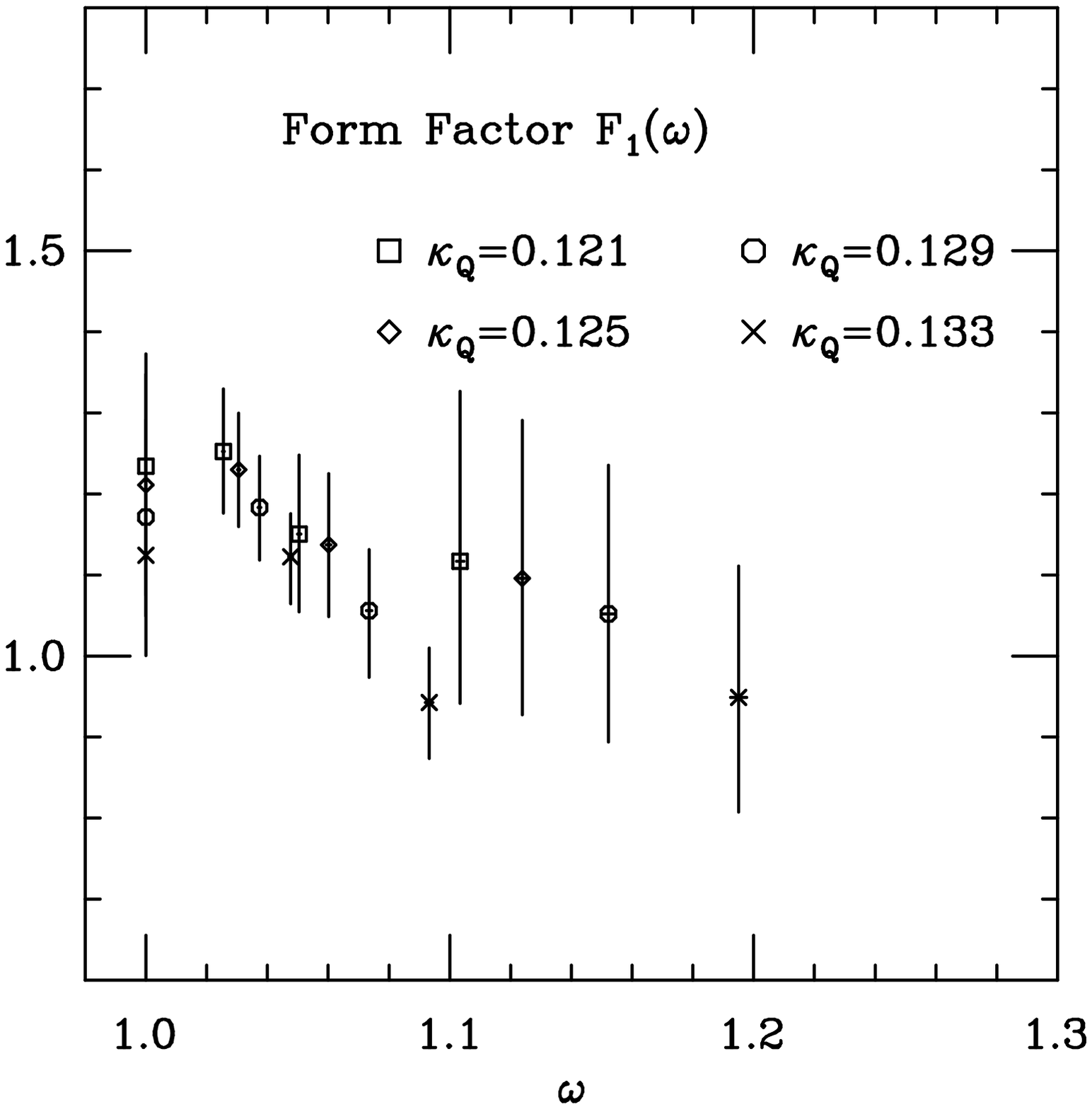}{90mm}}
\put(63,0){\ewxy{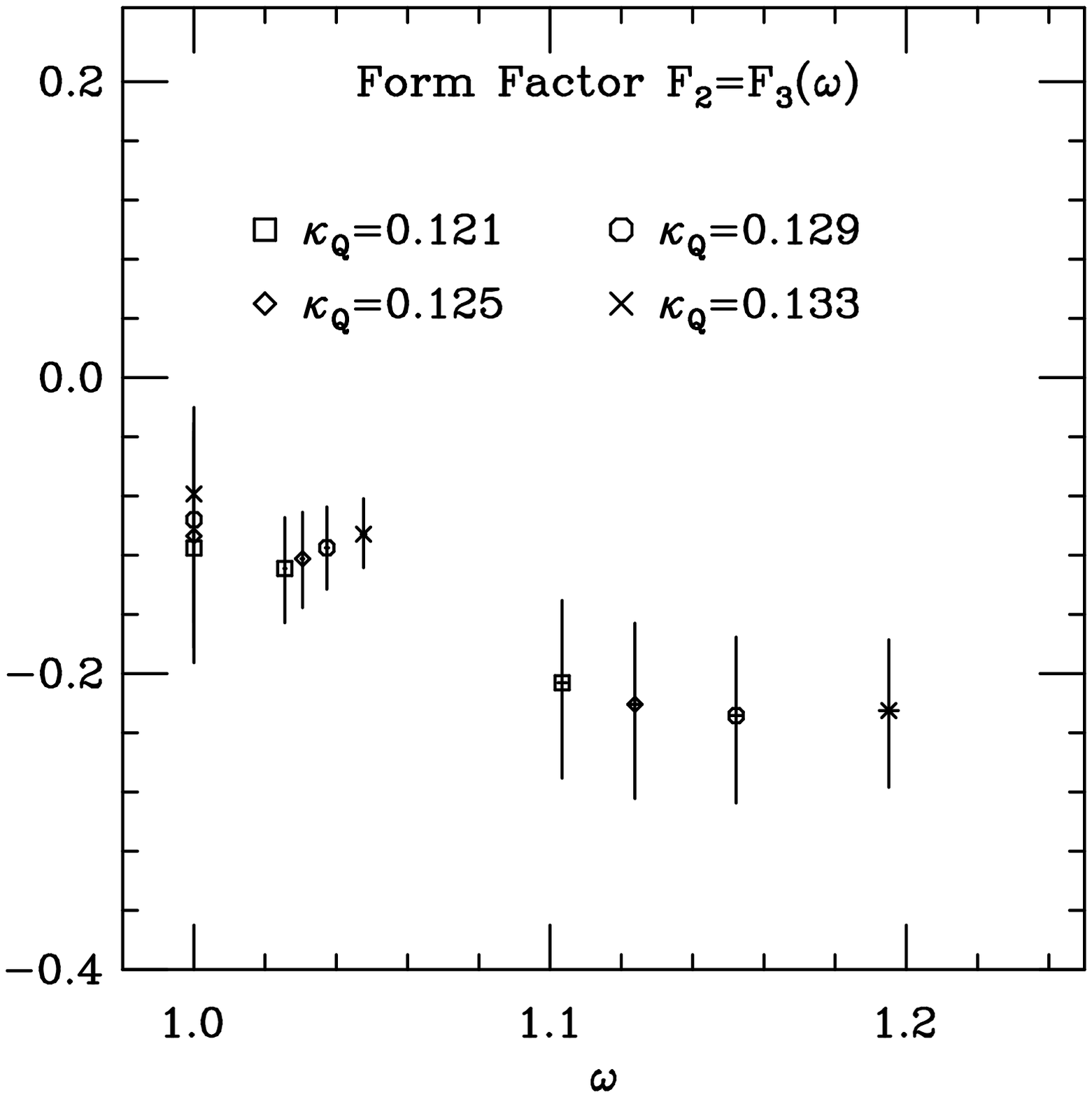}{90mm}}
\end{picture}
\caption{\em Form factors  $F_1$ and $F_2=F_3$, as obtained 
from degenerate quark transitions. Different symbols correspond to
different heavy-quark hopping parameters. The light-quark hopping
parameters are always \protect$\kappa_{l1}=\kappa_{l2}=0.14144\protect$.}
\label{fig:F_i}
\end{figure}

We end this subsection with a discussion of the determination of the
form factors $F_1, F_2$ and $F_3$ separately. 
The form factors $F_2$ and $F_3$ start at ${\cal{O}}(\alpha_s)$ in
perturbation theory and at next-to-leading order in the $1/m_Q$
expansion and they are thus expected to be small relative to
$F_1$ and
sensitive to the values of the heavy-quark masses (which in the present study
vary by almost a factor two: $m_q\in[1.48,2.38]$). In
Figure~\ref{fig:F_i}, we plot our estimates of
\be
F_{2,3}(\omega)=F^L_{2,3}(\omega)\,\frac{N_{\rm  sum}(1)}{\sum_i F^L_i(1)}.
\label{f2norm}
\ee
We observe that these form factors are small and negative as one would
expect from Eq.~(\ref{1-overm-exp}).
Our measurements of
\be
F_1(\omega)=F^L_1(\omega)\,\frac{N_{\rm  sum}(1)}{\sum_i F^L_i(1)}
\label{f1norm}
\ee
for degenerate transitions are also presented in Figure~\ref{fig:F_i}.
In order to decrease the number of parameters in the fit,
we restricted the analysis to degenerate transitions, for which
one can use the symmetry relation $F_2(\omega)=F_3(\omega)$.  

\begin{figure}
\begin{picture}(140,100)
\put(10,0){\ewxy{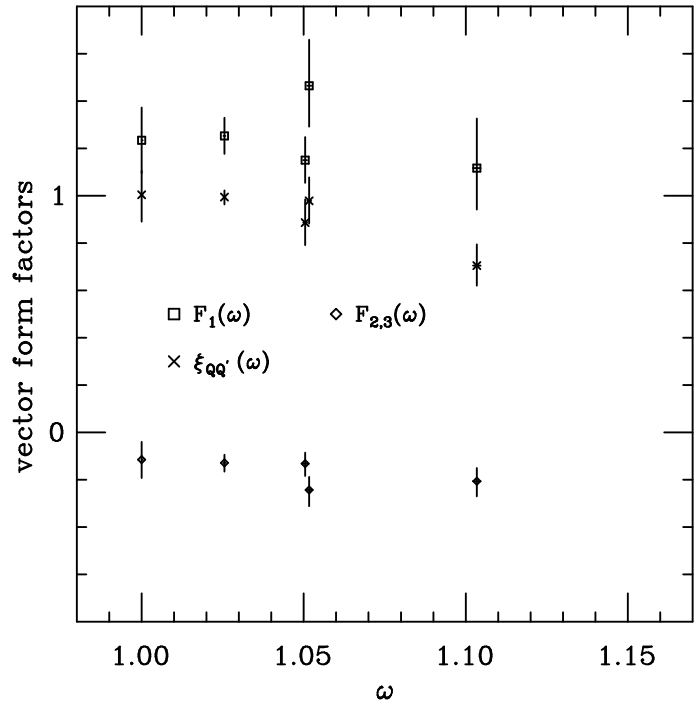}{130mm}}
\end{picture}
\caption{\em Example of the relative size of the normalized vector \ff.
  \protect$\kappa_{l1}=\kappa_{l2}=0.14144\protect$ and
  \protect$\kappa_Q=\kappa_{Q'}=0.121\protect$.  }
\label{figall}
\end{figure}

As anticipated above, the errors in the individual form factors are
larger than that in their sum.  To make this clear, we plot in
Figure~\ref{figall} the form factors $F_1, F_2$ and the sum, $F_1+2
F_2$, for the degenerate channel with heavy quark hopping parameter
$\k_Q=0.121$. It can be seen that the fluctuations in $F_1$ and $F_2$
partially compensate each other in the determination of the function
$\hat\xi_{QQ'}$

A more detailed discussion of the behavior of both $F_1$ and $F_{2,3}$
and of their relation with the Isgur-Wise function, 
will be presented in Section \ref{sec:hqdep}.



\section{The Isgur-Wise function}
\label{IW}

In this section we study the dependence of the function 
$\hat\xi_{QQ'}(\omega)$ on the masses of the initial and final 
heavy quarks so as to extract the corresponding Isgur-Wise function.
We also study its dependence on the velocity transfer $\omega$, and the mass
of the light quarks. We further attempt to estimate the size of $1/m_Q$
corrections.

\subsection{${\bf \hat\xi_{QQ'}(\omega)}$ as a function of
${\bf \omega}$ and of heavy-quark mass}
\label{sec:hqdep}

As discussed in Section \ref{teo}, both the axial form factor $G_1$
and the sum of the vector form factors $\sum_i F_i$, are protected
from $1/m_Q$ corrections at zero recoil (see Eq.~(\ref{normcond})).
Away from zero recoil, these quantities are no longer protected by
Luke's theorem and they suffer $1/m_Q$ corrections.
We recall their expansion in powers of the inverse heavy-quark mass,
given in Eqs.~(\ref{doppione}) and (\ref{newIW})
\bea
\frac{G_1(\omega)}{N_1^5(\omega)}&= &\frac{\sum_i F_i(\omega)}{N_{\rm sum}
(\omega)} = \hat\xi_{QQ'}(\omega)\nn \\
&=& \xi^{\rm ren}(\omega) 
+ \left(\frac{\bar\Lambda}{2m_Q} +
\frac{\bar\Lambda}{2m_{Q'}}\right)\left[
\frac{(\omega-1)}{(\omega+1)}\xi^{\rm ren}(\omega) 
+2\chi^{\rm ren}(\omega)\right]\nn\\
&&+\ord{1/m_{Q^{(')}}^2}.
\label{recall}
\eea
In this section we study the dependence of the \ff\  on $\omega$ in
order to extract some phenomenologically interesting quantities. In
particular, the slope of $\hat\xi_{QQ'}(\omega)$ at zero recoil can be
related to the slope of the physical form factors (see also Section
\ref{phenom}) through the correction coefficients given in
Eqs.~(\ref{doppione}) and (\ref{1-overm-exp}).  
The form factors, in turn, are needed in
the calculation of the decay rates and asymmetry parameters.

In order to obtain reliable estimates of phenomenological quantities,
we must learn how to extrapolate our data, obtained for initial and
final heavy quarks with masses around that of the
charm quark, to the physical $b\to c$ decays. HQET provides us
with the guide for this extrapolation, and it is important to understand
the r\^ole of the $1/m_Q$ corrections, present in the function
(\ref{recall}), and to check that higher-order corrections are small.

\begin{figure}
\begin{picture}(140,100)
\put(10,0){\ewxy{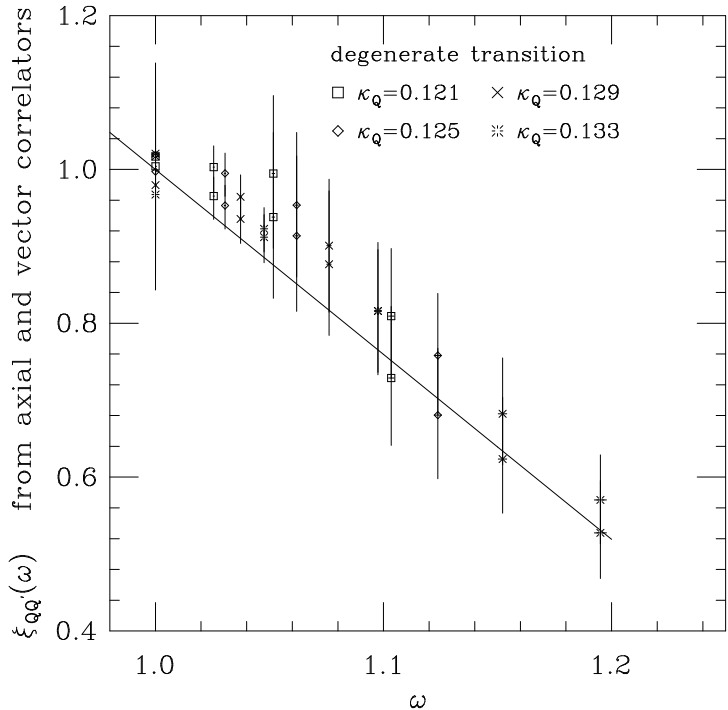}{130mm}}
\end{picture}
\caption{\em $\hat\xi_{QQ'}(\omega)$ from degenerate transitions, using vector
and axial current data. Different graphical symbols denote different heavy
quark masses, which appear to follow the same curve.}
\label{figFG}
\end{figure}

With the aim of reducing the statistical error we exploit the relation
(\ref{recall}), and fit the vector and axial data together. This is
correct up to terms of ${\cal{O}}(1/m_Q^2)$ and two-loop perturbative
corrections, which we neglect throughout this study.  Near $\omega =
1$ we expand $\hat\xi_{QQ'}(\omega)$ as a linear function of $\omega$
\be \hat\xi_{QQ'}(\omega)= 1 -\rho^2(\omega-1)+\ord{(\omega-1)^2}\ ,
\label{fitfunc}\ee
and we study whether there is any dependence of the slope, $\rho ^2$,
on the masses of the heavy quarks. Our results are obtained for the
set of initial- and final-state heavy-quark masses given in
Table~\ref{tab:Q_masses3pt}, but with fixed light-quark masses
around that of the strange ($\kappa_{l1}=\kappa_{l2}=0.14144$).  
It can be seen from
Figure~\ref{figFG} that there is no statistical evidence of a
dependence of $\hat\xi_{QQ'}(\omega)$ on the heavy-quark mass.  In
order to quantify this statement, we have fitted separately to the
function (\ref{fitfunc}), each of the four data sets corresponding to
degenerate transitions obtaining
\bea
\rho^2= 2.4\er{3}{3}  &\qquad {\rm at }\qquad &\kappa_Q=\kappa_{Q'}=0.121 
\nn \\
\rho^2= 2.4\er{3}{4}  &\qquad {\rm at }\qquad &\kappa_Q=\kappa_{Q'}=0.125 
\nn \\
\rho^2= 2.4\er{4}{4}  &\qquad {\rm at }\qquad &\kappa_Q=\kappa_{Q'}=0.129 
\nn \\
\rho^2= 2.4\er{4}{4}  &\qquad {\rm at }\qquad &\kappa_Q=\kappa_{Q'}=0.133 ,
\label{fourvalues}\eea
confirming that no dependence on the heavy-quark mass can be detected.

Our conclusion from this analysis is that our data for
$\hat\xi_{QQ'}(\omega)$ do not show any evidence of sizable $1/m_Q$
corrections, within our precision. So even though 
$\hat\xi_{QQ'}(\omega)$ is not, in principle, 
a universal function, it appears to be a good approximation to
the Isgur-Wise function. Thus, from now on, we will consider
$\hat\xi_{QQ'}(\omega)$ to be our estimate of the Isgur-Wise function,
$\xi^{\rm ren}(\omega)$.

For the slope of the Isgur-Wise function corresponding to light-quark 
masses around that of the strange we take the result in
Eq.~(\ref{fourvalues}), which will be used later to study the $1/m_Q$
corrections to the individual vector form factors.  This choice is
motivated by the fact that the analysis of the $1/m_Q$ corrections is
based on the data obtained from degenerate transitions.


\subsection{${\bf \xi^{\rm ren}(\omega)}$ as a function of the 
light-quark masses}

The Isgur-Wise function depends on the quantum numbers of the light
degrees of freedom, i.e. on the so-called ``brown muck''.  Previous
studies on the lattice~\cite{lpl} demonstrated that such a dependence
is not negligible in the case of mesons, where the brown muck contains
only one light quark.  Thus we might expect to measure an even
stronger dependence of the baryonic Isgur-Wise function on the
masses of the light quarks, and we investigate whether this is the
case in the present subsection.

\begin{figure}
\begin{picture}(140,80)
\put(-5,0){\ewdueup{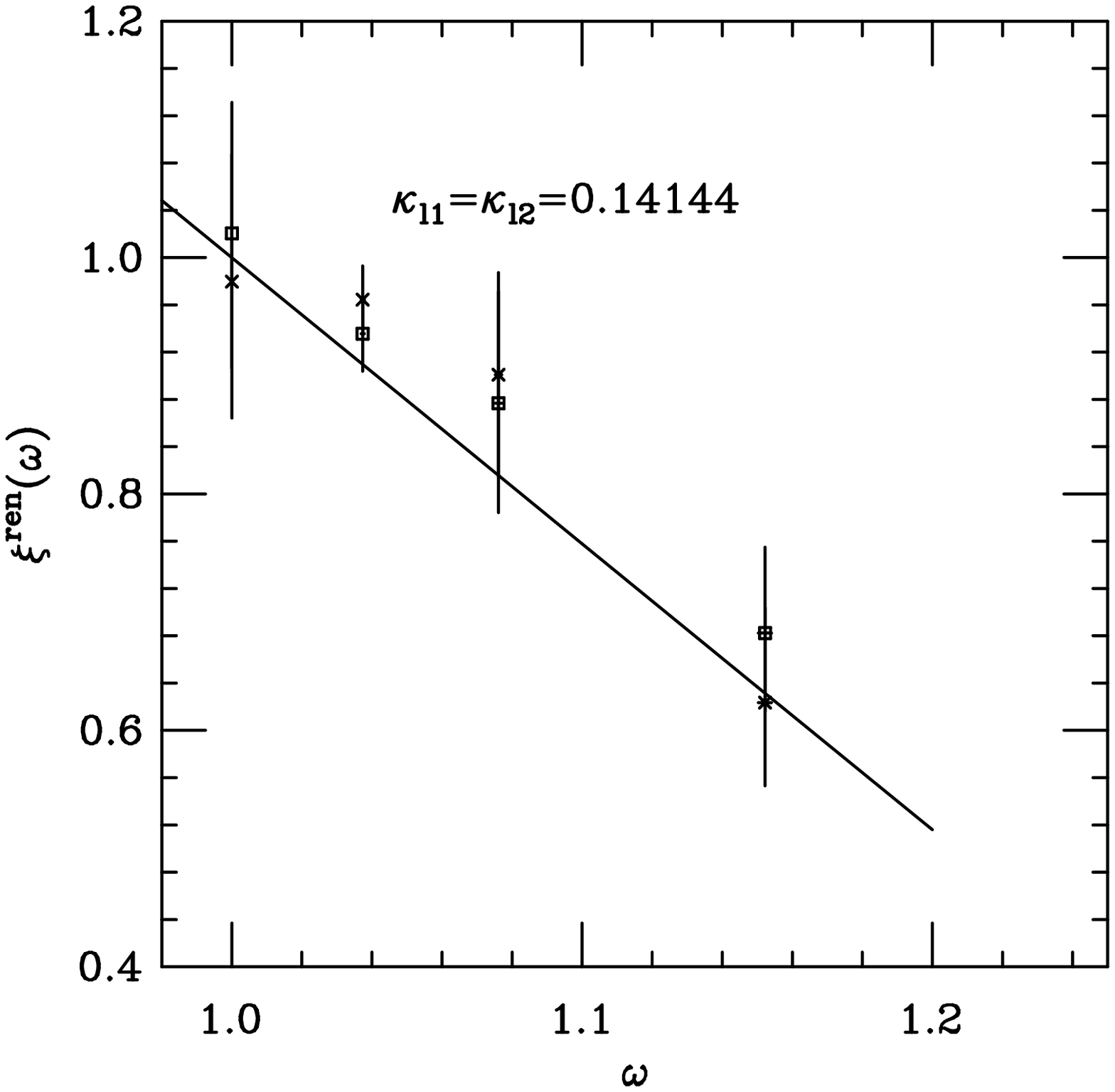}{90mm}}
\put(65,0){\ewdueup{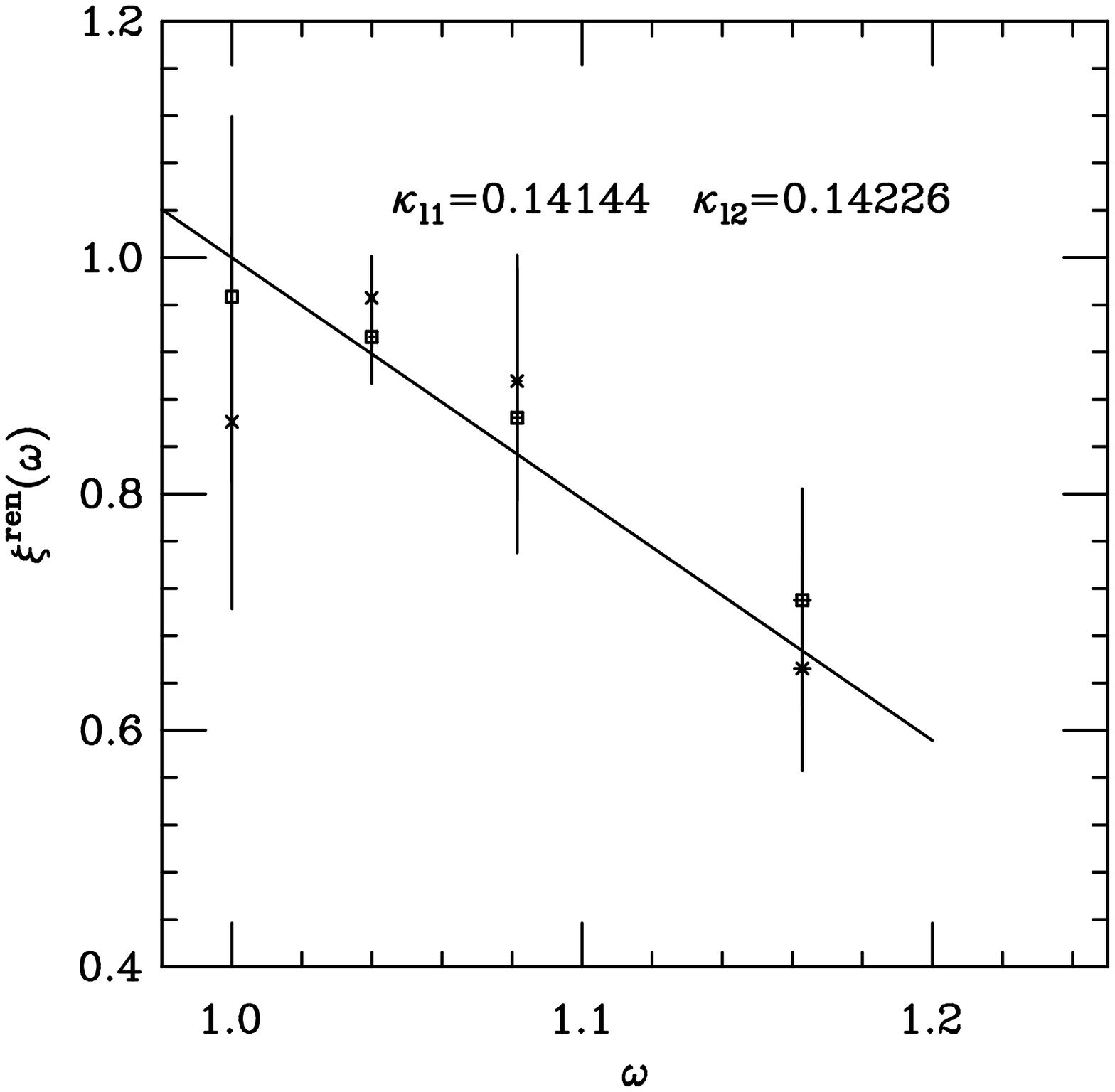}{90mm}}
\end{picture}

\begin{picture}(140,80)
\put(-5,0){\ewdueup{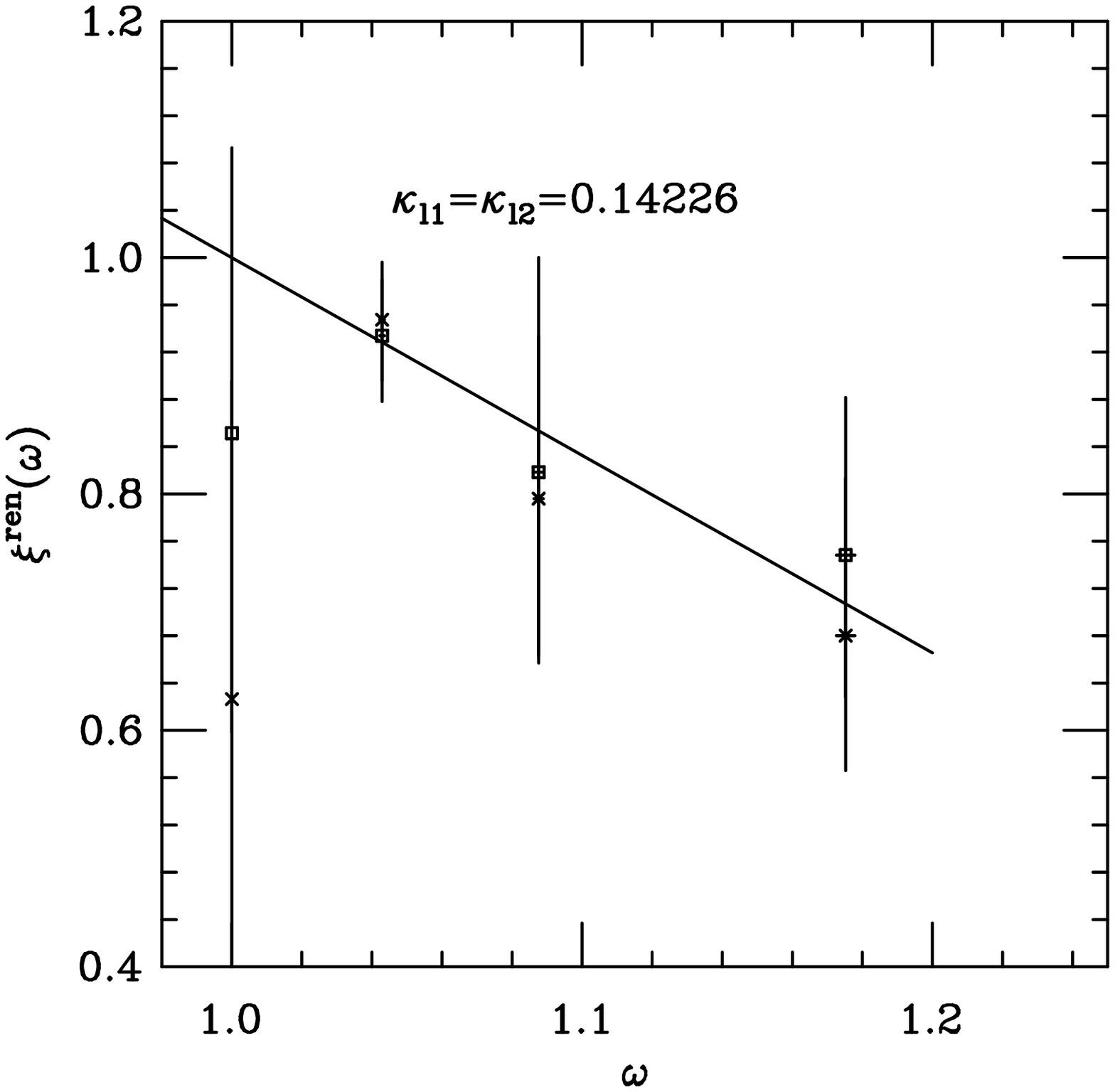}{90mm}}
\put(65,0){\ewdueup{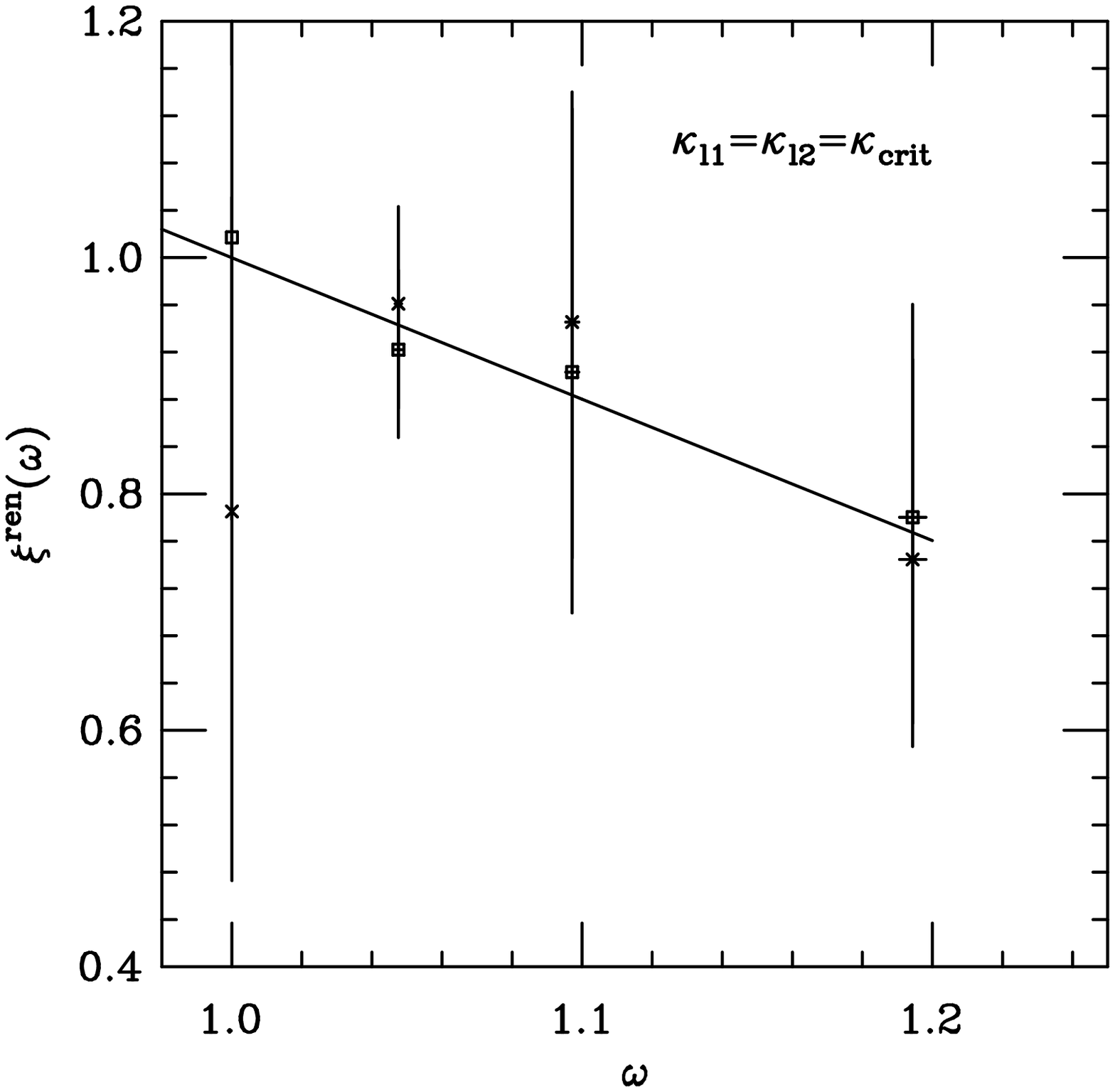}{90mm}}
\end{picture}
\caption{\em Plot of the Isgur-Wise function at fixed heavy-quark masses
  (\protect$ \kappa_{Q}=\kappa_{Q'}=0.129\protect$~) and various light
  quark masses, down to the chiral limit. Both vector (crosses) and axial
  (boxes) determinations were used for the fits.}
\label{figextra}\end{figure}

We study the dependence of $\xi^{\rm ren}(\omega)$ on the light quarks
by keeping the masses of the heavy quarks fixed at
$\kappa_Q=\kappa_{Q'}=0.129$ and letting the light quarks take the three
values listed in Table~\ref{tab:Q_masses3pt}.  By simultaneously fitting
$G_1(\omega)$ and $\sum_i F_i(\omega)$ to Eq.~(\ref{fitfunc}), we
find:
\bea
\rho^2=2.4\er{4}{4} &\ \ \ {\rm at}\ \ \  &\kappa_{l1}=0.14144,
\ \ \kappa_{l2}=0.14144\nn\\
\rho^2=2.0\er{5}{5} &\ \ \ {\rm at}\ \ \  &\kappa_{l1}=0.14144,\ \ 
\kappa_{l2}=0.14226\nn\\
\rho^2=1.7\er{6}{8} &\ \ \ {\rm at}\ \ \  &\kappa_{l1}=0.14226,
\ \ \kappa_{l2}=0.14226
\label{threelight}\eea
The comparison between these fits and the data is shown in
Figures~\ref{figextra}, for all three combinations of light-quark
masses and also for the results extrapolated to the chiral limit.

In order to obtain the slope $\rho^2$ in the chiral limit, 
we extrapolate the three estimates of 
both $\omega$ and $\xi^{\rm ren}(\omega)$ obtained with 
$\kappa_Q=\kappa_{Q'}=0.129$ and with the combinations of light-quark
masses in Eq.~(\ref{threelight}).
We assume that $\omega$ and $\xi^{\rm ren}(\omega)$
depend linearly  on the sum of the two light-quark masses,
that is
\be
\omega(\kappa_Q,\kappa_{l1},\kappa_{l2}) 
 =  \omega(\kappa_Q) 
      + C\left(\frac{1}{2\kappa_{l1}} + \frac{1}{2\kappa_{l2}} 
      - \frac{1}{\kappa_{\rm crit}} \right) 
\label{degenerate}\ee
and similarly for $\xi^{\rm ren}$.
This assumption is supported by our results for the spectrum presented
in Ref.~\cite{noi}.  The results of the extrapolation to the chiral
limit, which are relevant for the semileptonic decay
$\Lambda_b\to\Lambda_c l \bar{\nu}$, are presented in Table~\ref{tabextra}
and in Figure~\ref{figextra}.  Our best estimate of the slope of the
renormalized Isgur-Wise function for the $\Lambda_b$ baryon is
\be
\rho^2=1.2+0.8-1.1\ .
\label{rholam}
\ee

\begin{table}
\begin{center}
\begin{tabular}{c|lll}
$\kappa_{l1}/\kappa_{l2}$    & \ \ \ \ \ \ \ $\omega$ \ \ \ \ \ \ \ \ \ & 
$\xi^{\rm ren}$ (A)\ \ \ \ \ \ 
&$\xi^{\rm ren}$ (V)\ \ \ \\
\hline
$0.14144/0.4144$ & 1.0&$1.02\err{11}{11}$ & $0.98\err{11}{12}$\\
& $1.0373\er{4}{5}$ & $0.94\er{3}{3}$ & $0.96\er{3}{3}$\\
& $1.0761\er{9}{9}$ & $0.88\er{9}{9}$ & $0.90\er{8}{8}$\\
& $1.152\er{2}{2}$ & $0.68\er{7}{6}$ & $0.62\er{8}{7}$\\
\hline
$0.14144/0.14226$ & 1.0&$0.97\err{15}{16}$ & $0.86\err{16}{17}$\\
& $1.0399\er{5}{5}$ & $0.93\er{3}{4}$ & $0.97\er{3}{4}$\\
& $1.081\er{1}{1}$ & $0.86\err{12}{11}$ & $0.90\err{10}{10}$\\
& $1.163\er{2}{2}$ & $0.71\er{9}{9}$ & $0.65\er{9}{8}$\\
\hline
$0.14226/0.14226$ &1.0 & $0.85\err{24}{25}$ &$0.63\err{27}{27}$\\
&$1.0429\er{6}{6}$ &$0.93\er{5}{6}$ &$0.95\er{5}{5}$\\
&$1.088\er{1}{1}$ & $0.82\err{18}{15}$ &$0.80\err{13}{14}$\\
&$1.175\er{3}{3}$&$0.75\err{13}{12}$ & $0.68\err{13}{11}$\\
\hline
Chiral/Chiral &1.0 &$1.02\err{23}{24}$  &$0.79\err{26}{32}$\\
&$1.0475\er{9}{9}$&$0.92\er{7}{7}$ & $0.96\er{8}{9}$\\
&$1.097\er{2}{2}$&$0.90\err{22}{20}$ & $0.95\err{19}{19}$\\
&$1.194\er{4}{4}$&$0.78\err{17}{17}$ & $0.74\err{16}{15}$\\
\hline
Chiral/Strange & 1.0 & $1.03\err{18}{19}$ & $0.87\err{22}{22}$\\
&$1.0439\er{8}{8}$ & $0.93\er{5}{6}$ & $0.96\er{6}{6}$\\
&$1.090\er{2}{2}$ & $0.90\err{17}{16}$ & $0.92\err{16}{15}$\\
&$1.179\er{3}{4}$ & $0.75\err{13}{13}$ & $0.70\err{12}{12}$
\end{tabular}
\end{center}
\caption{\em Estimates of \protect{$\xi^{\rm ren}(\omega)$} from both 
axial (A) and vector (V) 
form factors, at $\kappa_Q=\kappa_{Q'}=0.129$ and for
three combinations of the light-quark masses, as well as at the physical 
limits. The results at $\omega =1$ correspond to those from the momentum
channel $(1,0,0)\to (1,0,0)$. \label{tabextra}}
\end{table}

Using the functional form (\ref{degenerate}), we also obtain the
values of $\omega$ and $\xi^{\rm ren}(\omega)$ for the semileptonic
decay $\Xi_b \to \Xi_c l \bar{\nu}$, interpolating to one strange and
extrapolating to one chiral quark. The numerical results are also
presented in Table~\ref{tabextra}.  Our best estimate for the slope
for the $\Xi_b$ baryon is 
\be \rho^2=1.5\er{7}{9}\ .  
\label{rhoxi}
\ee

In this exploratory study we have only used a very limited set of
light-quark masses, and hence our conclusions on the dependence of the
Isgur-Wise function on these masses are rather weak, and the
extrapolation to the chiral limit is not very precise. This should be
remedied in future simulations.



\subsection{${\cal{O}}(1/m_Q)$ corrections.}
\label{sec:1overm}

In this section we attempt to extract a value of $\bar\Lambda$
for the $\Lambda$ baryon, from the study of the $1/m_Q$ corrections
in the vector \ff\   $F_1(\omega)$ and $F_{2,3}(\omega)$. We start by
recalling the relevant expressions given in Section \ref{teo}
\bea
F_1(\omega) &=& N_1(\omega)\hat\xi_{QQ'}(\omega)+\ord{1/
m_{Q^{(')}}^2} \label{F12}\\
F_2(\omega)&=&F_3(\omega)=N_2(\omega)\hat\xi_{QQ'}(\omega)+\ord{1/
m_{Q^{(')}}^2}\nn
\eea
the last equality being valid in the limit of equal heavy-quark
masses, to which the present discussion is restricted. Also in this
limit, we have
\bea
N_1(\omega)&=&\hat C_1(\bar\omega)\left[1+\frac{2}{\omega+1}\ 
\frac{\bar\Lambda}{m_Q}\right]\nn\\
N_2(\omega)&=&\hat C_2(\bar\omega)\left[1+\frac{\omega-1}{\omega+1}\ 
\frac{\bar\Lambda}{m_Q}\right]-
\hat C_1(\bar\omega)\frac{1}{\omega+1}\ \frac{\bar\Lambda}{m_Q}.
\label{N12}\eea
Combining Eqs.~(\ref{F12}) and (\ref{N12}) and using the functions
$\hat\xi_{QQ'}$ which were determined from the fits to the
dominant \ff\  $G_1$ and $\sum F_i$ (see Section \ref{sec:hqdep}), one
can view the \ff\  $F_1$ and $F_2$ as functions of $\bar\Lambda$ alone.
In fact, Eqs.~(\ref{F12}) and (\ref{N12}), together with the
coefficient functions $\hat{C}_i(\hat \omega)$, evaluated at one
loop order in perturbation theory, can be considered as our definition(s)
of the binding energy.  As mentioned in Section \ref{teo}, these form
factors are sufficiently sensitive to $\bar\Lambda$ for us to attempt
an estimate $\bar\Lambda$ from the lattice data.  

To the order we are working in, the power corrections can be expressed
as powers of the inverse quark mass or of the inverse mass of any
hadron containing the heavy quark, provided that the same prescription
is used in the evaluation of the coefficient functions
$\hat{C}_i^{(5)}$. We have decided to use the inverse quark mass,
defined as
\be
m_Q= M_{\Lambda_{Q}} -\bar\Lambda\ .
\label{mbaryon}
\ee

\begin{figure}
\begin{picture}(140,80)
\put(-7,0){\ewxy{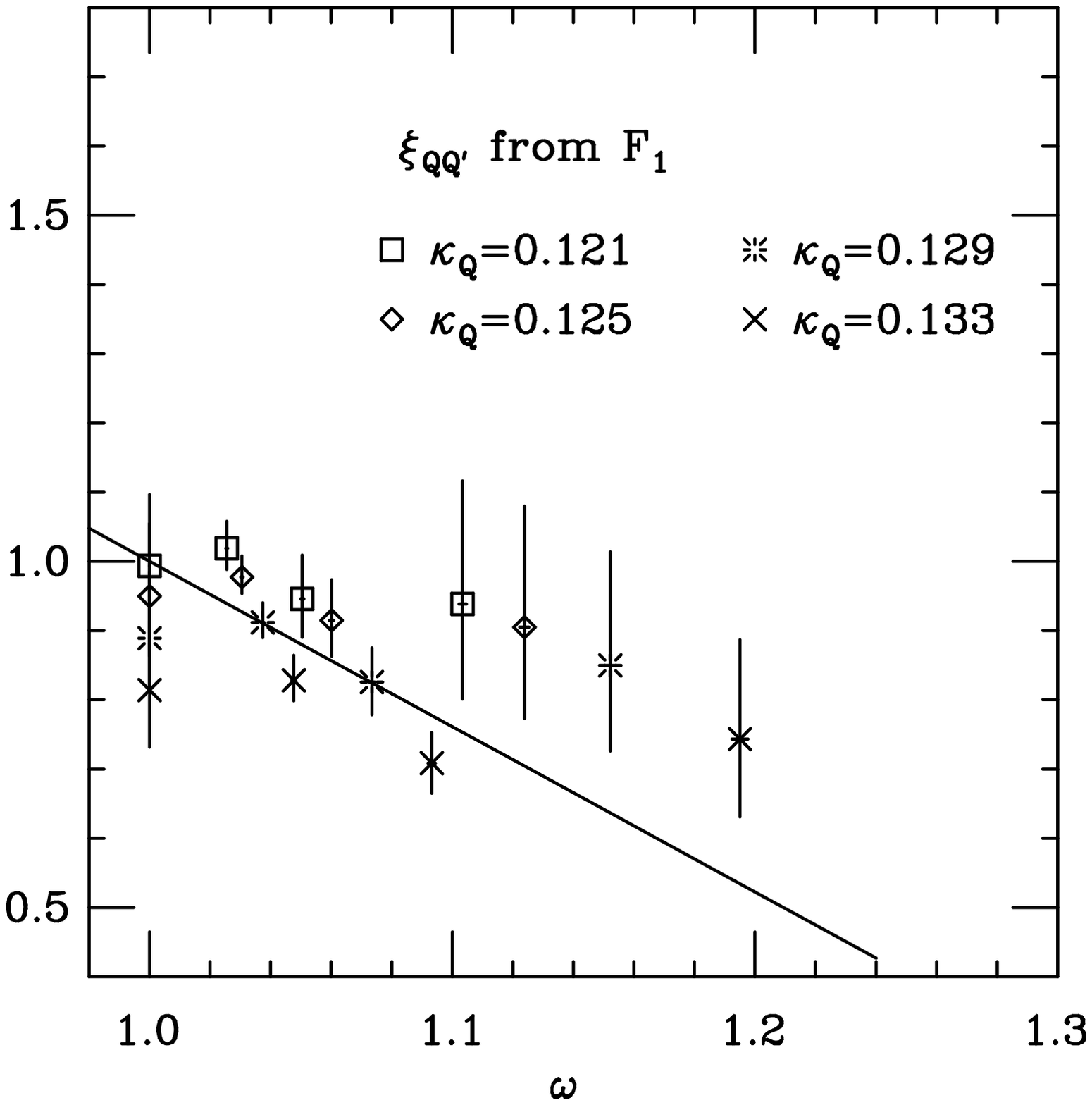}{90mm}}
\put(63,0){\ewxy{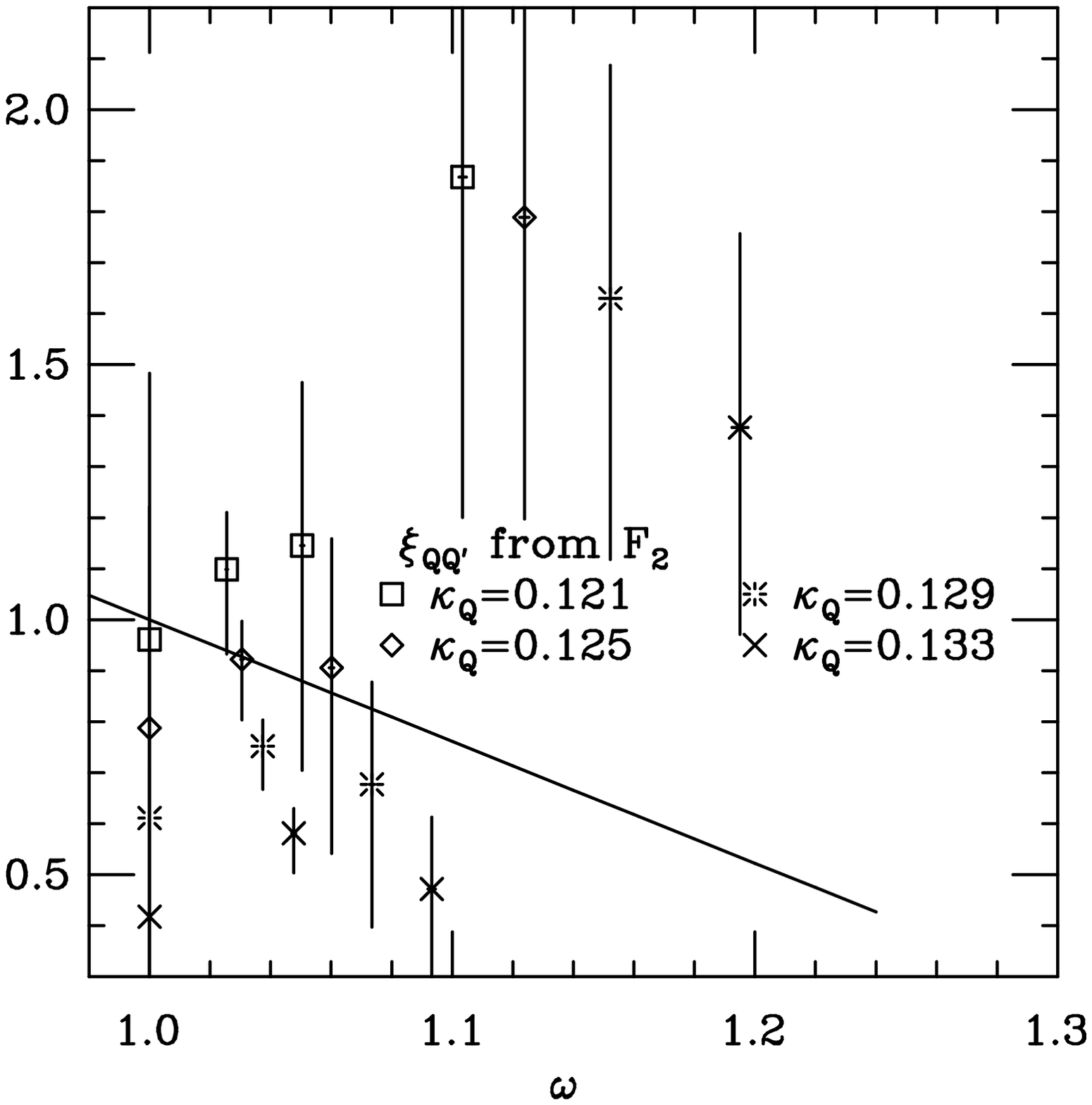}{90mm}}
\end{picture}
\caption{\em $\hat\xi_{QQ'}(\omega)$ as obtained from
  the \ff\  $F_1$ and $F_2=F_3$. The points are compared
  with the best fit of \ $\hat\xi_{QQ'}(\omega)$ obtained previously
  from the dominant \ff, $G_1$ and $\sum F_i$.}
\label{fig:F_iren}
\end{figure}

The analysis is performed using our data for degenerate initial and
final heavy quarks ($\kappa_Q=\kappa_{Q'}= 0.121, 0.125, 0.129$ and
0.133) for fixed light-quark masses,
$\kappa_{l1}=\kappa_{l2}=0.14144$.
The fit to $F_1(\omega)$ is good, and we obtain
\be
\bar\Lambda = 0.75\err{10}{13}\er{5}{6}\ \mbox{GeV}\qquad{\rm with}\qquad 
\chi^2_{dof}=1.0.
\label{bl1}
\ee
The less certain fit to $F_2(\omega)$ confirms this value with unexpected
precision, given the statistical errors affecting individual points,
\be
\bar\Lambda = 0.74\err{10}{11}\er{5}{5} \ \mbox{GeV}\qquad{\rm with}\qquad 
\chi^2_{dof}=1.6.
\label{bl2}\ee
In both Eqs.~(\ref{bl1}) and (\ref{bl2}) the first error is statistical
and the second is due to the uncertainty in the value of the lattice
spacing, see Eq.~(\ref{eq:ainv}).
Using these values of $\bar\Lambda$ we evaluate the coefficients 
$N_{1,2}(\omega)$ and estimate the 
function $\hat\xi_{QQ'}(\omega)$ 
from the \ff\   $F_1$ and $F_2$ using Eqs.~(\ref{F12}).
These estimates are presented in Tables~\ref{tab:IWF1} and \ref{tab:IWF2},
and are compared with the functions $\hat\xi_{QQ'}(\omega)$ obtained 
using the dominant \ff, $G_1$ and $\sum F_i$,  in Figure~\ref{fig:F_iren}.

\begin{table}
\begin{center}
\begin{tabular}{c|cc|cc}
$\kappa_Q\to\kappa_{Q'}$&$\omega$&$\hat\xi_{QQ'}$&$\omega$&
$\hat\xi_{QQ'}$\\
\hline
$0.121\to 0.121$
&$1.0$&$0.99\err{10}{9}$
&1.026&$  1.02\err{4}{3}$\\
$0.125\to 0.125$
&$1.0$&$  0.95\err{10}{9}$
&1.030&$  0.98\er{3}{2}$\\
$0.129\to 0.129$
&$ 1.0$&$  0.89\err{10}{8}$
&1.037&$  0.91\er{3}{2}$\\
$0.133\to 0.133$
&$1.0$&$  0.81\err{10}{8}$
&1.048&$  0.83\er{3}{3}$\\
\hline
$0.121\to 0.121$
&$1.050$&$  0.95\er{6}{6}$
&$ 1.103$&$  0.94\err{18}{14}$\\
$0.125\to 0.125$
&$1.060$&$  0.91\er{6}{5}$
&$1.124$&$  0.90\err{17}{13}$\\
$0.129\to 0.129$
&$1.073$&$  0.83\er{5}{5}$
&$1.152$&$  0.85\err{16}{12}$\\
$0.133\to 0.133$
&$1.093$&$  0.71\er{4}{4}$
&$1.195$&$  0.74\err{14}{11}$
\end{tabular}
\end{center}
\caption{\em $\hat\xi_{QQ'}$ as obtained from the
\ff\   $F_1$ for degenerate heavy quark transitions using Eqs.
(\protect\ref{F12}) and (\protect\ref{N12}) 
and the value of $\bar\Lambda$ given
in Eq.~(\protect\ref{bl1}). \label{tab:IWF1}}
\end{table}

\begin{table}
\begin{center}
\begin{tabular}{c|cc|cc}
$\kappa_Q\to\kappa_{Q'}$&$\omega$&$\hat\xi_{QQ'}$&$\omega$&$\hat\xi_{QQ'}$\\
\hline
$0.121\to 0.121$
&$1.0$&$  0.96\err{52}{70}$
&$1.026$&$  1.10\err{11}{17}$\\
$0.125\to 0.125$
&$1.0$&$  0.79\err{43}{58}$
&$1.030$&$  0.92\err{7}{12}$\\
$0.129\to 0.129$
&$1.0$&$  0.61\err{37}{45}$
&$1.037$&$  0.75\err{52}{85}$\\
$0.133\to 0.133$
&$1.0$&$  0.42\err{30}{35}$
&$1.048$&$  0.58\er{5}{8}$\\
\hline
$0.121\to 0.121$
&$1.050$&$  1.15\err{32}{44}$
&$1.103$&$  1.87\err{59}{66}$\\
$0.125\to 0.125$
&$1.060$&$  0.91\err{25}{36}$
&$1.124$&$  1.79\err{53}{60}$\\
$0.129\to 0.129$
&$ 1.073$&$  0.68\err{20}{28}$
&$1.152$&$  1.63\err{45}{51}$\\
$0.133\to 0.133$
&$   1.094$&$  0.47\err{14}{22}$
&$  1.195$&$  1.38\err{38}{40}$
\end{tabular}
\end{center}
\caption{\em $\hat\xi_{QQ'}$ as obtained from the
\ff\   $F_2=F_3$ for degenerate heavy quark transitions using Eqs.
(\protect\ref{F12}) and (\protect\ref{N12}) 
and the value of $\bar\Lambda$ given
in Eq.~(\protect\ref{bl2}). \label{tab:IWF2}}
\end{table}

Let us comment on the results for the $\Lambda$-baryon binding energy.
\begin{itemize}

\item 

As expected, the form factors $F_1(\omega)$ and $F_2(\omega)$ have a
significant dependence on the masses of the heavy quarks, a dependence
which can be partially understood in terms of the $1/m_Q$ corrections
in Eqs.~(\ref{F12}) and (\ref{N12}) with the values of $\bar\Lambda$ in
Eqs.~(\ref{bl1}) and (\ref{bl2}).

\item
We have neglected all higher-order $1/m_Q$ corrections. These could
alter the results of Eqs.~(\ref{bl1}) and (\ref{bl2}) by as much as
30\%.

\item 
The separation of $1/m_Q$ corrections from discretisation errors is
difficult. Although we have used an improved action to reduce these
errors,  and have normalized the form factors by $Z_V$ defined by
Eq.~(\ref{zvdef}), the stability of the results as the lattice spacing
and action are varied should be checked in future simulations.  The
consistency of the two results (\ref{bl1}) and (\ref{bl2}) may be
evidence that discretisation errors are relatively small.  

\item 

Although we have only performed this study with light quarks with
masses corresponding to $\kappa_{l1}=\kappa_{l2}=0.14144$, we can
nevertheless calculate $\bar\Lambda_{\Lambda}$ and $\bar\Lambda_{\Xi}$
by using
\bea 
\bar\Lambda -\bar\Lambda_{\Lambda}&=& M_{\rm
Baryon}-M_{\Lambda}+{\cal{O}}(\frac{1}{m_Q}) \nn\\
\qquad\mbox{and}&&\\
\bar\Lambda -\bar\Lambda_{\Xi}&=& M_{\rm
Baryon}-M_{\Xi}+{\cal{O}}(\frac{1}{m_Q})\nn
\ , 
\eea
where $\bar\Lambda$ and $M_{\mbox{Baryon}}$ are the binding energy and mass
of the baryons which we study in our computation. In this way we find:
\be
\bar\Lambda_{\Lambda} = 0.37\err{11}{12}\ {\rm GeV}\qquad {\rm
 and}\qquad \bar\Lambda_{\Xi} = 0.50\err{11}{13}\ {\rm GeV}\ ,
\label{ourlambda}\ee
where statistical and systematic errors have been combined in 
quadrature. 
The value of $\bar\Lambda_{\Lambda}$ would correspond to a mesonic
binding energy of $\bar\Lambda_{\rm mes}\simeq 80\pm 120\, {\rm  MeV}$,
in agreement, within the large error, with the result quoted in
Ref.~\cite{CG}. The central values of the quark masses obtained using
Eq.~(\ref{ourlambda}) are different from those used throughout this
paper which were obtained from Eq.~(\ref{formulamass}). We have checked
that this difference does not noticeably affect the correction
coefficients $\hat C_i^{(5)}$, and hence our estimates of physical
quantities.

\item 
In addition to the systematic uncertainties arising from lattice
artefacts, it must be remembered that the computation of power
corrections in general is a complicated subject (for a recent review
see \cite{ms}). In the present case, following standard procedure, we
are trying to quantify $1/m_Q$ corrections when we are ignorant of the
${\cal{O}}(\alpha_s^2)$ terms in the perturbation series for the
coefficient functions. Although, within our errors, we have found no
inconsistencies, we cannot be sure that the values of $\bar\Lambda$
would not change significantly if higher-order perturbative terms were
included, or that the values would agree with other definitions of
$\bar\Lambda$.

\end{itemize}

\section{Phenomenological implications}
\label{phenom}

In this section we use the results for the baryonic Isgur-Wise
function and for the baryon binding energy $\bar\Lambda$ computed
above, to obtain the physical \ff\  corresponding to the semileptonic
decays:
\be
\Lambda_b \to \Lambda_c + e \bar{\nu} \qquad {\rm and}\qquad
\Xi_b \to \Xi_c + e \bar{\nu} \ .
\ee
We also derive the expressions for the decay rates near zero recoil and
use them to make quantitative predictions. We will discuss in detail
the decay of the $\Lambda_b$; that of the $\Xi_b$ is very similar.

\subsection{Physical form factors}

The physical form factors can be reconstructed from the computed Isgur-Wise 
function $\xi^{\rm ren}(\omega)$, using the relations in
Eq.~(\ref{doppione}).  In the following we will use the result,
discussed in Section \ref{sec:hqdep}, that the function
$\hat\xi_{QQ'}(\omega)$ is effectively independent of the mass of the
heavy quark. Therefore, the \ff\  depend on the masses of the heavy
quarks only through the correction coefficients
$N_i^{(5)}(\omega)=N_i^{(5)}( \omega,m_Q,m_{Q'})$, which are
calculated using expressions (\ref{1-overm-exp}). The short-distance 
coefficients, $\hat{C}^{(5)}_i(\bar\omega,m_b,m_c)$, were
computed for $\Lambda_{QCD}=250$ MeV, $n_f=4$ and by fixing the quark
masses to the values
\be
m_b=4.8 \ {\rm GeV}, \qquad
m_c=1.45 \ {\rm GeV}.
\ee
The factors $\hat{C}_i^{(5)}$ depend on the
quark masses very mildly. On the other hand, the correction coefficients
$N_i^{(5)}$ are very sensitive to the quark masses, because they
contain $1/m_Q$ corrections.  In this case, we have expressed $m_b$
and $m_c$ as
\be
m_b=M_{\Lambda_b}-\bar\Lambda \qquad {\rm and}\qquad 
m_c=M_{\Lambda_c}-\bar\Lambda 
\ee
with $M_{\Lambda_b}=5.64$ GeV and $M_{\Lambda_c}=2.285$ GeV, respectively.
We report our estimates of the correction coefficients $N_i^{(5)}$ for
various values of $\omega$, for the $\Lambda$ and $\Xi$ decays, 
in Table~\ref{tab:N}.

\begin{table}
\begin{center}
\begin{tabular}{c|c|cccccc}
decay &$\omega$&$N_1$&$N_2$&$N_3$&$N^{5}_1$&$N^{5}_2$&$N^{5}_3$\\
\hline
&1.0 &$1.28\er{6}{6}$&$-0.19\er{4}{4}$&$-0.06\er{2}{1}$&$0.99$&
$-0.24\er{5}{4}$&$0.09\er{2}{2}$\\
$\Lambda_b\to$&1.1 &$1.25\er{6}{5}$&$-0.18\er{4}{4}$&$-0.06\er{1}{1}$&$0.97
$&$-0.23\er{5}{4}$&$0.08\er{2}{2}$\\
$\Lambda_c   $&1.2 &$1.21\er{5}{5}$&$-0.17\er{4}{3}$&$-0.05\er{1}{1}$&$0.95
$&$-0.21\er{4}{4}$&$0.08\er{2}{2}$\\
              &1.3 &$1.15\er{5}{5}$&$-0.16\er{4}{3}$&$-0.05\er{1}{1}$&$0.91
$&$-0.19\er{4}{3}$&$0.07\er{1}{1}$\\
\hline
          &1.0 &$1.32\er{5}{6}$&$-0.23\er{4}{4}$&$-0.07\er{1}{1}$&$0.99$&
$-0.28\er{5}{4}$&$0.10\er{2}{2}$\\
$\Xi_b\to$&1.1 &$1.30\er{5}{6}$&$-0.22\er{4}{4}$&$-0.07\er{1}{1}$&$0.97$&
$-0.26\er{5}{4}$&$0.10\er{2}{2}$\\
$\Xi_c   $&1.2 &$1.26\er{5}{5}$&$-0.20\er{4}{3}$&$-0.06\er{1}{1}$&$0.95$&
$-0.24\er{4}{4}$&$0.09\er{1}{2}$\\
          &1.3 &$1.19\er{4}{5}$&$-0.18\er{4}{3}$&$-0.06\er{1}{1}$&$0.91$&
$-0.22\er{4}{3}$&$0.08\er{1}{1}$\\
\end{tabular}
\end{center}
\caption{\em Correction  factors needed to relate the \ff\  at the physical 
limit with the Isgur-Wise function. \label{tab:N}}
\end{table}

It is convenient to expand the physical \ff\  in $\omega - 1$, near zero
recoil:
\be F_i(\omega,m_b,m_c) = \eta_i^V
-\tilde{\rho}^V_i(\omega-1) ,
\qquad G_i(\omega,m_b,m_c) = \eta_i^A
-\tilde{\rho}^A_i(\omega-1);
\label{physff}\ee
where the normalisations $\eta_i^{V,A}$ and the new slopes are related
to the coefficients $N_i^{(5)}$ and to the slope of the Isgur-Wise
function by
\begin{equation}
\eta_i^V = N_i(1,m_b,m_c),\hspace{0.15in} \eta_i^A = N_i^5(1,m_b,m_c),
\end{equation}
\begin{equation}
\tilde{\rho}_i^V= \rho^2\,N_i(1, m_b, m_c) - 
\left.\ds{
\frac{d N_i(\omega,m_b,m_c)}{d \omega}}\right|_{\omega=1}\ ,
\end{equation}
and
\begin{equation}
\tilde{\rho}_i^A =  \rho^2\,N_i^5(1, m_b, m_c) - 
\left.\ds{\frac{d N_i^5(\omega,m_b,m_c)}{d \omega}}\right|_{\omega=1}.
\end{equation}

Our results for $\tilde\rho_i^{V,A}$ are presented in
Table~\ref{tab:newslope}.
We observe that this procedure has the effect of taking us back to the
\ff\  for physical $b\to c$ decays from the Isgur-Wise function,
which, in turn, was determined by dividing the lattice data for the
\ff\ (for unphysical quark masses) by the coefficients
$N_i^{(5)}$. Clearly, most of the uncertainty in the factors
$N_i^{(5)}$, due to their dependence on $\bar\Lambda$ and the quark
masses, is now cancelled, as it should be for any physical quantity.
To quantify this statement, we report the result of the following
exercise. We have measured the ratio of the slopes of
the form factor $G_1(\omega)$, at the
chiral limit, letting $\bar\Lambda$ vary from 200 to 600 MeV, and
changing the coefficients $N_i^{(5)}$ accordingly. We obtain
\be
\frac{\tilde\rho_1^A(\bar\Lambda=200)}{\tilde\rho_1^A(\bar\Lambda=600)}
=1.00\err{12}{8}\ ,
\ee
where the relatively large error is largely due to the extrapolation to the
chiral limit.

\begin{table}
\begin{center}
\begin{tabular}{c|c|cccccc}
decay&$ - $&$F_1$&$F_2$&$F_3$&$G_1$&$G_2$&$G_3$\\
\hline
$\Lambda_c\to\Lambda_c$&$\tilde{\rho}$&$1.8\errr{0.9}{1.5}$&$-0.4\er{2}{1}$&
$-0.10\er{7}{4}$& $1.3\errr{0.8}{1.2}$&$-0.4\er{3}{2}$&$0.16\err{6}{10}$\\
\hline
$\Xi_c\to\Xi_c$&$\tilde{\rho}$&$2.3\errr{0.9}{1.4}$&$-0.5\er{2}{2}$&$-0.15
\er{7}{4}$&$1.6\errr{0.6}{1.0}$&$-0.6\er{3}{2}$&$0.22\err{7}{10}$\\
\end{tabular}
\end{center}
\caption{\em Slope of the physical \ff\  near zero recoil.
\label{tab:newslope}}
\end{table}

\subsection{Decay rates}
\label{subsec:decay}

Following Refs.~\cite{DESY} and \cite{ivanov}, we define the helicity
amplitudes, in terms of the physical form factors (\ref{physff}), in
the velocity basis:
\bea
H^{V,A}_{\frac{1}{2},0} &=&\ds{\frac{
\sqrt{2\Mb\Mc(\omega\mp 1)}}{\sqrt{\Mb^2+\Mc^2-
2\Mb\Mc\omega}} }\times\label{helampl}
\\ 
&\times&\left((\Mb\pm\Mc)F_1^{V,A}\pm\Mc(\omega\pm 1)
F_2^{V,A}\pm\Mb(\omega\pm 1)F_3^{V,A}\right)\nn\\
H^{V,A}_{\frac{1}{2},1} &=& -2\sqrt{\Mb\Mc(\omega\mp 1)} F_1^{V,A}\nn
\eea
where, for brevity, we set
\be
F_i^{V}=F_i(\omega), \qquad F_i^{A}=G_i(\omega).
\ee
and where the upper sign corresponds to $V$ and the lower one to $A$.

The helicity amplitudes $H_{\lambda_c,\lambda_W}^{V,A}$ carry information
about the helicity of the current ($\lambda_W=0 $ for a longitudinally
polarised $W$ and $\lambda_W=\pm 1$ for transversely polarised one),
and of the daughter baryon $\Lambda_c$ ($\lambda_c=\pm 1/2$).  
The missing amplitudes can be
computed by means of the relations
\be
H_{-\lambda_c,-\lambda_W}^{V,A}=\pm
H_{\lambda_c,\lambda_W}^{V,A} .
\ee
For convenience, we also define:
\be
H_{\lambda_c,\lambda_W} =
H_{\lambda_c,\lambda_W}^{V}+H_{\lambda_c,\lambda_W}^{A}\ .
\ee

Differential decay rates can then be evaluated:
\bea \frac{d\Gamma_T}{d\omega} &=&
\frac{G_F^2}{(2\pi)^3}\, |V_{cb}|^2\,
\frac{q^2 \Mc^2\sqrt{(\omega^2-1)}}{12 \Mb}\left( 
|H_{1/2,1}|^2 + |H_{-1/2,-1}|^2
\right),\nn\\
\frac{d\Gamma_L}{d\omega} &=& \frac{G_F^2}{(2\pi)^3}\, |V_{cb}|^2\,
\frac{q^2 \Mc^2\sqrt{(\omega^2-1)}}{12 \Mb}
\left(|H_{1/2,0}|^2 + |H_{-1/2,0}|^2\right),
\label{partialrate}
\eea
where $\Gamma_T$ and $\Gamma_L$ are the contributions to the rate from 
transversely and longitudinally polarised $W$'s respectively, 
and whose sum is
\bea
\frac{d\Gamma}{d\omega} &=& \frac{G_F^2}{(2\pi)^3} |V_{cb}|^2
\frac{q^2 \Mc^2\sqrt{(\omega^2-1)}}{12 \Mb}
\label{totalrate}\\
&\times& \left( 
|H_{1/2,1}|^2 + |H_{-1/2,-1}|^2 +
|H_{1/2,0}|^2 +  |H_{-1/2,0}|^2  \right)\nn
\ .
\eea

As can be seen from Eqs.~(\ref{partialrate}) and (\ref{totalrate}),
these quantities can be estimated, near zero recoil, using our results
for the \ff, in a model-independent way. In complete analogy with
what is done for $\bar B\to D^{(*)}\ell\bar\nu$
decays, we define a form factor, ${\cal B(\omega)}$, which reduces,
in the heavy-quark limit,
to the Isgur-Wise function, $\xi^{\rm ren}(\omega)$, defined
in Eq.~(\ref{iwdef}). In terms of this form factor,
the rate of Eq.~(\ref{totalrate}) is:
\bea
\frac{d\Gamma}{d\omega} & = &\frac{G_F^2}{4\pi^3} |V_{cb}|^2
\Mc^3(\Mb-\Mc)^2\sqrt{\omega^2-1}\nn\\
&&\times\l(\frac{\omega+1}{2}\r)\ 
\frac{1+r^2-2 r(2\omega+1)/3}{(1-r)^2}\nonumber\\
&& \times \l(1+\l(\frac{\omega-1}{\omega+1}\r)\frac{1+r^2-2 r(2\omega-1)/3}
{1+r^2-2 r(2\omega+1)/3}\r)|{\cal B}(\omega)|^2
\ ,\label{rateff}
\eea
with $r=\Mc/\Mb$. Near $\omega=1$, the form factor can be
expanded as
\be
{\cal B}(\omega) = G_1(1)\l\{1-\rho^2_{\cal B}\,(\omega-1)+
\ord{(\omega-1)^2}\r\}
\ee
and the results of Tables \ref{tab:N} and \ref{tab:newslope} 
can be used to determine
$G_1(1)$ and $\rho^2_{\cal B}$. We find, combining errors in
quadrature, a slope
\be
\rho^2_{\cal B} = 1.1\pm 1.0
\label{rateffslopelam}
\ee
for $\Lambda_b\to\Lambda_c\ell\bar\nu$ decays
and
\be
\rho^2_{\cal B} = 1.4\pm 0.8
\label{rateffslopexi}
\ee
for $\Xi_b\to\Xi_c\ell\bar\nu$ decays. These values are the ones
that should be compared to the slopes obtained by performing
fits to experimental results for $d\Gamma/d\omega$
versus $\omega$, for $\omega$ near 1, when such results become available.
The results of Eqs.~(\ref{rateffslopelam}) and (\ref{rateffslopexi}) 
can also be compared to the slopes we found 
for the corresponding Isgur-Wise functions (Eqs.~(\ref{rholam}) 
and (\ref{rhoxi})).
These two sets of slopes are virtually indistinguishable, especially
given the size of our present errors. 

For both decays we find $G_1(1)=0.99$ which is just
$1+\delta_1^5(1)\alpha_s$, as it should be at the level of precision
at which we are working (see Eqs.~(\ref{c1pertexp}) and (\ref{normcond})).  
In principle, though, $G_1(1)$ receives
also $1/m_Q^2$ corrections and higher-order perturbative corrections,
both of which are beyond the precision reached in the present paper.

\bigskip

We now turn to integrated rates. 
The physical limit for
$\omega$ extends up to $\omega\simeq 1.43$, which is beyond the range
of velocity transfer accessible to us ($\omega\in[1.0,1.2]$) in the
present simulation. We thus define the partially-integrated decay
rate,
\be
\Gamma^{\rm part}_i(\omega_{\rm max}) =\int_1^{\omega_{\rm max}} d\omega 
\,\frac{d \Gamma_i}{d\omega}
\label{superior}
\ee
as a function of the upper limit of integration, for each of the
rates $i=T,L$ and $T+L$.

In Table~\ref{tabrate}, we present our results for the quantities
$\Gamma_i^{\rm part}(\omega_{\rm max})$ for several values of
$\omega_{\rm max}$.  For the case of the $\Lambda_c$, the $\Lambda_b$
and the $\Xi_c$ we have used the experimental values for the masses,
whereas, for the $\Xi_b$, which is as yet undiscovered, we have used
the value computed in our previous paper on heavy baryon spectroscopy
\cite{noi}: 
\be 
M_{\Xi_c}=2.47\ {\rm GeV} \ [{\rm Exp}]\qquad
M_{\Xi_b}=5.76\ {\rm GeV}\qquad\ [{\rm Latt}].  
\ee

\begin{table}
\begin{center}
\begin{tabular}{c|l|ccccc}
decay&$\omega_{\rm max}$&1.1& 1.15&1.20 &1.25 &1.30 \\
$\Lambda_b\to\Lambda_c$&$\Gamma^{\rm part}$&$ 0.57\er{9}{7}$&
$ 0.98\err{25}{18}$&$ 1.4\er{5}{4}$&$ 1.8\err{9}{7} $&$2.2\errr{1.4}{1.2}$\\
\hline
$\Xi_b\to\Xi_c$&$\omega_{\rm max}$&1.1&1.15& 1.20&1.25 &1.30  \\
&$\Gamma^{\rm part}$&$0.66\er{7}{8} $&$1.1\er{2}{2}$&$1.6\er{4}{5}
$&$1.9\er{7}{8}$&$2.2\errr{1.2}{1.3}$\\
\hline
decay&$\omega_{\rm max}$&1.1& 1.15&1.20 &1.25 &1.30 \\
$\Lambda_b\to\Lambda_c$&$\Gamma_L^{\rm part}$&$ 0.23\er{3}{2}$&
$ 0.44\er{8}{6}$&$ 0.71\err{17}{13}$&$ 1.0\er{3}{2} $&$1.4\er{5}{4}$\\
\hline
$\Xi_b\to\Xi_c$&$\omega_{\rm max}$&1.1& 1.15&1.20 &1.25&1.30 \\
&$\Gamma_L^{\rm part}$&$ 0.28\er{2}{3}$&$0.54\er{7}{8}$&
$0.86\err{14}{16}$&$1.2\er{2}{3}$&$1.7\er{4}{4}$\\
\hline
decay&$\omega_{\rm max}$&1.1& 1.15&1.20 &1.25 &1.30 \\
$\Lambda_b\to\Lambda_c$&$\Gamma_T^{\rm part}$&$ 0.34\er{6}{4}$&
$ 0.53\err{16}{14}$&$ 0.7\er{3}{3}$& 0.8\er{6}{5}& -\\
\hline
$\Xi_b\to\Xi_c$&$\omega_{\rm max}$&1.1&1.15& 1.20&1.25 &1.30 \\
&$\Gamma_T^{\rm part}$&$ 0.38\er{5}{5}$&$0.58\err{13}{15}$&
$0.7\er{3}{3}$& - & -\\
\end{tabular}
\end{center}
\caption{\em Partial decay rates, in units of $|V_{cb}|^2\ 10^{13}\,s^{-1}$, 
  for the $\Lambda$ and $\Xi$ semileptonic decays for various values
  of $\omega_{\rm max}$.  The transverse decay rate
  is very sensitive to quadratic terms in $(\omega-1)$, and the
  predictions for $\omega> 1.2$ are no longer reliable.
\label{tabrate}}
\end{table}

\begin{figure}
\begin{picture}(140,100)
\put(10,0){\ewxy{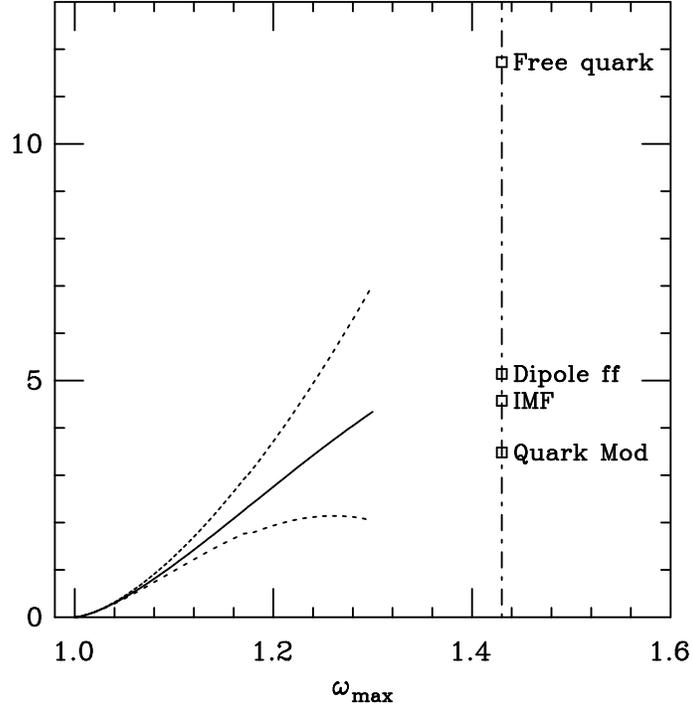}{130mm}}
\end{picture}
\caption{\em Partially-integrated 
decay rate for the process $\Lambda_b\to\Lambda_c +
  l \bar{\nu}$, 
  in units of $10^{10}\,s^{-1}$, as a function of the limit of
  integration $\omega_{\rm max}$. A comparison with several model
  estimates is shown at the end-point~\protect\cite{DESY,ivanov}. The solid
curve corresponds to our central values for $\tilde\rho$ and the dotted
curves to the errors on these values.}
\label{fig:rate}
\end{figure}

At present, a direct comparison of our results with experiments is not
possible.  Although the semileptonic decay of $\Lambda_b$ has
been observed by various experiments \cite{LEPSEMIL}, a measurement of
the decay rate is not yet available.  The problem of determining the
rate of the $\Lambda_b$ and $\Xi_b$ semileptonic decays has been
addressed theoretically, making use of different models and approaches
(e.g. infinite momentum frame (IMF), dipole form factors in Ref.~\cite{DESY}
and quark model in Ref.~\cite{ivanov}), by several authors. Their
predictions for the rate, integrated up to the end-point, are
reported in Figure~\ref{fig:rate}, and compared with the function
$\Gamma(\omega_{\rm max})$. To evaluate this function we
have assumed $|V_{cb}|=0.044$.

Finally, we note that many other interesting quantities could be
computed and confronted with future experiments, e.g. asymmetry
parameters (see for example \cite{Nm2}) and the ratio of the
longitudinal to transverse rates. However, all these quantities are
sensitive to either $1/m_Q^2$ effects or to terms proportional to
$(\omega-1)^2$. Both of these effects are beyond the precision reached
in the present study.

\section*{Conclusions}

During the past few years lattice simulations have been applied very
successfully to weak decays of heavy mesons, and in this work we have
extended these techniques to heavy baryons. We have presented an
extensive lattice study of the semileptonic 
$\Lambda_b\to\Lambda_c+ l\bar{\nu}$ and $\Xi_b\to\Xi_c +l\bar{\nu}$ decays,
resulting in predictions for the decay rates, for values of the
velocity transfer up to about $\omega = 1.2$. We have developed the
formalism necessary for extracting the decay amplitudes, demonstrated
the feasibility of obtaining phenomenologically interesting results,
and presented the first set of predictions. We anticipate that the
application of lattice QCD to studies of the forthcoming experimental
data on heavy-baryon decays will be an active area of phenomenology
during the coming years.

HQET is an important tool in the application of lattice QCD to
weak decays of heavy baryons, as it is also for mesons. In order to
keep lattice artefacts reasonably small, we are forced to perform the
computations with heavy quarks with masses not much larger than that
of the charm quark, and then to extrapolate the results to physical 
$b\to c$ decays.  HQET provides us with a sound theoretical
formalism for performing this extrapolation.

Perhaps the weakest feature of our study was the inability to
determine the behaviour of the decay amplitudes with the mass of the
light quarks with sufficient precision. We did perform the
computations for three combinations of masses for the two light
quarks, which allowed us to attempt an extrapolation to the physical
$\Lambda$ (with two almost massless valence quarks) and $\Xi$ (with
one strange and one almost massless quark) baryons. However, in
performing these extrapolations the statistical errors are
amplified, and one of the priorities of a future simulation should be
to generate substantial datasets for a larger set of light-quark masses.
For example, it will be very interesting to check whether the slope
of the Isgur-Wise function decreases as the masses of the light quarks
are reduced, as expected theoretically. In the present simulation we
have a hint that this is the case, but the statistical errors are too large
to draw a definitive conclusion.

\section*{Acknowledgements}

We are indebted to the staff of Edinburgh Parallel Computing Centre
for the provision of services on the Meiko Computing Surface and on
the Cray T3D.  LL and JN thank the Theory Group in Southampton for its
kind hospitality during the early stages of this project. This
research was supported by the UK Science and Engineering Research
Council under grant GR/G~32779, by the Particle Physics and Astronomy
Research Council under grant GR/J~21347 and by the Engineering and
Physical Sciences Research Council under grant GR/K~41663. CTS and DGR
acknowledge the Particle Physics and Astronomy Research Council for
its support through the award of a Senior Fellowship and an Advanced
Fellowship respectively. JN acknowledges support from DGES under
contract PB95-1204. OO is supported by JNICT under grant
BD/2714/93. We also acknowledge partial support by the EU contracts
CHRX-CT92-0051 and ERBCHBICT-930877.


\appendix

\section{The use of extended interpolating operators}
\label{app1}
  
In order to enhance the signal for the baryonic correlation functions,
the light and heavy quark propagators have been computed using the
Jacobi smearing method \cite{SMEARING}.  Since smearing is not a
Lorentz-invariant operation, it alters the transformation properties
of the correlation functions, particularly at non-zero momentum. In
this Appendix we present the formalism required to extract the
form factors from two- and three-point correlation functions computed
using extended (smeared) interpolating operators for the baryons.

Consider the local operator ${\cal O}(x)$ defined in Eq.~(\ref{oqdef}),
\be
{\cal O}_{\rho}(x)= \epsilon_{abc}\,
\Big(l_1^{aT}(x){\cal C}\gamma_5 l_2^b(x)\Big)
Q_{\rho}^c(x)\ ,
\label{J1}\ee
where $\rho$ is a spinor index. Here we have suppressed the index $Q$ in 
labelling the operator. ${\cal O}$ has non-zero overlap with spin
1/2 states, such the $\Lambda-$baryon:
\be
\langle 0| {\cal O}_{\rho}(0) |\vec{p}, r\rangle = Z u_{\rho}^{(r)}(\vec{p}).
\label{J2}
\ee
where $r$ is the polarisation index.\footnote{Our spinors are
normalized such that $u^{(r)\dagger}u^{(s)}=
v^{(r)\dagger}v^{(s)}=(E/m)\,\delta^{rs}$.}  The ket in Eq.~(\ref{J2})
represents a heavy $\Lambda$ state (e.g. $\Lambda_b$ or $\Lambda_c$).
The amplitude $Z$ is a Lorentz scalar.

The smeared baryonic operator can be written as
\bea
{\cal O}^s_{\rho}(\vec{x},t)&=&\epsilon_{abc}\,
\sum_{\vec{y},\vec{z},\vec{w}}
f(|\vec{y}-\vec{x}|) f(|\vec{z}-\vec{x}|) f(|\vec{w}-\vec{x}|)\times\nn\\
&\times& \Big(l_1^{aT}(\vec{y},t){\cal C}\gamma_5\,l_2^b(\vec{z},t)\Big)
Q^c_{\rho}(\vec{w},t)\ .
\eea
Because the smearing is performed only in the spatial directions, 
Lorentz symmetry is lost and 
only spatial translations, rotations, parity and time reversal survive.
Therefore, the overlap
of the operator ${\cal O}^s_{\rho}(\vec{x},t)$ with the state 
$| \vec{p}, r\rangle $
is given by the more general expression
\bea
\langle 0| {\cal O}_{\rho}^{s}(0) |\vec{p}, r \rangle &=& \left[
\Big(Z_1(|\vec{p}|) + Z_2(|\vec{p}|) 
\gamma_0 \Big) u^{(r)}(\vec{p})\right]_{\rho}\nn\\
\langle 0| \overline{{\cal O}}_{\rho}^{s}(0) |\vec{p}, r \rangle &=& \left[
\bar{v}^{(r)}(\vec{p})\Big(
Z_1(|\vec{p}|) - Z_2(|\vec{p}|) \gamma_0
\Big)\right]_{\rho}
\label{J2s}
\eea
where the amplitudes $Z_1$ and $Z_2$ may depend on the magnitude of
the three-momentum of the state $|\vec{p},r \rangle$, in accord
with the restricted symmetries of the system.


\subsection{Smeared two-point functions}
\label{seca1}

We now study the consequences of the above discussion in the case of
smeared source and sink (SS) 
two-point functions.  Using Eq.~(\ref{J2s}), we can derive the
general expression for the two-point function at large values of $t$
and $(T-t)$~\footnote{Note that we are using anti-periodic boundary
  conditions in time.}
\bea
&&G_{\rho\sigma}^{SS}(t,\vec{p})=\sum_{\vec{x}}e^{-i\vec{p}\cdot\vec{x}}
\langle 
{\cal O}_{\rho}(\vec{x},t) \bar{\cal O}_{\sigma}(\vec{0},0)\rangle 
 \\
\nn\\
&&= \sum_{|\vec{q},r\rangle} \sum_{\vec{x}} 
\frac{m}{E(\vec{q})}e^{-i\vec{p}\cdot\vec{x}}
\left[e^{-E(\vec{q})t+i\vec{q}\cdot\vec{x}}
\langle 0| {\cal O}_{\rho}(\vec{0},0)| \vec{q},r\rangle
\langle \vec{q},r|\overline{\cal O}_{\sigma}(\vec{0},0)|0\rangle- \right.\nn\\
&&\qquad\qquad -e^{-E(\vec{q})(T-t)-i\vec{q}\cdot\vec{x}}\left.
\langle 0|\overline{\cal O}_{\sigma}(\vec{0},0)| \vec{q},r\rangle
\langle \vec{q},r| {\cal O}_{\rho}(\vec{0},0)|0\rangle \right]
\nn\\
&&= \sum_{r}\frac{m }{E(\vec{p})} 
\left\{\left[ e^{-E(\vec{p})t}
\Big( Z_1(|\vec{p}|) + Z_2(|\vec{p}|)\gamma_0\Big)
u^{(r)}(\vec{p})\right.\right.\times\nn\\
&&\qquad\qquad\qquad\times\left.\bar{u}^{(r)}(\vec{p})
\Big( Z_1(|\vec{p}|) + Z_2(|\vec{p}|)\gamma_0\Big) \right]_{\rho\sigma}
\nn \\
&&\qquad\qquad-\left[ e^{-E(\vec{p})(T-t)}
\Big( Z_1(|\vec{p}|) - Z_2(|\vec{p}|)\gamma_0\Big)
v^{(r)}(-\vec{p})\right.\times\nn\\
&&\qquad\qquad\qquad\left.\times\left.\bar{v}^{(r)}(-\vec{p})
\Big( Z_1(|\vec{p}|) - Z_2(|\vec{p}|)\gamma_0\Big) \right]_{\rho\sigma}
\right\}
\ .
\nn
\eea

We find it convenient to write the spin matrix $G^{SS}_{\rho\sigma}$
in terms of the parameters
\be
Z_s=Z_1+Z_2,\qquad\qquad \alpha=(Z_1-Z_2)/(Z_1+Z_2) \ :
\label{redefinition}
\ee 
rather than $Z_1$ and $Z_2$:
\bea
G^{ss}_{\rho\sigma}(t,\vec{p})&=&Z_s^2(|\vec{p}|) e^{-E(\vec{p})t}  \left\{
\left[
\frac{E+m-\alpha^2(E-m)}{4E}\one +\right.\right.\\
&&\qquad\qquad\left.+\frac{E+m+\alpha^2(E-m)}{4E}\gamma_0 -
\frac{2\alpha}{4E}\vec{p}\cdot\vec{\gamma} \right] \nn \\
&-& e^{-E(\vec{p})(T-t)} \left[
\frac{E+m-\alpha^2(E-m)}{4E}\one -\right.\nn\\
&&\qquad\qquad-\left.\left.\frac{E+m+\alpha^2(E-m)}{4E}\gamma_0 -
\frac{2\alpha}{4E}\vec{p}\cdot\vec{\gamma} \right]\right\}.\nn
\eea
Local operators, and full 4-dimensional cubic symmetry, correspond
to the case $\alpha=1$, i.e. $Z_2=0$.

\subsection{Smeared three-point functions for the study of the  
  semileptonic decay of the $\Lambda_b$.} 

We now present the expressions for the smeared-smeared three-point
functions from which the form factors are extracted.  The $V-A$ weak
current $J_\mu$ is of course a local operator; it is the
interpolating operators for the baryons in Eq.~(\ref{main3p}) which are
now smeared. In the forward half of the lattice ($F$), for large t, so
that only the lightest state contributes, we obtain
\bea
C^F(t_x,t_y) &=& K(t_x,t_y) 
\left\{ \Big[ E +M +\alpha(M-E) \Big]\one\right. +\label{SSF}
\\
&&\qquad+\Big[ E +M -\alpha(M-E)
  \Big]\gamma_0 + \Big[ (1-\alpha)\vec{p}\cdot\vec{\gamma}\gamma_0 \Big]
-\nn\\ 
&&\qquad-\left.\Big[ (1+\alpha)\vec{p}\cdot\vec{\gamma} \Big] \right\} 
\times \nn\\
&\times&
\left\{ \left[ F_1^L(\omega)\gamma_{\mu} + F_2^L (\omega) v_{\mu}^{\prime}
+ F_3^L (\omega) v_{\mu}\right]\right.- \nn\\
&&\qquad\qquad -\left.\left[\left(
      G_1^L(\omega)\gamma_{\mu} + G_2^L (\omega) v_{\mu}^{\prime} + G_3^L
      (\omega) v_{\mu}\right)\gamma_5 \right]\right\} \times\nn \\
&\times&\left\{ \Big[ E' +M' +\beta(M'-E') \Big]\one + \Big[ E' +M'
  -\beta(M'-E') \Big]\gamma_0\right.-\nn\\
&&\qquad\qquad-\left.\Big[
  (1-\beta)\vec{p'}\cdot\vec{\gamma}\gamma_0 \Big] -
  \Big[ (1+\beta)\vec{p'}\cdot\vec{\gamma} \Big] \right\} \nn
\eea
whereas in the backward half ($B$), for large $T-t$,
\bea
C^B(t_x,t_y)  &=& K(T-t_x,T-t_y)  
\left\{ \Big[ E +M +\alpha(M-E) \Big]\one\right. -\label{SSB}
\\
&&\qquad+    \Big[ E +M -\alpha(M-E) \Big]\gamma_0 -
        \Big[ (1-\alpha)\vec{p}\cdot\vec{\gamma}\gamma_0 \Big] -\nn\\
&&\qquad-\left. \Big[ (1+\alpha)\vec{p}\cdot\vec{\gamma} \Big] \right\} 
\times \nn \\
&\times&
\left\{ 
\left[ F^L_1(\omega)\gamma_{\mu} - F^L_2 (\omega) \tilde v_{\mu}^{\prime} - 
F^L_3 (\omega) \tilde v_{\mu}\right]\right. +\nn\\
&&\qquad\qquad+\left.\left[\left( G_1^L(\omega)\gamma_{\mu} + 
G_2^L (\omega) \tilde v_{\mu}^{\prime} + 
G_3^L (\omega) \tilde v_{\mu}\right)\gamma_5 \right]
\right\} \times\nn \\
&\times&\left\{ \Big[ E' +M' +\beta(M'-E') \Big]\one -
        \Big[ E' +M' -\beta(M'-E') \Big]\gamma_0\right. +\nn\\
&&\qquad\qquad+\left.\Big[ (1-\beta)\vec{p'}\cdot\vec{\gamma}\gamma_0 \Big] -
        \Big[ (1+\beta)\vec{p'}\cdot\vec{\gamma} \Big] \right\} \nn
\eea
where $\tilde v\equiv (E,-\vec{p})/M$ (similarly for $\tilde v'$)
and where in both these equations the normalisation factor is given by:
\be
K(t_x,t_y) = \frac{Z_sZ'_s}{16 EE'} e^{-E(t_x-t_y)} e^{-E't_y}.
\ee
Since $t_x = 24=T/2$ in our simulations, the dependence on
$t_x$ is the same in both halves of the lattice.
In Eqs.~(\ref{SSF}) and (\ref{SSB}) $\alpha$ and $\beta$ are the
functions defined in (\ref{redefinition}), for the final and initial
particle respectively. They are identical only for degenerate
transitions when $|\vec{p}|=|\vec{p'}|$.  For local operators, 
$\alpha = \beta = 1$ and expressions~(\ref{SSF}-\ref{SSB}) reduce to
Eq.~(\ref{local}). 

Expressions (\ref{SSF}) and (\ref{SSB}) are used in Section \ref{ana}
to extract the values of the form factors from the correlation
functions. Further clarification of the procedures which are used is
pressented in the following appendix.


\section{On the analysis of three-point correlators}
\label{app2}

In this appendix we discuss in greater detail the analysis procedure
used to extract the six form factors from the correlation functions.
Such an analysis is complicated both by the very large number of
non-vanishing components of the three-point correlation function,
which has both spinorial and Lorentz indices, and by the effect of the
smearing on the baryonic operators discussed in the previous appendix.
It proved to be convenient to restrict the analysis to those components
which are proportional to large, and hence precisely measured, 
kinematical coefficients, as we now explain.

For illustration we consider here the forward half of the lattice; the
extension to the backward half is straightforward.  We consider two
typical cases.
\begin{itemize}
\item{\bf Example 1}: Coefficient of the form factor $F_1^L$
for $\mu=0$.

  We rewrite the expression in Eq.~(\ref{SSF}), which is a $4\times 4$
  matrix in spinor space, in terms of $2\times 2$ matrices~\footnote
{The following representation for the gamma matrices is used here:
\be
\gamma_0=\left(\ba{cc}1 &0\\ 0 &-1\ea\right);\ \ \ 
\gamma_i=\left(\ba{cc}0 &\sigma_i\\-\sigma_i&0\ea\right)\ \
{\rm with }\ i=1,2,3.
\ee}:
\bea
  \frac{{\rm Vector}_0}{K(t_x,t_y)}
  &=& 4\left[(E +M)(E'+M') +\vec{p}\cdot \vec{p}^{\,\,\prime}\right]
(\one+\gamma_0)  \nn\\
  &+& 4\alpha\beta\left[-(E -M)(E'-M') -\vec{p}\cdot \vec{p}^{\,\,\prime}
\right](\one-\gamma_0)  \nn\\
  &+&  4ip_ip_j'\sigma^{ij}(\one+\gamma_0) -4\alpha\beta ip_ip_j'\sigma^{ij}
(\one-\gamma_0) \nn\\
  &+& 4\alpha\left[ (M-E)p_i- (E'+M')p_i\right]\gamma_i (\one+\gamma_0)\nn \\
  &-& 4\beta\left[ (M+E)p_i- (E'-M')p_i\right]\gamma_i (\one-\gamma_0),
\label{separation}
\eea
where Vector$_0$ denotes the time component of the three point
correlation function with the vector current.
So 
the matrix structure shared by the coefficients
of the vector form factors is of the form:
\be
\frac{{\rm Vector}_0}{K(t_x,t_y)} \propto
\left(\ba{cc}
\left(\ba{c}
\\ \ \  1\ \  \\ \\
\ea\right) &
\left(\ba{c}
\\ \ \  \beta\ \  \\ \\ \ea\right) \\
\left(\ba{c}
\\ \ \ \alpha\ \  \\ \\ \ea\right) &
\left(\ba{c}
\\  \ \alpha\beta \  \\ \\ \ea\right) 
\ea\right)
\label{submatrix}\ee
where each of the submatrices is a $2\times 2$ matrix.

\item{\bf Example 2}: Coefficient of the form factor $G_1^L$
for $\mu=k=1,2,3$. 

A similar manipulation of Eqs.~(\ref{SSF}) for the axial current
leads to
\bea
\frac{{\rm Axial}_k}{K(t_x,t_y)}
&=& \Big\{ 4\left[(M+E)(M'+E')\right]\gamma_k (\one+\gamma_0) 
\label{separationA}\\
&+&4\alpha\beta\left[(M-E)(M'-E')\right]\gamma_k (\one-\gamma_0)  \nn\\
&+& 4\alpha\left[(M-E)p_i'\gamma_k\gamma_i -
(M'+E')p_i\gamma_i\gamma_k\right](\one+\gamma_0)  \nn\\
&+& 4\beta\left[(M+E)p_i'\gamma_k\gamma_i +(E'-M')p_i\gamma_i\gamma_k\right]
(\one-\gamma_0)  \nn\\
&&\mbox{\hspace{-1cm}}
+ 4\left[p_ip_j'\gamma_i\gamma_k\gamma_j\right](\one+\gamma_0) 
+ 4\alpha\beta\left[p_ip_j'\gamma_i\gamma_k\gamma_j\right](\one-\gamma_0)\Big\}
\gamma_5\nn
\ ,
\eea
which also has the structure of (\ref{submatrix}).
\end{itemize}

The relative sizes of the components
of the matrices (\ref{separation}) and (\ref{separationA}) are as follows:

\begin{itemize}

\item the submatrices proportional to the identity
are proportional to $MM'$ (plus small terms quadratic in the momenta),
and are therefore {\em large} and precisely determined.

\item the submatrices proportional to $\alpha$ and $\beta$
are proportional to terms like $M p_i'$ or
$M'p_i$ and are therefore {\em medium}-sized. Furthermore
they have an additional statistical uncertainty due to the presence
of the amplitude factors $\alpha$ or $\beta$.

\item the submatrices proportional to $\alpha\beta$ are
proportional to $(E'-M')(E-M)$ or other terms which are quadratic
in the momenta, and are therefore {\em small}.

\end{itemize}

Finally, we note that, as a general feature, the vector current will
give large contributions for $\mu = 0$ and the axial current when $\mu
= i$, $i=1,2,3$.



\section{Determination of $Z_V$}
\label{sec:zv}

In this appendix we discuss the normalisation of the lattice vector
current, used for extracting the vector form factors. The
normalisation factor, $Z_V$ has previously been determined from matrix
elements between heavy meson states, using the same configurations,
and for the same values of the quark masses as those used in the
present study~\cite{lpl}, and we present the results below. We start
however with a determination of $Z_V$ from matrix elements between
heavy baryon states.

$Z_V$ can be measured from the correlators (\ref{main3p}), computed
for degenerate initial and final heavy quarks ($Q=Q'$). For instance,
with Dirac index $i=1,2$, we define
\be
Z_V(t) ={\cal K} \frac{\l[C_2^Q(\vec{p},T/2)\r]_{ii}}{\l[C(\vec{p},
\vec{q}=\vec{0},
T/2,t)^{Q\rightarrow Q}_{\mu=0}\r]_{ii}} =
\frac{1}{F^L_1(1)+F^L_2(1)+F^L_3(1)}
\label{zvc2c3}\ee
with
\be
{\cal K} = \frac{ (E+M)}{2 E}\frac{2 E}{E+M-\alpha^2(M-E) }
\ee
and where $C^Q_2$ was defined in Eq.~(\ref{defC2}) and $t$ is taken
in $(0,T/2)$. $Z_V$ is then obtained by fitting $Z_V(t)$ to a constant
in the plateau region. 

We have computed the necessary two-
and three-point correlators for two different forward channels:
$\vec{p}=\vec{0}$ and $\vec{p} = (p_{\mbox{min}},0,0)$.  
From the correlation functions with the baryon at rest we obtain:
\be
\ba{lccr}
Z_V & = & 0.898\er{3}{2} \hspace{0.25in} {\rm at\ }&\kappa_{l1} =
\kappa_{l2} = 0.14144,\ \kappa_Q=0.133\\ 
Z_V & = & 0.924\er{2}{2} \hspace{0.25in} {\rm at\ }&\kappa_{l1} =
\kappa_{l2} = 0.14144,\ \kappa_Q=0.129\\ 
Z_V & = & 0.948\er{2}{2} \hspace{0.25in} {\rm at\ }&\kappa_{l1} =
\kappa_{l2} = 0.14144,\ \kappa_Q=0.125\\ 
Z_V & = & 0.970\er{1}{2} \hspace{0.25in} {\rm at\ }&\kappa_{l1} =
\kappa_{l2} = 0.14144,\ \kappa_Q=0.121
\ea .
\label{nostraZV}
\ee
The statistical errors are very small (although not as small as for
meson states), which is not unexpected since we are studying the
effects of the charge operator.

For $Z_V$ measured from correlation functions with 
$\vec{p}=(p_{\rm min},0,0)$, the statistical errors are too
large for us to make any comparison with the results in (\ref{nostraZV}).
In this case we find:
\be
\ba{lccr}
Z_V & = & 0.88\err{15}{10} \hspace{0.25in} {\rm at\ }&\kappa_{l1} =
\kappa_{l2} = 0.14144,\ \kappa_Q=0.133\\ 
Z_V & = & 0.90\err{14}{11} \hspace{0.25in} {\rm at\ }&\kappa_{l1} =
\kappa_{l2} = 0.14144,\ \kappa_Q=0.129\\ 
Z_V & = & 0.94\err{14}{11} \hspace{0.25in} {\rm at\ }&\kappa_{l1} =
\kappa_{l2} = 0.14144,\ \kappa_Q=0.125\\ 
Z_V & = & 0.97\err{15}{12} \hspace{0.25in} {\rm at\ }&\kappa_{l1} =
\kappa_{l2} = 0.14144,\ \kappa_Q=0.121\ .
\ea
\label{nostraZVp}
\ee

We now compare the results for $Z_V$ obtained between baryonic states
(Eq.~(\ref{nostraZV})\,), and those obtained in Ref.~\cite{lpl}  
from matrix elements between pseodoscalar
states using the same configurations and quark masses:
\be
\ba{lccr}
Z_V & = & 0.8913\er{2}{1} \hspace{0.25in} {\rm at\ }&\kappa_{l} = 0.14144,\ 
\kappa_Q=0.133\\
Z_V & = & 0.9177\er{3}{2} \hspace{0.25in} {\rm at\ }&\kappa_{l} = 0.14144,\ 
\kappa_Q=0.129\\
Z_V & = & 0.9428\er{4}{2} \hspace{0.25in} {\rm at\ }&\kappa_{l} = 0.14144,\ 
\kappa_Q=0.125\\
Z_V & = & 0.9659\er{6}{3} \hspace{0.25in} {\rm at\ }&\kappa_{l} = 0.14144,\ 
\kappa_Q=0.121 \ .
\ea 
\label{lplZV}
\ee
The agreement of the results obtained for $Z_V$ using mesonic and
baryonic correlation functions to within less than 1\% is reassuring.

The variation of the values of $Z_V$ with the mass of the heavy quark
in (\ref{nostraZV}) and (\ref{lplZV}) is an effect of the discretisation
errors, due to the fact that the quark $Q$ is heavy. To see this more
clearly we compare the results with those obtained between light
pseudoscalar mesons (with degenerate valence quarks): 
\bea
Z_V & = & 0.8314(4) \hspace{0.25in} {\rm at\ }\kappa = 0.14144\nonumber\\
Z_V & = & 0.8245(4) \hspace{0.25in} {\rm at\ }\kappa = 0.14226\nonumber\\
Z_V & = & 0.8214(6) \hspace{0.25in} {\rm at\ }\kappa = 0.14262\ .
\label{eq:zvjonivar}\eea 
The results in (\ref{eq:zvjonivar}) were obtained on a subset of 10 gluon
configurations \cite{jonivar}. The dependence on the masses of the
light quarks is seen to be very mild, and the reuslts are in
consistent with the expectations from one-loop perturbation theory
\cite{borrelli}:
\be
Z_V = 1 - 0.10 g^2 + {\cal{O}}(g^4)\simeq 0.83\mbox{   at   }\beta=6.2
\label{eq:zvpert}
\ee
when evaluated using the boosted value of the coupling constant, 
obtained from the mean field resummation of tadpole diagrams\cite{lm}. 

The results for $Z_V$ in Eq.~(\ref{nostraZV}) and (\ref{lplZV}),
obtained using heavy baryon and meson states, differ from those
obtained with light mesons, (\ref{eq:zvjonivar}), by about 10-20\% for
the range of quark masses used in our simulations. This difference is
a good indication of the size of mass-dependent discretisation errors
in our calculation; it is consistent with our expectation that they
should be of $\ord{\alpha_s am_Q }$ and $\ord{a^2 m_Q^2}$. However, as
explained in Section~\ref{ana}, the errors in the computed
form factors are expected to be considerably smaller, because we
normalise all of them by $Z_V$ also determined between heavy baryon
states.




\def \ajp#1#2#3{Am. J. Phys. {\bf#1}, #2 (#3)}
\def \apny#1#2#3{Ann. Phys. (N.Y.) {\bf#1}, #2 (#3)}
\def \app#1#2#3{Acta Phys. Polonica {\bf#1}, #2 (#3)}
\def \arnps#1#2#3{Ann. Rev. Nucl. Part. Sci. {\bf#1}, #2 (#3)}
\def \cmts#1#2#3{Comments on Nucl. Part. Phys. {\bf#1}, #2 (#3)}
\def \cn{Collaboration}
\def \cp89{{\it CP Violation,} edited by C. Jarlskog (World Scientific,
Singapore, 1989)}
\def \efi{Enrico Fermi Institute Report No. EFI}
\def \f79{{\it Proceedings of the 1979 International Symposium on Lepton and
Photon Interactions at High Energies,} Fermilab, August 23-29, 1979, ed. by
T. B. W. Kirk and H. D. I. Abarbanel (Fermi National Accelerator Laboratory,
Batavia, IL, 1979}
\def \hb87{{\it Proceeding of the 1987 International Symposium on Lepton and
Photon Interactions at High Energies,} Hamburg, 1987, ed. by W. Bartel
and R. R\"uckl (Nucl. Phys. B, Proc. Suppl., vol. 3) (North-Holland,
Amsterdam, 1988)}
\def \ib{{\it ibid.}~}
\def \ibj#1#2#3{~{\bf#1}, #2 (#3)}
\def \ichep72{{\it Proceedings of the XVI International Conference on High
Energy Physics}, Chicago and Batavia, Illinois, Sept. 6 -- 13, 1972,
edited by J. D. Jackson, A. Roberts, and R. Donaldson (Fermilab, Batavia,
IL, 1972)}
\def \ijmpa#1#2#3{Int. J. Mod. Phys. A {\bf#1}, #2 (#3)}
\def \ite{{\it et al.}}
\def \jpb#1#2#3{J.~Phys.~B~{\bf#1}, #2 (#3)}
\def \lkl87{{\it Selected Topics in Electroweak Interactions} (Proceedings of
the Second Lake Louise Institute on New Frontiers in Particle Physics, 15 --
21 February, 1987), edited by J. M. Cameron \ite~(World Scientific, Singapore,
1987)}
\def \ky85{{\it Proceedings of the International Symposium on Lepton and
Photon Interactions at High Energy,} Kyoto, Aug.~19-24, 1985, edited by M.
Konuma and K. Takahashi (Kyoto Univ., Kyoto, 1985)}
\def \mpla#1#2#3{Mod. Phys. Lett. A {\bf#1}, #2 (#3)}
\def \nc#1#2#3{Nuovo Cim. {\bf#1}, #2 (#3)}
\def \np#1#2#3{Nucl. Phys. {\bf#1}, #2 (#3)}
\def \PDG{Particle Data Group, L. Montanet \ite, \prd{50}{1174}{1994}}
\def \pisma#1#2#3#4{Pis'ma Zh. Eksp. Teor. Fiz. {\bf#1}, #2 (#3) [JETP Lett.
{\bf#1}, #4 (#3)]}
\def \pl#1#2#3{Phys. Lett. {\bf#1}, #2 (#3)}
\def \pla#1#2#3{Phys. Lett. A {\bf#1}, #2 (#3)}
\def \plb#1#2#3{Phys. Lett. B {\bf#1}, #2 (#3)}
\def \pr#1#2#3{Phys. Rev. {\bf#1}, #2 (#3)}
\def \prc#1#2#3{Phys. Rev. C {\bf#1}, #2 (#3)}
\def \prd#1#2#3{Phys. Rev. D {\bf#1}, #2 (#3)}
\def \prl#1#2#3{Phys. Rev. Lett. {\bf#1}, #2 (#3)}
\def \prp#1#2#3{Phys. Rep. {\bf#1}, #2 (#3)}
\def \ptp#1#2#3{Prog. Theor. Phys. {\bf#1}, #2 (#3)}
\def \rmp#1#2#3{Rev. Mod. Phys. {\bf#1}, #2 (#3)}
\def \rp#1{~~~~~\ldots\ldots{\rm rp~}{#1}~~~~~}
\def \si90{25th International Conference on High Energy Physics, Singapore,
Aug. 2-8, 1990}
\def \slc87{{\it Proceedings of the Salt Lake City Meeting} (Division of
Particles and Fields, American Physical Society, Salt Lake City, Utah, 1987),
ed. by C. DeTar and J. S. Ball (World Scientific, Singapore, 1987)}
\def \slac89{{\it Proceedings of the XIVth International Symposium on
Lepton and Photon Interactions,} Stanford, California, 1989, edited by M.
Riordan (World Scientific, Singapore, 1990)}
\def \smass82{{\it Proceedings of the 1982 DPF Summer Study on Elementary
Particle Physics and Future Facilities}, Snowmass, Colorado, edited by R.
Donaldson, R. Gustafson, and F. Paige (World Scientific, Singapore, 1982)}
\def \smass90{{\it Research Directions for the Decade} (Proceedings of the
1990 Summer Study on High Energy Physics, June 25--July 13, Snowmass,
Colorado),
edited by E. L. Berger (World Scientific, Singapore, 1992)}
\def \tasi90{{\it Testing the Standard Model} (Proceedings of the 1990
Theoretical Advanced Study Institute in Elementary Particle Physics, Boulder,
Colorado, 3--27 June, 1990), edited by M. Cveti\v{c} and P. Langacker
(World Scientific, Singapore, 1991)}
\def \yaf#1#2#3#4{Yad. Fiz. {\bf#1}, #2 (#3) [Sov. J. Nucl. Phys. {\bf #1},
#4 (#3)]}
\def \zhetf#1#2#3#4#5#6{Zh. Eksp. Teor. Fiz. {\bf #1}, #2 (#3) [Sov. Phys. -
JETP {\bf #4}, #5 (#6)]}
\def \zpc#1#2#3{Zeit. Phys. C {\bf#1}, #2 (#3)}
\def \zpd#1#2#3{Zeit. Phys. D {\bf#1}, #2 (#3)}


\end{document}